\documentclass[11pt]{article}
\pdfoutput=1

\usepackage{jcappub}
\usepackage{mathrsfs,amssymb,amsmath,verbatim,mathtools,graphicx,bm,booktabs,tabularx,multirow,relsize,framed,orcidlink,comment,dcolumn,physics,enumitem}
\usepackage{xcolor}
\usepackage[T1]{fontenc}
\usepackage[utf8]{inputenc}
\usepackage{epstopdf}
\usepackage{aas_macros}
\usepackage{etoolbox}
\usepackage{latexsym}
\usepackage{longtable}
\usepackage{xltabular}
\usepackage{tabularx}
\usepackage{makecell}
\usepackage[table]{xcolor}
\usepackage{caption}
\usepackage{enumerate}
\usepackage{tensor}
\usepackage{url}
\usepackage{float}
\usepackage{nicefrac}
\usepackage{hyperref}
\usepackage{soul}
\definecolor{MidnightBlue}{RGB}{25, 25, 112}
\definecolor{MidnightBlueLight}{RGB}{239,239,251}
\definecolor{linkcolor}{rgb}{0.0,0.3,0.5}
\hypersetup{colorlinks=true, citecolor=linkcolor,
            linkcolor=linkcolor, urlcolor=blue, linktocpage=true}
\usepackage{acronym}

\acrodef{3g}[3G]{third-generation}

\acrodef{bbh}[BBH]{binary black hole}
\acrodefplural{bbh}[BBHs]{binary black holes}
\acrodef{bns}[BNS]{binary neutron star}
\acrodefplural{bns}[BNSs]{binary neutron stars}
\acrodef{nsbh}[NSBH]{neutron star-black hole}
\acrodefplural{nsbh}[NSBHs]{neutron star-black holes}
\acrodef{cbc}[CBC]{compact binary coalescence}
\acrodefplural{cbc}[CBCs]{compact binary coalescences}
\acrodef{ccsn}[CCSN]{core-collapse supernova}
\acrodefplural{ccsn}[CCSNe]{core-collapse supernov\ae{}}
\acrodef{imbh}[IMBH]{intermediate-mass black hole}
\acrodefplural{imbh}[IMBHs]{intermediate-mass black holes}
\acrodef{et}[ET]{Einstein Telescope}
\acrodef{ce}[CE]{Cosmic Explorer}
\acrodef{gw}[GW]{gravitational wave}
\acrodefplural{gw}[GWs]{gravitational waves}
\acrodef{ns}[NS]{neutron star}
\acrodefplural{ns}[NSs]{neutron stars}
\acrodef{bh}[BH]{black hole}
\acrodefplural{bh}[BHs]{black holes}
\acrodef{snr}[SNR]{signal-to-noise ratio}
\acrodefplural{snr}[SNRs]{signal-to-noise ratios}
\acrodef{sgwb}[SGWB]{stochastic gravitational-wave background}
\acrodefplural{sgwb}[SGWBs]{stochastic gravitational-wave backgrounds}
\acrodef{pls}[PLS]{Power-law integrated Sensitivity}
\acrodef{pbh}[PBH]{primordial black hole}
\acrodefplural{bbh}[BBHs]{binary black holes}
\acrodef{nr}[NR]{Numerical Relativity}
\acrodef{eos}[EoS]{Equation of State}
\acrodefplural{eos}[EoS]{Equations of State}
\acrodef{em}[EM]{electromagnetic}
\acrodef{cgw}[CGW]{Continuous Gravitational-Wave}
\acrodefplural{cgw}[CGWs]{Continuous Gravitational-Waves}
\acrodef{psd}[PSD]{power spectral density}
\acrodef{asd}[ASD]{amplitude spectral density}
\acrodefplural{psd}[PSDs]{power spectral densities}
\acrodef{fim}[FIM]{Fisher information matrix}
\acrodefplural{fim}[FIMs]{Fisher information matrices}
\acrodef{EoS}[EoS]{equation of state}
\acrodef{lwa}[LWA]{long-wavelength approximation}
\acrodef{lf}[LF]{low frequency}
\acrodef{hf}[HF]{high frequency}
\acrodef{IC}[IC]{intermediate-case}
\acrodef{WC}[WC]{worst-case}
\acrodef{FC}[FC]{filter cavity}
\acrodefplural{FC}[FCs]{filter cavities}
\acrodef{GRB}[GRB]{gamma-ray burst}
\acrodef{gr}[GR]{general relativity}

\usepackage[noabbrev]{cleveref} 
\crefname{section}{section}{sections}
\Crefname{section}{Section}{Sections}
\crefname{appendix}{appendix}{appendices}
\Crefname{appendix}{Appendix}{Appendices}
\crefname{footnote}{footnote}{footnotes}
\Crefname{footnote}{Footnote}{Footnotes}
\numberwithin{equation}{section}

\newcommand{\unige}{D\'epartement de Physique Th\'eorique, Universit\'e de Gen\`eve, 24 quai Ernest Ansermet, 1211 Gen\`eve 4, Switzerland}
\newcommand{\unihamburg}{Institut f\"ur Quantenphysik and Zentrum f\"ur Optische Quantentechnologien, Universit\"at Hamburg, Luruper Chaussee 149, 22761 Hamburg, Germany}
\newcommand{\gwsc}{Gravitational Wave Science Center (GWSC), Universit\'e de Gen\`eve, CH-1211 Geneva, Switzerland}
\newcommand{\mrs}{Aix-Marseille Universit\'e, Universit\'e de Toulon, CNRS, CPT, Marseille, France}
\newcommand{\GSSI}{Gran Sasso Science Institute (GSSI), I-67100 L’Aquila, Italy}
\newcommand{\GranSasso}{INFN, Laboratori Nazionali del Gran Sasso, I-67100 Assergi, Italy}
\newcommand{\jhu}{William H.\ Miller III Department of Physics and Astronomy, Johns Hopkins University, \\ 3400 N. Charles Street, Baltimore, Maryland, 21218, USA}
\newcommand{\mbi}{Marietta Blau Institute - Austrian Academy of Sciences, 1010 Vienna, Austria}
\newcommand{\UGent}{Universiteit Gent, B-9000 Gent, Belgium}
\newcommand{\sapienza}{Dipartimento di Fisica, Sapienza Università di Roma, Piazzale Aldo Moro 5, 00185, Roma, Italy}

\newcommand{\infnto}{INFN, Sezione di Torino, Via Pietro Giuria 1, 10125 Torino, Italy}
\newcommand{\Unito}{Dipartimento di Fisica, Università degli Studi di Torino, Via Pietro Giuria 1, 10125 Torino, Italy}
\newcommand{\torvergata}{Department of Physics, University of Rome Tor Vergata, via della Ricerca Scientifica 1, 00133, Rome, Italy}

\newcommand{\ita}{Institut f\"ur Theoretische Astrophysik, Zentrum f\"ur Astronomie, Universit\"at Heidelberg, Albert Ueberle Str. 2, D-69120 Heidelberg, Germany}
\newcommand{\infnNA}{INFN, Sezione di Napoli, I-80126 Napoli, Italy}
\newcommand{\unina}{Universit\`a di Napoli “Federico II”, I-80126 Napoli, Italy}

\newcommand{\unipd}{Dipartimento di Fisica e Astronomia ``G. Galilei'', Università degli Studi di Padova, via Marzolo 8, I-35131 Padova, Italy}
\newcommand{\infnpd}{INFN, Sezione di Padova, via Marzolo 8, I-35131 Padova, Italy}
\newcommand{\inafoab}{Istituto Nazionale di Astrofisica - Osservatorio Astronomico di Roma, Via Frascati 33, I-00040, Monteporzio Catone, Italy}
\newcommand{\ceico}{CEICO, Institute of Physics, Czech Academy of Sciences, Prague, Czechia}
\newcommand{\infngenova}{INFN, Sezione di Genova, Via Dodecaneso 33, 16146 Genova, Italy}

\graphicspath{figures/} 

\makeatletter
\def\@reportnumber{\@empty}
\newcommand{\reportnumber}[1]{\gdef\@reportnumber{#1}}

\renewcommand\ps@titlepage{%
  \renewcommand\@oddfoot{}%
  \renewcommand\@oddhead{%
    \if!\@reportnumber!\else%
      \hfill\raisebox{10pt}[\z@][\z@]{\@reportnumber}%
    \fi%
  }%
}
\makeatother

\begin{document}

\title{Assessing the Impact of Instrumental Requirements on the Scientific Performance of the Einstein Telescope}

\author[1]{Ulyana Dupletsa\,\orcidlink{0000-0003-2766-247X},}
\author[2]{Francesco Iacovelli\,\orcidlink{0000-0002-4875-5862},}
\author[3]{Mikhail Korobko\,\orcidlink{0000-0002-3839-3909},}
\author[4,5]{Valeria Sequino\,\orcidlink{0000-0002-8137-4797},}
\author[6]{Alessandro Agapito\,\orcidlink{0009-0005-9004-3163},}
\author[7,8,9]{Manuel Arca Sedda\,\orcidlink{0000-0002-3987-0519},}
\author[7,8]{Biswajit Banerjee\,\orcidlink{0000-0002-8008-2485},}
\author[10,11]{Nicolò Cibrario\,\orcidlink{0000-0003-3842-4493},}
\author[7,8]{Andrea Cozzumbo\,\orcidlink{0009-0004-2772-2692},}
\author[12]{Francesco Crescimbeni\,\orcidlink{0009-0001-4088-5443},}
\author[7,8]{Alessio Ludovico De Santis\,\orcidlink{0009-0005-4288-3758},}
\author[13,14]{Gabriele Franciolini\,\orcidlink{0000-0002-6892-9145},}
\author[15]{Yufeng Li\,\orcidlink{0000-0002-4135-074X},}
\author[6]{Michele Mancarella\,\orcidlink{0000-0002-0675-508X},}
\author[7,8,16]{Benedetta Mestichelli\,\orcidlink{0009-0002-1705-4729},}
\author[17,18]{Niccol\`o Muttoni\,\orcidlink{0000-0002-4214-2344},}
\author[7,8,9]{Lavinia Paiella\,\orcidlink{0009-0001-7605-991X},}
\author[19]{Ippocratis D. Saltas\,\orcidlink{0000-0002-0533-1285},}
\author[7,8]{Filippo Santoliquido\,\orcidlink{0000-0003-3752-1400},}
\author[7,8]{Pawan Tiwari\,\orcidlink{0000-0002-1414-2371},}
\author[7,8,9]{Cristiano Ugolini\,\orcidlink{0009-0005-9890-4722},}
\author[7,8]{Marica Branchesi\,\orcidlink{0000-0003-1643-0526},}
\author[20]{Archisman Ghosh\,\orcidlink{0000-0003-0423-3533},}
\author[7,8]{Jan Harms\,\orcidlink{0000-0002-7332-9806},}
\author[17,18]{Michele Maggiore\,\orcidlink{0000-0001-7348-047X},}
\author[21]{Fiodor Sorrentino\,\orcidlink{0000-0002-9605-9829}}

\affiliation[1]{\mbi}
\affiliation[2]{\jhu}
\affiliation[3]{\unihamburg}
\affiliation[4]{\unina}
\affiliation[5]{\infnNA}
\affiliation[6]{\mrs}
\affiliation[7]{\GSSI}
\affiliation[8]{\GranSasso}
\affiliation[9]{\inafoab}
\affiliation[10]{\infnto}
\affiliation[11]{\Unito}
\affiliation[12]{\sapienza}
\affiliation[13]{\unipd}
\affiliation[14]{\infnpd}
\affiliation[15]{\torvergata}
\affiliation[16]{\ita}
\affiliation[17]{\unige}
\affiliation[18]{\gwsc}
\affiliation[19]{\ceico}
\affiliation[20]{\UGent}
\affiliation[21]{\infngenova}

\emailAdd{ulyana.dupletsa@oeaw.ac.at}
\emailAdd{fiacovelli@jhu.edu}

\abstract{
We investigate the relationship between instrumental requirements and the scientific performance of the Einstein Telescope (ET), a third-generation (3G) gravitational-wave (GW) observatory. Different technical design choices result in distinct noise budgets, ultimately shaping the detector's scientific capabilities. To systematically assess and compare their impact, we define a comprehensive set of performance metrics spanning compact binary coalescence (CBC) detection and parameter estimation, as well as other sources, including stochastic GW backgrounds, isolated spinning neutron stars, and core-collapse supernov\ae{} (CCSNe).

We build a comparative reference framework that links degradations in specific noise contributions and frequency bands to losses in scientific capabilities. We consider a representative selection of technical parameters, such as coating and suspension temperatures, the filter cavity length in the low-frequency instrument, and the beam size in the high-frequency instrument. We evaluate how sensitivity variations across specific frequency bands affect different scientific objectives. 

We quantify how the sensitivity below $30$\,Hz impacts the detectability of massive and/or high-redshift sources and the reconstruction of long-duration CBC signals, affecting early warning and sky localization for binary neutron stars (BNSs). Sensitivity in the $30 - 450$\,Hz range governs most CBC parameter-estimation metrics, while high-frequency sensitivity above $\sim\!450$\,Hz predominantly impacts BNS post-merger studies and CCSN detectability, with modest effects on detection rates. Even with the most significant degradations considered, the ET science case remains robust overall. 

Our results provide a comprehensive benchmark linking scientific objectives to instrumental requirements, particularly important as the final design and infrastructure of 3G observatories are being defined.
}

\reportnumber{ET-0242A-26}

\maketitle
\flushbottom

\acresetall %
\section{Introduction}\label{sec:intro}

Modern \ac{gw} detectors like LIGO~\cite{LIGOScientific:2014pky}, Virgo~\cite{Virgo:2014yos}, and KAGRA~\cite{KAGRA:2020tym} opened a new window on our Universe, delivering direct observations of the \ac{gw} signal emitted by coalescing binary systems. With a catalog counting about 400 confident detections~\cite{LIGOScientific:2026wfs}, and more \ac{cbc} candidates soon to be published, \acp{gw} are emerging as a powerful tool for astrophysics, fundamental physics, and cosmology~\cite{KAGRA:2021duu,LIGOScientific:2025pvj,LIGOScientific:2026ctl,LIGOScientific:2021sio,LIGOScientific:2025rid,LIGOScientific:2017vwq,LIGOScientific:2021aug,LIGOScientific:2025jau,LIGOScientific:2026uyd}.

As current detectors approach their design sensitivity, further upgrades~\cite{fritschel2022postO5,LIGOScientific:2026sit} will face infrastructure and instrumental limitations, motivating the development of next-generation observatories such as the \ac{et}~\cite{Hild:2008ng, Punturo:2010zz, Hild:2010id} and \ac{ce}~\cite{Reitze:2019iox,Evans:2021gyd, Evans:2023euw}. These observatories are designed to achieve improvements in sensitivity and bandwidth, enabled by major technological advances and the construction of new large-scale infrastructures. This will enable a fundamental shift in the scientific potential, expanding the cosmological reach, increasing detection rates, and opening the possibility to detect phenomena not yet observed~\cite{Maggiore:2019uih,Evans:2021gyd,Branchesi:2023mws,Gupta:2023lga,ET:2025xjr}. 
Each year, \ac{et} will detect tens to hundreds of thousands of \acp{cbc} from \ac{bbh}, \ac{bns}, and \ac{nsbh} systems formed through different channels~\cite{ET:2025xjr,Borhanian:2022czq,Ronchini:2022gwk,Pieroni:2022bbh,Iacovelli:2022bbs,Begnoni:2025oyd}, compared to about 150 observations per year by current instruments~\cite{LIGOScientific:2025slb,LIGOScientific:2026wfs}. 
Many of these events will be detected with very high \ac{snr}, enabling precise reconstruction of source parameters, which will allow us to accurately infer their distribution~\cite{DeRenzis:2024dvx,Plunkett:2025mjr}, nature~\cite{Maggio:2021ans,Crescimbeni:2024cwh,Mastrogiovanni:2025ixe}, composition~\cite{Gupta:2022qgg,Iacovelli:2023nbv}, and formation pathways~\cite{Ng:2020qpk,Singh:2021zah}.
At the same time, high measurement precision will allow testing the theory of \ac{gr}~\cite{Perkins:2020tra,Bhagwat:2023jwv,Begnoni:2025mtz} and probing the expansion history of the Universe~\cite{Belgacem:2019tbw,Muttoni:2023prw,Tagliazucchi:2026dpr}. 
Crucially, \ac{et} is also expected to see \ac{gw} sources that have eluded direct detection so far, including continuous \acp{gw} from non-axisymmetric rapidly rotating \acp{ns} (among other sources); bursts from \acp{ccsn}, which are also prime multimessenger candidates~\cite{Muller:2026ofz, 2019BAAS...51c.122F, Zegarelli:2024ivy}; and \acp{sgwb} of both astrophysical and cosmological origin~\cite{Christensen:2018iqi, Caprini:2018mtu}. For a comprehensive review of the scientific targets and capabilities we refer the reader to Ref.~\cite{ET:2025xjr} and references therein.

Reaching this sensitivity level will require significant technological advancements. \ac{et} will have arms of $10 - 15$\,km, located deep underground, will exploit cryogenic cooling of its mirrors and suspensions, will use an order-of-magnitude-higher optical power in the arms and increased mirror masses, and will feature significant upgrades to every system, from lasers, suspensions, mirror materials to quantum noise suppression and detector control systems~\cite{fritschel2022postO5, ETconceptual2020}. Current science case scenarios are based on the best known estimates of all noises that will contribute to the sensitivity of the detector. It is foreseeable that some of these might deviate from current modeling, at least in a preliminary stage. Building on the commonly adopted metrics used to evaluate the performance of \ac{et} and \ac{ce} (see, e.g., Refs.~\cite{Branchesi:2023mws,ET:2025xjr, Gupta:2023lga}), we introduce a revised metric framework to evaluate the impact of specific design variations affecting different frequency ranges.

In this work, we create a framework that allows for tracking changes in the prospects of several significant science cases as the detector's sensitivity varies. We apply this framework to demonstrate the impact of changes to specific noise contributions arising from variations in representative design parameters. Furthermore, we present a ``traffic light'' system, where we analyze the impact of changes on sensitivity in a specific frequency range, independent of the details of which noise contributes to it. This will allow instrumental scientists to evaluate how any noise arising at a particular frequency would affect the scientific performance of \ac{et}. 

This work was conceived and carried out as part of the Einstein Telescope Organization (ETO)-driven process to define the \ac{et} design~\cite{taskforce}. The central question addressed here is how degradations from the baseline \ac{et} noise budget translate into losses across diverse science cases. We are interested not only in quantifying how instrumental changes affect scientific performance, but also in providing an inverse mapping from scientific objectives to detector sensitivity in different frequency bands~\cite{Davis:2025qjf}. Our goal is not to optimize the detector sensitivity itself or to present a preferred detector design. Instead, we aim at characterizing how scientific performance responds to changes in sensitivity, in order to comprehend design trade-offs. In this regard, our analysis serves as a diagnostic tool -- given a proposed sensitivity change, it allows one to quickly identify which science cases are most affected and which frequency ranges are most critical for the broader science case of \ac{et}. Note that, although we consider a broad and diverse range of scientific cases, our analysis does not prioritize any specific science case. %
The results presented here are intended as comparative diagnostics rather than absolute performance forecasts. Our goal is to quantify, within a controlled and uniform analysis framework, how variations in the assumed detector sensitivity propagate into a broad set of science metrics. For this reason, the same source populations, waveform models, detection thresholds, and methodology are used for all sensitivity curves and detector configurations. Some of the adopted metrics are approximate and may not provide faithful predictions in all regimes. Accordingly, these results should be read primarily as relative indicators of sensitivity-driven trends with respect to the chosen baseline. Within this scope, they provide a meaningful assessment of which science cases are most affected by specific instrumental degradations and which frequency ranges are most critical for the \ac{et} science case.

This paper is organized as follows. In Section~\ref{sec:noise_budget}, we present a detailed description of the \ac{et} noise budget considered in our analyses, how it varies when changing specific representative design parameters, and introduce our analysis considering degradations in individual frequency bins. In Section~\ref{sec:metrics_and_pops}, we describe the adopted metrics, separating them into general ones and population-based ones, and discuss the population models employed for the latter. In Section~\ref{sec:results}, we present the results of our analyses across all the chosen metrics for the selected noise budget degradations. Finally, Section~\ref{sec:discussion} and Section~\ref{sec:conclusions} are devoted to the discussion and conclusions, respectively.

\section{Instrumental constraints and ET noise budget}\label{sec:noise_budget}
Two detector geometries are currently being considered for \ac{et}: a 10\,km triangular ($\Delta$) configuration with three xylophone detectors, with each vertex hosting a central beam-splitter; and a double 15\,km L-shape design, with two xylophone detectors located at different geographical sites in Europe. These two configurations feature different arm lengths, $10$\,km for the former and $15$\,km for the latter, and different opening angles between the arms, $60^\circ$ vs $90^\circ$. When comparing these two configurations in the following, we will always refer to a 10-km $\Delta$ and 15-km 2Ls. This comparison should thus not be intended as an overall comparison of the two geometries, rather of the two configurations under active consideration by the \ac{et} Collaboration. For a comparison of the two geometries with the same arm length, we refer the reader to Ref.~\cite{Branchesi:2023mws}.

Independent of the adopted geometry, \ac{et} is conceived as a ``xylophone'' detector consisting of two independent nested %
co-located interferometers: \ac{lf}, optimized for \ac{gw} signals in the band $3\,{\rm Hz} \lesssim f \lesssim 30\,{\rm Hz}$ (see Appendix~\ref{app:minimum_freq} for a discussion on the role of the minimum frequency in our analysis), and \ac{hf}, ranging from $30\,{\rm Hz}$ up to several kHz. %
The two signals of the ET-LF and ET-HF are jointly processed to create a virtual broadband instrument~\cite{Hild:2009ns, Branchesi:2023mws}. ET-LF uses low optical power and cryogenic cooling to reduce the impact of thermal noise on sensitivity. ET-HF uses high power and operates at room temperature. 

The stationary performance of a \ac{gw} detector is defined by the noise sources that limit the ability to resolve signals. The sensitivity curve represents, for each frequency, the minimum \ac{gw} strain the detector can measure, expressed in terms of the amplitude spectral density. Current sensitivity is computed using the pyGWINC~\cite{2020ascl.soft07020R} software, developed independently in the context of LIGO design, and modified to meet \ac{et}-specific criteria. The models for the noises include a limited set of parameters, but nonetheless provide sufficient detail to serve as the benchmark for judging the impact of the changes on the scientific case. 

\begin{figure}[tbp]
    \centering
    \includegraphics[width=0.85\textwidth]{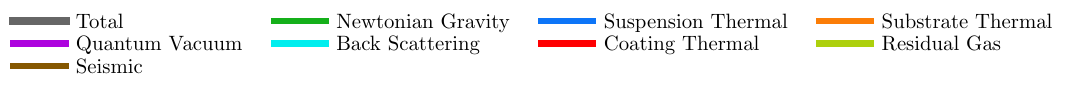}
    \vspace{0.5em}
     \begin{minipage}{0.48\linewidth}
        \includegraphics[width=1\linewidth]{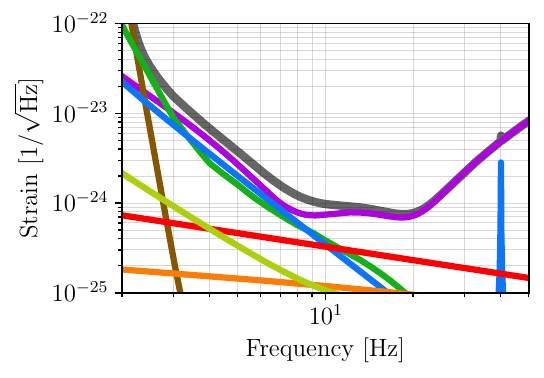}
    \end{minipage}
     \begin{minipage}{0.48\linewidth}
        \includegraphics[width=1\linewidth]{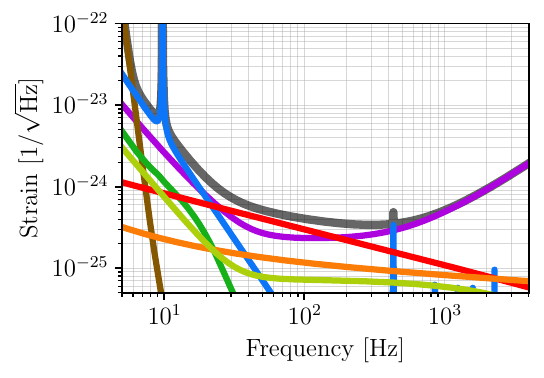}
    \end{minipage}
    \begin{minipage}{0.48\linewidth}
         \includegraphics[width=1\linewidth]{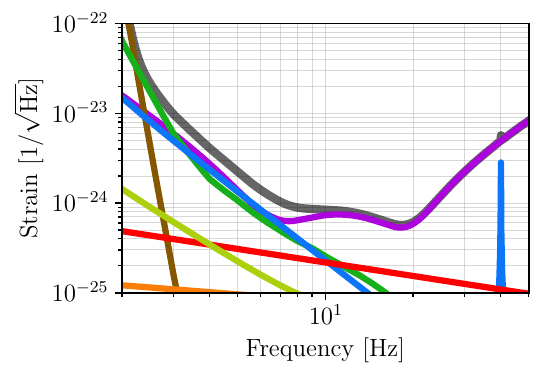}
    \end{minipage}
    \begin{minipage}{0.48\linewidth}
        \includegraphics[width=1\linewidth]{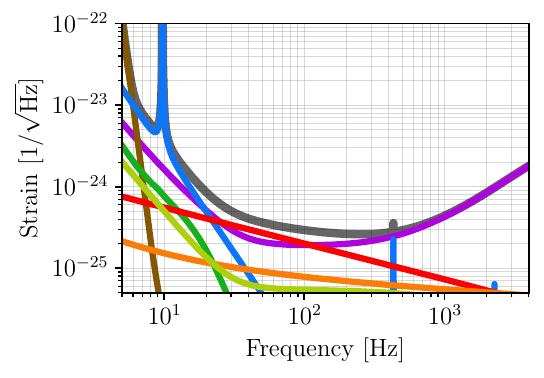}
    \end{minipage}
    \caption{Noise contributions for ET-LF (left) and ET-HF (right), considering a%
    detector with 10 km (top) and 15 km (bottom) arm lengths. The effects of the number of detectors and the opening angle between the arms in the $\Delta$ and 2L configurations are %
    taken into account in the science case analysis presented in this paper in the detectors' response evaluation.}
        \label{fig:noiseContrTotal}
\end{figure}
\Cref{fig:noiseContrTotal} shows that the major contributions to ET-LF are dominated by seismic, Newtonian, quantum, and suspension thermal noise, while those for ET-HF arise from coating thermal noise and quantum noise. Note that these sensitivity curves represent differential strain noise and are independent of the opening angle of the two arms. The direction of the arms is considered in the detailed response calculation of the detectors when computing the scientific metrics.

A realistic detector design is a tightly coupled system in which the performance of each subsystem, including its associated noise, depends strongly on others. Consequently, changing a single detector parameter would generally modify several noise contributions simultaneously. It could also affect detector controllability, commissioning procedures and the design of the required scientific and civil infrastructure. In this study, we do not include these full system-level interdependencies. Instead, we consider an isolated increase in a particular noise source, or noise in a total noise level within a specific frequency band. 
We represent this increase by a single parameter. This parametrization should not be interpreted as directly adjustable detector parameter: in a realistic design, achieving such change would likely imply a substantial redesign and re-optimization of the detector as a whole, which is outside the scope of this study.

The analysis in this paper considers three illustrative operational scenarios: Baseline, \ac{IC}, and \ac{WC}. We emphasize that the IC and WC scenarios are defined solely for the purposes of this study. They are not intended to represent a realistic projection of technological development or the operational conditions expected for \ac{et}. Rather, they provide examples for assessing how instrumental constraints affect the science case and for demonstrating the workflow that connects such constraints to their scientific consequences.  We chose four representative noise contributions to study this impact: ET-LF quantum noise, ET-LF coating thermal noise, ET-LF suspension thermal noise, and ET-HF coating thermal noise. This choice allows us to capture the effects of noise contributions with sufficiently distinct and non-trivial frequency dependencies while keeping the scope of the study manageable.

\textbf{Quantum noise} arises from quantization of the light field in the detector, which leads to measurement (shot) noise in the detected signal and quantum radiation-pressure noise due to fluctuating amplitude of the laser field acting on the mirrors~\cite{Caves:1980pha,Caves:1981hw,Danilishin:2012fa}. In ET-LF, this radiation pressure is further exploited to create an optomechanical resonance boosting the sensitivity~\cite{Buonanno:2001kr,Harms:2003hn}. Together with frequency-dependent quantum squeezing~\cite{Kimble:2000gu,Virgo:2023wes,LIGOO4Detector:2023wmz,Jia:2024iqe}, this allows to reach exceptional performance at low frequencies~\cite{2025Galax..13...11K}. Quantum noise is very fragile to any source of quantum decoherence~\cite{Schnabel2017}. Optical loss in filter cavities, used to create frequency-dependent phase rotation on the squeezed field, is among the key contributions to quantum squeezing degradation~\cite{Kwee:2014vba}. The effective loss scales with length: longer cavities allow to reduce decoherence. The baseline level of 5\,km is optimized for achieving highest quantum advantage while constraining the total length, which is a significant driver of the overall detector design. IC and WC values have direct impact on the loss of quantum squeezing and thus on low-frequency sensitivity, particularly in the $5 - 10$\,Hz frequency range~\cite{2025Galax..13...11K}.

\textbf{Coating thermal noise} arises from thermal motion of molecules in the material, which translates into effective change of the mirror surface distance, and, averaged over the beam size, contributes directly to the phase noise on the reflected laser field~\cite{Harry:2001iw}. This noise depends explicitly on the temperature of the coating and the beam size (due to the averaging effect). Coating thermal noise contributes both to LF and HF operation. For ET-LF, cryogenic operation is envisioned with a baseline temperature of 10\,K. While the IC temperature of 20\,K may potentially be accommodated without major changes to the detector design, the WC temperature of 70\,K would likely require substantial changes to the test-mass and suspension designs, potentially to the material choice, and to the overall cryogenic infrastructure. For ET-HF, cryogenic operation is not envisaged. Its coating thermal noise could nevertheless increase if a smaller beam size were adopted, because this would reduce the spatial averaging over the mirror surface.

\textbf{Suspension thermal noise} has the same fundamental origin: thermally driven microscopic motion in the suspension material~\cite{Koroveshi:2023fwk}. This motion exerts a fluctuating force on the test mass, producing a displacement that appears as a phase shift on the optical field. The mechanical response of the suspension chain strongly filters this force, giving the resulting thermal noise a pronounced frequency dependence with its dominant contributions at very low frequency. As with the coating thermal noise for ET-LF, we parametrize this increase through the cryogenic temperature. The practical detector-design implications of such a change are beyond the scope of this study.

\begin{figure}[tb]
    \centering
     \begin{minipage}{0.48\linewidth}
        \includegraphics[width=1\linewidth]{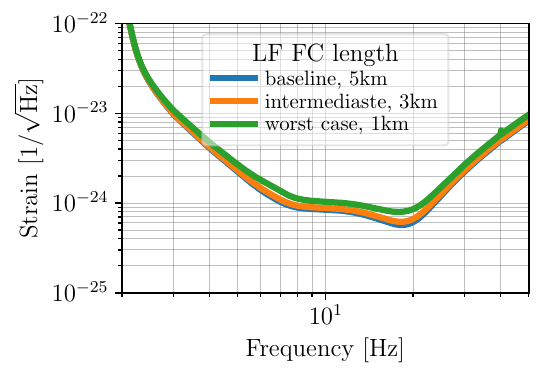}
    \end{minipage}
     \begin{minipage}{0.48\linewidth}
        \includegraphics[width=1\linewidth]{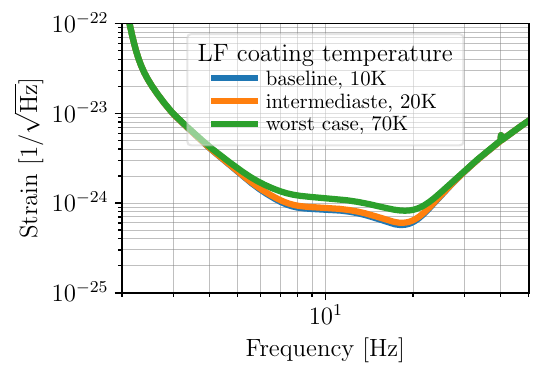}
    \end{minipage}
    \begin{minipage}{0.48\linewidth}
        \includegraphics[width=1\linewidth]{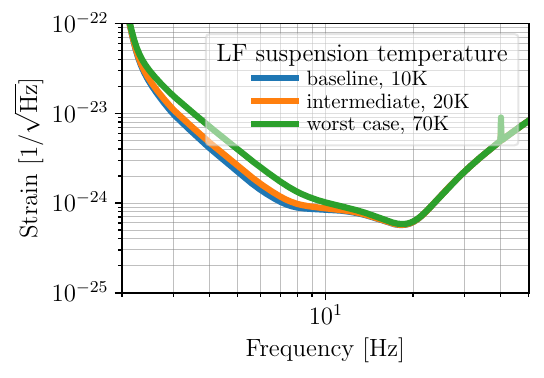}
    \end{minipage}
    \begin{minipage}{0.48\linewidth}
        \includegraphics[width=1\linewidth]{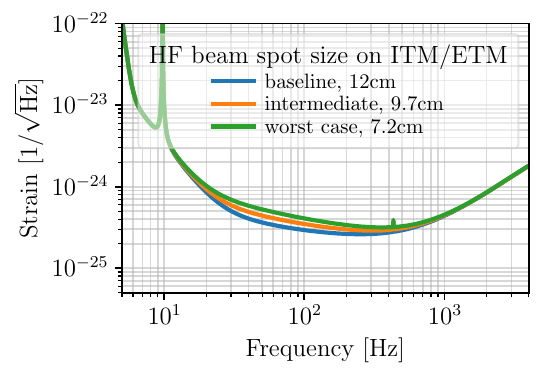}
    \end{minipage}
    \caption{
    Sensitivity scenarios for different noise contributions considered in \Cref{tab:parameters}, for a 15\,km detector as an example. \textit{Top left:} impact of quantum noise degradation of ET-LF, parametrized by the filter cavities length; \textit{top right:} impact of coating thermal noise degradation of ET-LF, parametrized by coating temperature; \textit{bottom left:} impact of suspension thermal noise degradation of ET-LF, parametrized by coating temperature; \textit{bottom right:} impact of coating thermal noise degradation of ET-HF, parametrized by the beam size on the mirrors.}
        \label{fig:noise_scenarios}
\end{figure}

The parameter values adopted in this analysis are schematized in \Cref{tab:parameters} and the resulting noise budgets are shown in \Cref{fig:noise_scenarios}.

\begin{table}[tb]
\caption{Interferometer design parameter values used in the analysis. We report the numbers for the chosen noise contributions in the baseline configuration, the intermediate and the worst-case scenarios.
}
    \centering
    \setlength{\tabcolsep}{2.5pt} 
    \begin{tabular}{lccc}
    \toprule
    \bf{Design parameter} & \bf{Baseline} & \bf{Intermediate-Case (IC)} & \bf{Worst-Case (WC)} \\
    \midrule
    ET-LF filter cavity length & 5\,km  & 3\,km & 1\,km \\
    \rowcolor{MidnightBlueLight} ET-LF coating temperature & 10\,K & 20\,K & 70\,K \\
    ET-LF suspension temperature & 10\,K & 20\,K & 70\,K\\
    \rowcolor{MidnightBlueLight} ET-HF beam size & 12\,cm & 9.6\,cm (80\%) & 7.2\,cm (60\%) \\
    \bottomrule
    \end{tabular}
    \label{tab:parameters}
\end{table}

\begin{figure}[tbp]
    \centering
    \includegraphics[width=0.9\textwidth]{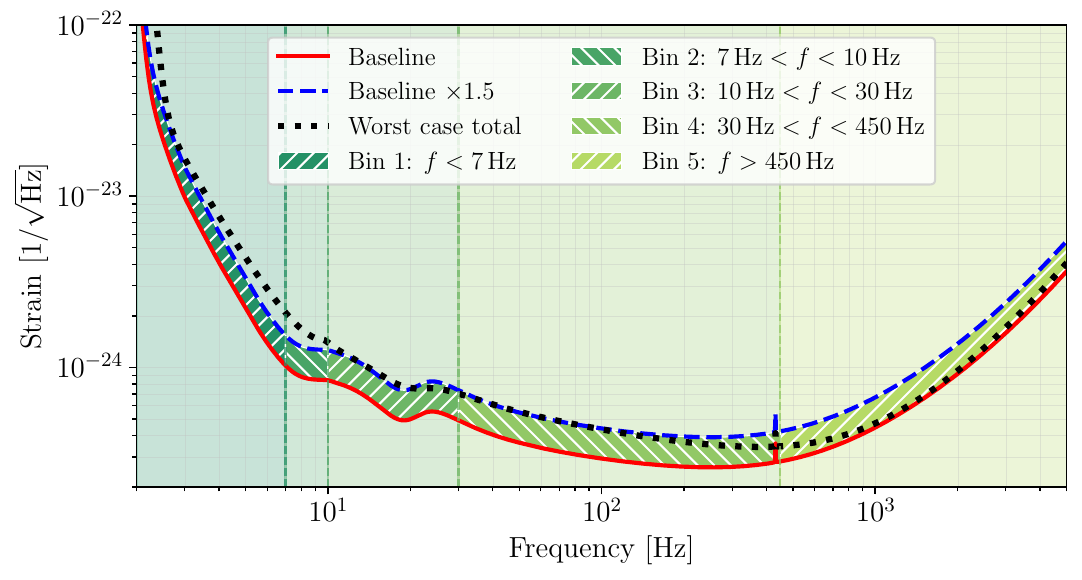}
    \caption{We show the five different sensitivity curves obtained by degrading the amplitude of the Baseline for 15\,km arm length by a factor of 1.5 in each frequency bin delimited by the green vertical lines. For reference, we display the Baseline sensitivity (in red) and the overall Baseline $\times 1.5$ curve (in blue). The different shades of green represent the binned sensitivities: in each selected bin, the sensitivity is degraded from the Baseline to the Baseline $\times 1.5$, while in the remaining bins it remains equal to the Baseline. For comparison, we also include the Worst case total scenario (in black).} 
    \label{fig:binned_psd_alternative}
\end{figure}

We further introduce a ``traffic light'' system, considering a sensitivity degradation by a factor of 1.5 in amplitude, agnostic of the specific noise source in a given frequency bin. This sets an example of how a particular threshold sensitivity degradation that could be considered as a ``red light'' from the instrument perspective impacts the science case. The particular degradation factor is chosen for illustration purposes to approximately match the total WC degradation over a broad frequency band. The frequency bins are chosen to refer to various physical sub-systems dominating the sensitivity. We list some major drivers of the noise budget in each bin (this list is non-exclusive):

\begin{itemize}
    \item \textbf{$\bm{f < 7}$\,Hz (ET-LF)}: parameters of the suspension chain; seismic and acoustic environment; material parameters and cryogenic cooling; quantum noise reduction subsystem; optical power in the arms; control noises (not considered in the current model).
    \item \textbf{$\bm{7-10}$\,Hz (ET-LF)}: impact on quantum noise through optical power in the arms and quality of filter cavities; impact on thermal noise and cryogenic cooling.
    \item \textbf{$\bm{10-30}$\,Hz (ET-LF)}: impact of the optical tuning and optical bandwidth of ET-LF on quantum noise.
    \item \textbf{$\bm{30-450}$\,Hz (ET-HF)}: parameters of the suspension chain; seismic and acoustic environment; optical power in the arms; quantum noise reduction subsystem.
    \item \textbf{$\bm{f > 450}$\,Hz (ET-HF)}: optical power in the arms; detection bandwidth of the detector; quantum noise reduction subsystem.
\end{itemize}
The chosen frequency bins and rescaled noise curve are illustrated in \Cref{fig:binned_psd_alternative}.

This binned analysis was performed for the 15 km, 2L configuration only; analogous results are expected for the $\Delta$ geometry in terms of relative comparisons of the metrics within one configuration. 

For both \ac{et} configurations, we used the same settings as in Ref.~\cite{Branchesi:2023mws}, for definiteness. We place the triangular \ac{et} in Sardinia at the Sos Enathos site, while for the 2L configuration, one detector is also in Sardinia and the other in the Meuse-Rhine Euroregion. Regardless of the configuration, we consider all detectors to have the same noise budget. In the 2L case, we considered the two interferometers misaligned by $45^{\circ}$. For details on the detectors' position and relative orientation, refer to Ref.~\cite{Branchesi:2023mws}.\footnote{For the 2L configuration, we expect that, choosing the more recently proposed Lusatia site instead of the one in the Meuse-Rhine Euroregion would have a limited impact on our analysis, given the similar baseline ($\sim\!1200$\,km) between the two detectors. Choosing the Lusatia site instead of the Sardinia one would negatively affect the parameter estimation, since the baseline in this case is halved.} We provide a short comparison between the sensitivities presented in Ref.~\cite{Branchesi:2023mws} and the Baseline used in this paper in Appendix~\ref{app:sens_curves}. Finally, in our analyses we neglect correlated noise between the detectors, mostly relevant for ET-LF. The relevant correlated components differ for the two configurations: in the triangular design they include common Newtonian and magnetic noise, while in the 2L design seismic/Newtonian noise is effectively uncorrelated between sites given their distance (except for large earthquakes), with only global magnetic fields (Schumann resonances) being correlated~\cite{Janssens:2024jln}.

\section{Metrics to assess detector performance}\label{sec:metrics_and_pops}

Given the breadth of its scope, assessing \ac{et}'s scientific output and reducing it to simple numbers is not straightforward, if feasible at all. For example, forecasts for \ac{cbc} populations will inevitably depend on our still limited astrophysical knowledge and thus carry significant uncertainties. More importantly, all science cases are extremely valuable, and prioritizing one over another should be avoided. Furthermore, present-day forecasts rely on our current understanding of possible \ac{gw} sources, but \ac{et} has the potential to be a discovery machine, revealing unexpected, transformative events. With these caveats in mind, we will evaluate the scientific performance of different experimental sensitivities using a broad, computationally efficient, yet meaningful set of metrics relevant across the full frequency spectrum, divided into general metrics and population metrics. General metrics are largely independent of specific astrophysical assumptions and characterize the fundamental sensitivity and reach of the detector across different signal morphologies and frequency bands. They provide a robust baseline for comparing different detector configurations. Population metrics, on the other hand, depend on the adopted astrophysical model and source distributions, thus providing estimates of detection rates and parameter estimation capabilities for specific source classes. While these may depend sensitively on uncertainties related to our incomplete understanding of source populations, they still offer valuable insights into \ac{et}'s potential for addressing key astrophysical questions. Furthermore, the scope of our work is a relative comparison with respect to the Baseline sensitivity. Consequently, population and waveform modeling uncertainties, while important to highlight, play a secondary role in our analysis. In what follows, we give an overview of the considered metrics.

\begin{description}[align=left]
\item[General metrics:] We consider a set of metrics independent of any assumptions regarding the source distribution, including:
    \begin{description}[align=left]
        \item[Detection horizons:] We report the maximum redshift out to which a non-spinning \ac{cbc} source with an optimal sky position and orientation can be detected with a given \ac{snr}, as a function of its source-frame total mass. We focus on equal-mass non spinning binaries. The horizon represents the limiting distance for detection and scales with the detector's sensitivity across different frequency bands and the width of the frequency band itself.
        \item[\ac{snr} in the post-merger phase of \ac{bns} systems:] \ac{et}'s high-frequency sensitivity will allow us to observe and characterize the post-merger signal of mergers involving two \acp{ns}, which is tightly linked to the internal structure of the objects and the remnant. We therefore compute the \ac{snr} in the post-merger only for some selected \ac{nr} simulations, spanning different \acp{eos} and masses for the two objects. The post-merger signal contains information about the hot \ac{ns} \acp{eos}, in particular the maximum allowed \ac{ns} mass~\cite{Bauswein:2013jpa,Koppel:2019pys,Agathos:2019sah}, and gives complementary information to that extracted from the inspiral phase.
        \item[Power-law integrated sensitivity to \acp{sgwb}:] A common metric we use to assess the capability of a detector network to detect a \ac{sgwb} with power-law behavior through a cross-correlation search is the \ac{pls} curve. %
        \item[Minimum detectable ellipticity for pulsars:] One of the metrics we will use to assess the detectability of \acp{cgw} emitted from non-axisymmetric pulsars is the minimum detectable ellipticity for broadband all-sky searches, assuming a search pipeline adapted to the \texttt{FrequencyHough}~\cite{2014PhRvD..90d2002A}. The detectability depends on the integration time and the pulsar's rotational frequency, along with the detector sensitivity.
        \item[Maximum detectability distance for \ac{ccsn} signals:] An as-yet elusive and extremely interesting class of events detectable through their \ac{gw} emission are \acp{ccsn}. To assess the detectability for these kinds of systems we report the maximum distance out to which some selected signals from 3D simulations would be observable, assuming a perfect recovery template. We note that the \ac{gw} emission from \acp{ccsn} is stochastic and depends on the stellar structure and properties, making these signals particularly challenging to detect with template-based searches. For this reason, our results should be regarded as optimistic if taken as absolute values, but they still offer a meaningful comparative benchmark.
    \end{description}
\item[Population metrics:] We also consider a set of metrics depending on specific representative choices for the distribution of \ac{cbc} sources across different classes, including:
    \begin{description}[align=left]
        \item[Number of detections of \acp{bbh} and \acp{bns}:] For two different classes of sources and different formation scenarios for \acp{bbh}, we report the number of detectable sources with a cut on their \ac{snr}. The detection rates rely completely on the assumed representative astrophysical population models.\footnote{Due to the high computational cost of running analyses on two detector configurations and for a large set of noise curves, we do not compute detection rates and statistical uncertainties on the source parameters for \ac{nsbh} systems. We expect the detection rates prospects for this class of systems to fall between the rates we have for \acp{bbh} and \acp{bns}; see, e.g., Ref.~\cite{Iacovelli:2022bbs}.}
        \item[Parameter estimation uncertainties for \acp{bbh} and \acp{bns}:] For systems generated by Population I and II progenitors, we report the statistical uncertainties achievable on the source parameters through a \ac{fim} analysis. We emphasize that the \ac{fim} results should be interpreted with some caveats~\cite{Vallisneri:2007ev,Rodriguez:2013mla}: the Fisher approximation has some inherent limitations and cannot capture multimodalities, or handle strong parameter correlations, or prior-boundary effects~\cite{Iacovelli:2022bbs,Dupletsa:2024gfl,Pandey:2024mlo,Santoliquido:2025aiq, Santoliquido:2025lot, Negri:2026clm}. Its quantitative accuracy can therefore depend on the detector network geometry and on the degree of independent information available in the data; in particular, it may be less reliable for configurations with nearly colocated instruments. We use the \ac{fim} results primarily as relative indicators of how parameter-estimation performance changes under modifications of the sensitivity curve, rather than as absolute uncertainty forecasts. In this case, this approach can yield meaningful results for population studies within a reasonable time for the large sets of sensitivities and configurations we consider.
        \item[\ac{bns} pre-merger alert and localization:] Due to its low-frequency sensitivity, \ac{et} will have improved capabilities of delivering early alerts to \ac{bns} mergers, which may emit an \ac{em} counterpart. This is of pivotal importance for multi-messenger studies, as it gives the opportunity to point timely telescopes at the event and increase the chances of counterpart identification. We thus provide the numbers of detections at different times before the merger, as well as the sky localizations at that time. Moreover, early warnings could enable real-time \ac{em} observations of the merger and its immediate aftermath, potentially shedding light on the counterpart emission processes through observation of the bright multi-wavelength prompt and the early afterglow. 
        \item[Number of detections of \acp{cgw}:] Using a specific catalog of sources, we compute the number of detectable \ac{cgw} signals for fixed values of the ellipticity, assuming a targeted search strategy.
    \end{description}
\end{description}

\noindent In Section~\ref{subsec:cbcpops} we will describe our specific choices for the distributions of masses, spins, and redshifts of the chosen classes of \ac{cbc} sources. The luminosity distances are then computed from the redshifts assuming a \textsc{Planck18} cosmology~\cite{Planck:2018vyg} as implemented in the \texttt{astropy} package~\cite{2022ApJ...935..167A}, which is used throughout this work.
The remaining parameters for the \ac{cbc} population analyses, namely sky position angles ($\alpha$ and $\delta$), inclination angle ($\iota$), polarization angle ($\psi$), time and phase at coalescence ($t_c$ and $\Phi_c$) are drawn from uninformative uniform distributions; i.e., uniform on the unit sphere for $\alpha$ and $\delta$, uniform in cosine for $\iota$, uniform in $[0,\,\pi]$ and $[0,\,2\pi]$ for $\psi$ and $\Phi_c$, respectively, and uniform in GPS time over one year time for $t_c$.

\subsection{Populations of compact binaries}\label{subsec:cbcpops}
In the following sections, we provide all the simulation details of the adopted \ac{cbc} population sources, namely \acp{bns} and \acp{bbh}. For the latter, we consider \acp{bbh} from Population I and II stars, and as-yet-unobserved \acp{bbh} from putative populations of Population~III (Pop.~III) stars and \acp{pbh}.

In \Cref{tab:pop_properties}, we summarize the main properties of the adopted \ac{cbc} populations. In particular, we report the local rate and the total number of astrophysical events for one year of observations. 

\begin{table}[t]
\centering
\caption{Summary of main properties of the used \ac{cbc} populations. For each type, we provide the local merger rate,  $R_0$,\textsuperscript{*} and the total number of astrophysical mergers, $N_{\rm events}$, corresponding to one year of observation. We also provide the relevant references. [\textsuperscript{*}For Population~III (Pop.~III) events the merger rate is estimated at $z \sim 15$.]
}
\label{tab:pop_properties}
\setlength{\tabcolsep}{6.2pt} 
\begin{tabularx}{\textwidth}{c c c X}
\toprule
\bf{Population} & $\bm{R_0\,[{\rm\bf Gpc}^{-3}\,{\rm\bf yr}^{-1}]}$ & $\bm{N_{\rm events}}$ & \multicolumn{1}{c}{\bf{Main features \& references}} \\
\midrule
\multirow{2}{*}{BBH (IMBH)}

& \multirow{2}{*}{$\sim\!12$ ($\sim\!0.04$)} 
& \multirow{2}{*}{$\sim\!1.3 \times 10^5 $ ($\sim\!400$)} %
& {\sc b-pop}~\cite{ArcaSedda:2018cyl, Sedda:2020vwo, Sedda:2021vjh}, isolated + dynamical channels \\

\rowcolor{MidnightBlueLight}
\multirow{2}{*}{BNS} 
& \multirow{2}{*}{$\sim\!107$} 
& \multirow{2}{*}{$\sim\!3.5 \times 10^5$} %
& \textsc{sevn}~\cite{Iorio:2022sgz,Costa:2023xsz} + \textsc{cosmo}$\mathcal{R}$\textsc{ate} \cite{Santoliquido:2020axb,Santoliquido:2020bry,Santoliquido:2023wzn}, Gaussian mass distribution, small spins, $\alpha=1$ \\

\multirow{2}{*}{Pop.~III} 
& \multirow{2}{*}{$\sim\!200^*$} 
& \multirow{2}{*}{$\sim\!3.4 \times 10^3$} %
& \textsc{fastcluster}~\cite{Mapelli:2021syv,Mapelli:2021gyv} + \textsc{sevn} \cite{Iorio:2022sgz,Costa:2023xsz} + \textsc{ cosmo}$\mathcal{R}$\textsc{ate}~\cite{Santoliquido:2020axb,Santoliquido:2020bry,Santoliquido:2023wzn}, isolated + dynamical channels \\

\rowcolor{MidnightBlueLight}
\multirow{2}{*}{PBH} 
& \multirow{2}{*}{$\sim\!89$} 
& \multirow{2}{*}{$\sim\!1.9\times 10^6$} %
& Wide mass range in $[10^{-2}, 10^3] M_\odot$, small spins~\cite{Franciolini:2023opt} \\

\bottomrule
\end{tabularx}
\end{table}

\subsubsection{Binary neutron stars}
\label{sec:bns}

\begin{figure}[tbp]
    \centering
    \includegraphics[width=0.5\linewidth]{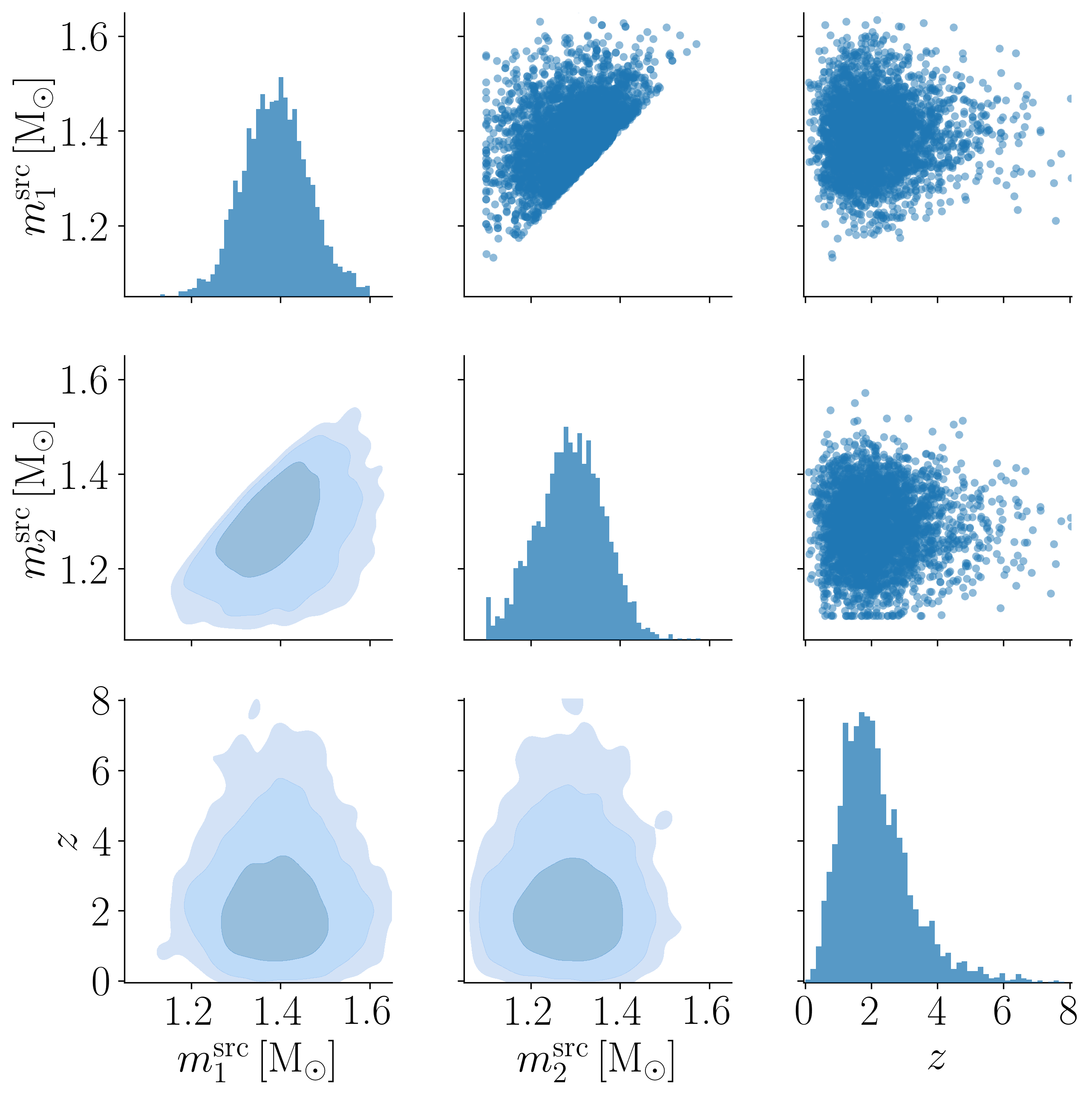}
    \caption{Marginalized one-dimensional (diagonal panels) and two-dimensional (bottom panels) distributions, together with individual samples (top panels), of the source-frame primary and secondary masses and redshift for the adopted \ac{bns} merger population. Contours correspond to the 68\%, 95\%, and 99\% credible intervals. Further details are provided in Section~\ref{sec:bns}.}
    \label{fig:bns}
\end{figure}

The \ac{bns} mergers were simulated using the population-synthesis code~{\sc sevn}~\cite{Spera:2018wnw, Mapelli:2019ipt, Iorio:2022sgz}. We employed the fiducial setup in which all {\sc sevn} parameters follow their default configuration, as detailed in Ref.~\cite{Iorio:2022sgz}, and adopted a common-envelope ejection efficiency of $\alpha = 1$. 

To compute the redshift-dependent merger rate density, we made use of {\sc cosmo}$\mathcal{R}${\sc ate}~\cite{Santoliquido:2020bry,Santoliquido:2020axb}. This tool models the cosmic star-formation history and metallicity evolution according to Ref.~\cite{Madau:2016jbv}, assuming a metallicity dispersion of $\sigma_Z = 0.2$. The masses of the neutron-star components were randomly sampled from two Gaussian distributions, following Ref.~\cite{Loffredo:2024gmx}. Each Gaussian is centered at $1.33\,M_\odot$ with a standard deviation of $0.09\,M_\odot$, consistent with fits to observed Galactic \acp{bns}~\cite{Ozel:2015fia,Kiziltan:2013oja,Ozel:2016oaf}. This model is, however, in tension with current \ac{bns} merger observations, which favor a broader mass distribution~\cite{KAGRA:2021duu}, a preference driven mainly by GW190425~\cite{LIGOScientific:2020aai}. We assumed slowly rotating neutron stars with aligned spins uniform in the range $-0.05 < \chi_{1,2} < 0.05$, consistent with the expectation that NSs in binaries have negligible spins~\cite{Burgay:2003jj}. The tidal deformabilities of the two components, $\Lambda_1$ and $\Lambda_2$, were computed from the component source-frame masses according to the \texttt{PCP(BSK24)} \ac{EoS}\footnote{The data were obtained from the \href{https://compose.obspm.fr/eos/253}{\textsc{CompOSE}} database~\cite{Typel:2013rza}. For more information, please see the references provided there.}, consistent with the latest constraints from the LIGO--Virgo--KAGRA observations~\citep{LIGOScientific:2018cki}.

\Cref{fig:bns} shows the distributions of masses and redshifts for the simulated population of \ac{bns} mergers.

\subsubsection{Binary black holes}
\label{sec:bbh}

\begin{figure}[tb]
    \centering
    \includegraphics[width=0.95\linewidth]{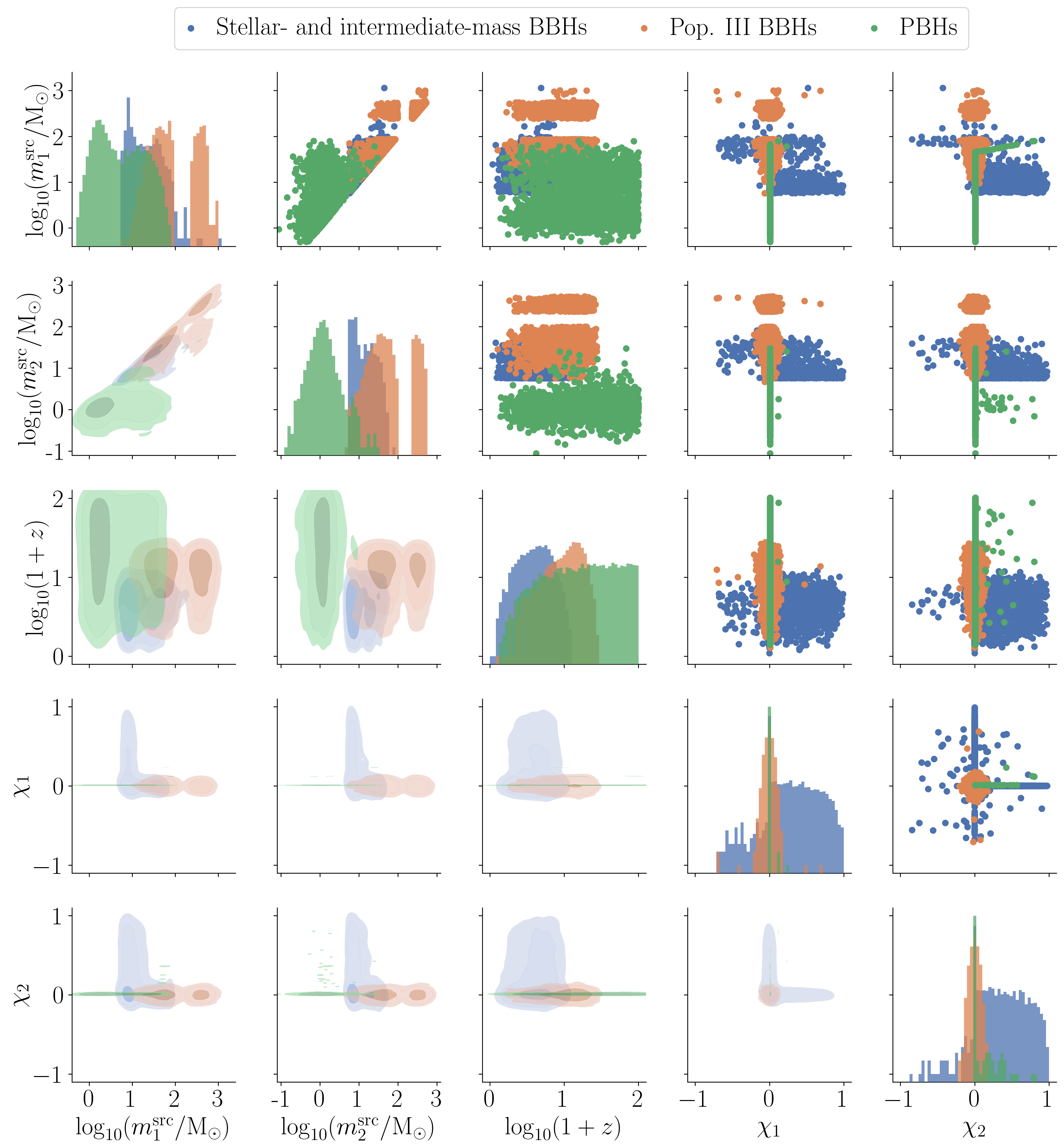}
    \caption{Marginalized one-dimensional (diagonal panels) and two-dimensional (bottom panels) distributions, together with individual samples (top panels), of the source-frame primary and secondary masses, redshift, and the first and second aligned spins for the adopted \ac{bbh} merger populations. Blue contours refer to \acp{pbh}, orange to stellar- and intermediate-mass \acp{bbh}, while green contours correspond to Pop.~III \acp{bbh}. Contours represent the 68\%, 95\%, and 99\% credible intervals. Further details are provided in Section~\ref{sec:bbh}.}
    \label{fig:bbh}
\end{figure}

\paragraph{Stellar- and intermediate-mass BBHs} The stellar- and intermediate-mass \ac{bbh} populations were simulated with the {\sc b-pop} code~\citep[Binary merger POPulations;][]{ArcaSedda:2018cyl, Sedda:2020vwo, Sedda:2021vjh}. {\sc b-pop} is a semi-analytic software that models \ac{bbh} mergers forming from isolated stellar binaries (IBs) and from dynamical interactions in young clusters (YCs), globular clusters (GCs), and nuclear clusters (NCs). The initial \ac{bh} properties (mass, formation time, and natal kicks) are drawn from catalogs of single and binary stars evolved with {\sc sevn}, following the prescriptions of Ref.~\cite{Iorio:2022sgz}. The catalogs account for metallicities in the interval $Z=0.0002 -0.03$. BHs are then distributed across the different environments according to the cosmic star-formation history and metallicity evolution~\cite{Madau:2016jbv}.

The fraction of stellar mass forming in clusters follows Ref.~\cite{Madau:2016jbv}, rescaled by factors $f_{\rm YC}=0.01$ and $f_{\rm NC}=0.0005$ for YCs and NCs, respectively, while the star-formation history of GCs follows Ref.~\cite{2019MNRAS.482.4528E}.

{\sc b-pop} accounts for a wide range of initial cluster masses and sizes, as well as their long-term structural evolution. The code also models the formation of binaries through dynamical encounters and the growth of \acp{bh} through stellar collisions and interactions.

\ac{bh} natal spins are assigned according to their formation history~\cite{Bavera:2020inc}. \acp{bh} formed from single stars and first-born \acp{bh} in binaries are assumed to have negligible spin~\cite{Fuller:2019sxi}, while second-born \acp{bh} and \acp{bh} produced by stellar collisions are assigned a uniform spin-magnitude distribution between 0 and 1. For dynamically formed binaries, spin vectors are isotropically oriented. For IBs, the polar angles $\theta_1$ and $\theta_2$ are extracted from a cumulative distribution $P(\theta_i|n) = [(\cos\theta_i +1)/2]^{n +1}$~\cite{ArcaSedda:2018cyl}, where $n$ determines the degree of alignment of the spins. For this work, we assumed $n_\theta=8$, which implies that in $55\%\, (20\%)$ of systems the two polar angles differ by less than $20\%\, (5\%)$; see Ref.~\cite{Sedda:2020vwo, Sedda:2021vjh}.

Multiple-generation mergers in clusters are treated using numerical-relativity fitting formulae to compute the remnant mass and spin~\cite{Jimenez-Forteza:2016oae} and the associated gravitational recoil~\cite{Campanelli:2007cga, Gonzalez:2006md, Lousto:2007db, Lousto:2012su}.

\Cref{fig:bbh} shows the resulting distributions of masses, redshifts and spins of the simulated stellar- and intermediate-mass \ac{bbh} mergers.

\paragraph{Population III BBHs}

The \ac{bbh} mergers generated from Pop.~III stars were simulated considering both the dynamical and the isolated formation channels. For the dynamical channel, we used the semi-analytic code {\sc fastcluster}~\cite{Mapelli:2021gyv,Vaccaro:2023cwr,Torniamenti:2024uxl}, which follows the hardening of binaries and their gravitational-wave driven evolution while self-consistently modeling the main physical processes governing the host star cluster~\cite{Mestichelli:2024djn}. The input catalogs of single and binary black holes, along with the isolated \ac{bbh} population, were produced with {\sc sevn}~\cite{Iorio:2022sgz,Costa:2023xsz}. As initial conditions, we adopted a log-flat initial mass function~\cite{Stacy:2012iz,Susa:2014moa,Hirano:2015wxa,Chon:2021jlx,Tanikawa:2020cca,Jaura:2022sny,Prole:2021nym} and the orbital-parameter distributions from Ref.~\cite{Sana:2012px}.  

We considered two formation channels within clusters: the evolution of primordial binaries (Orig) and three-body interactions among single black holes (Dyn). The initial spin magnitudes $\chi_1$ and $\chi_2$ of first-generation \acp{bbh} were drawn from a Maxwellian distribution with $\sigma_{\chi}=0.5$, truncated at $\chi=1$. This simplified model reproduces the main features of the LIGO--Virgo population~\cite{KAGRA:2021duu}. Spin-orientation angles $\theta_1$ and $\theta_2$ were sampled isotropically, as dynamical encounters efficiently erase any initial alignment~\cite{Rodriguez:2016vmx}. For each \ac{bbh} merger, the mass and spin of the remnant, which may undergo further hierarchical mergers, were computed following Ref.~\cite{Jimenez-Forteza:2016oae}. We assumed that both high-mass (HM) and low-mass (LM) clusters formed at $z = 30$. Their initial masses, $M_{\rm cl}$, were drawn from log-normal distributions with means $\langle\log_{10}(M_{\rm cl}/M_{\odot})\rangle = 5.6$ (HM) and 4.3 (LM) and a standard deviation $\sigma_{\rm M}=0.4$.

The merger rate density of Pop.~III \acp{bbh} was computed as:
\begin{equation}
\begin{aligned}
\mathcal{R}(z) = f_{\mathrm{ISO}}\mathcal{R}(z|\mathrm{ISO}) &+ f_{\rm HM}\,\big[f_{\rm Orig}\,\mathcal{R}(z|\mathrm{HM,Orig}) + f_{\rm Dyn}\,\mathcal{R}(z|\mathrm{HM,Dyn})\big] \\
&+ f_{\rm LM}\,\big[f_{\rm Orig}\,\mathcal{R}(z|\mathrm{LM,Orig}) + f_{\rm Dyn}\,\mathcal{R}(z|\mathrm{LM,Dyn})\big]\;,
\end{aligned}
\end{equation}
where $\mathcal{R}(z|\mathrm{ISO})$, $\mathcal{R}(z|\mathrm{HM,Orig})$, $\mathcal{R}(z|\mathrm{HM,Dyn})$, $\mathcal{R}(z|\mathrm{LM,Orig})$, and $\mathcal{R}(z|\mathrm{LM,Dyn})$ were evaluated using {\sc cosmo}$\mathcal{R}${\sc ate}~\cite{Santoliquido:2020axb,Santoliquido:2020bry,Santoliquido:2023wzn}, assuming the Pop.~III star formation rate density from Ref.~\cite{Hartwig:2022lon}. We adopted relative fractions $f_{\rm ISO} = 0.5,\, f_{\rm HM} = 0.25$,\, $f_{\rm LM} = 0.75$, and $f_{\rm Orig} = f_{\rm Dyn} = 0.5$. \Cref{fig:bbh} shows the resulting distributions of merger masses and redshifts for the simulated Pop.~III \ac{bbh} population. The original publication of the catalogs we use can be retrieved from \textsc{Zenodo}~\raisebox{-1pt}{\href{https://doi.org/10.5281/zenodo.17629260}{\includegraphics[width=9pt]{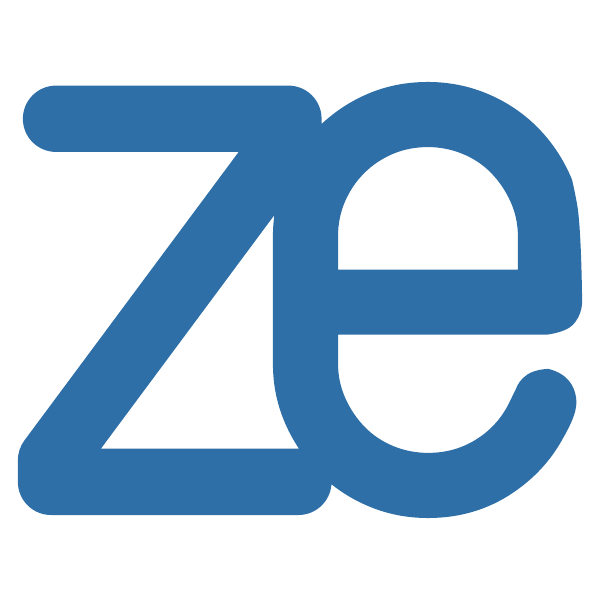}}}~\cite{mestichelli_2025_17629260}.

\paragraph{Primordial BBHs} \Acp{pbh} may originate from the collapse of large primordial inhomogeneities during the radiation era~\cite{Zeldovich:1967lct,Hawking:1974rv,Chapline:1975ojl,Carr:1975qj}. Depending on the model, their masses can span a large range of values~\cite{Ivanov:1994pa,GarciaBellido:1996qt,Ivanov:1997ia,Blinnikov:2016bxu}, including masses accessible to \ac{3g} detectors such as \ac{et} (see Ref.~\cite{Carr:2020gox} for a review).

The \ac{pbh} merger rate is expected to be dominated by binaries assembled in the early Universe, once pairs decouple from the Hubble flow well before matter-radiation equality~\cite{Nakamura:1997sm,Ioka:1998nz}. Dynamical formation in late-time halos, via gravitational capture or three-body interactions~\cite{Kritos:2022ggc}, is also possible~\cite{Ali-Haimoud:2017rtz,Raidal:2017mfl,Vaskonen:2019jpv,DeLuca:2020jug, Raidal:2024bmm}. 
For \ac{pbh} binaries, the merger rate density grows monotonically with redshift as
${\cal R}_\text{\tiny PBH}(z)\propto t_H^{-34/37}$~\cite{Ali-Haimoud:2017rtz,Raidal:2018bbj,DeLuca:2020qqa}, where $t_H$ is the age of the Universe, and the rate extends to redshift $z \gtrsim 10^{3}$. This behavior is predicted by the dynamics of \ac{pbh} pairs decoupling from the Hubble expansion before structure formation. 

Since no definitive evidence for \acp{pbh} currently exists, forecasts must rely on model assumptions. Here, we build the reference \ac{pbh} population following Ref.~\cite{Franciolini:2023opt}, which adopts the \ac{pbh} abundance saturating the upper limits from GWTC--3~\cite{LIGOScientific:2021djp} and from the current non-observation of subsolar-mass mergers~\cite{Nitz:2022ltl,LIGOScientific:2022hai}. 
Following Ref.~\cite{Franciolini:2022tfm}, we model the \ac{pbh} population as arising from an enhanced primordial curvature power spectrum confined between two characteristic scales; see Section~II.E of Ref.~\cite{Franciolini:2022tfm} for a detailed description. The resulting mass function incorporates an up-to-date treatment of the QCD epoch and its influence on both the collapse threshold and the critical-scaling spectrum (see also Refs.~\cite{Jedamzik:1996mr,Byrnes:2018clq,Musco:2023dak,Escriva:2022bwe}). 
This model is specified by four hyper-parameters: the PBH contribution to the dark matter abundance $f_\text{\tiny PBH}$, the local spectral index $n_s$ (unrelated to the large-scale CMB tilt), and the minimum and maximum masses $(M_\text{\tiny S}, M_\text{\tiny L})$ associated with the horizon masses that delimit the formation window. For the forecasts presented here, we adopt
$
f_\text{\tiny PBH} = 10^{-2.8},\;
n_s = 0.84, \; 
M_\text{\tiny S} = 10^{-1.6}\,M_\odot, \;
M_\text{\tiny L} = 10^3 M_\odot,
$
which corresponds to the representative upper limit on \ac{pbh} abundance allowed by the current GWTC--3 constraints. 
With these parameters, the local \ac{pbh} merger rate is ${\cal R}_\text{PBH}(z=0)=89\,\text{Gpc}^{-3}\text{yr}^{-1}$, integrated over the mass range $m_{1,2} \in [10^{-2},10^{3}]\,M_\odot$. Consistent with the assumption of negligible PBH mass growth adopted in Ref.~\cite{Franciolini:2022tfm}, spin magnitudes are generated assuming a small accretion efficiency and using the accretion model described in Refs.~\cite{DeLuca:2020bjf,DeLuca:2020fpg,DeLuca:2020qqa}. For definiteness, we impose a cutoff redshift $z_{\text{\tiny cutoff}}\approx23$ for \ac{pbh} binary accretion (potentially compatible with GW231123~\cite{DeLuca:2025fln}), while spin orientations are drawn from independent and uniform distributions.

\Cref{fig:bbh} shows the distributions of masses, redshifts, and aligned spins for the simulated population of primordial \ac{bbh} mergers.

\section{Results}\label{sec:results}

In the following sections, we report the results of our analyses, emphasizing the comparative performance of the degradations in sensitivity considered with respect to the reference results obtained with the Baseline sensitivity. We refer the reader to Section~\ref{sec:metrics_and_pops} for the general description of the adopted metrics and populations.

\subsection{CBC metrics: detection horizons}\label{subsec:dethor}

\begin{figure}[tbp]
    \centering    \includegraphics[width=1.\linewidth]{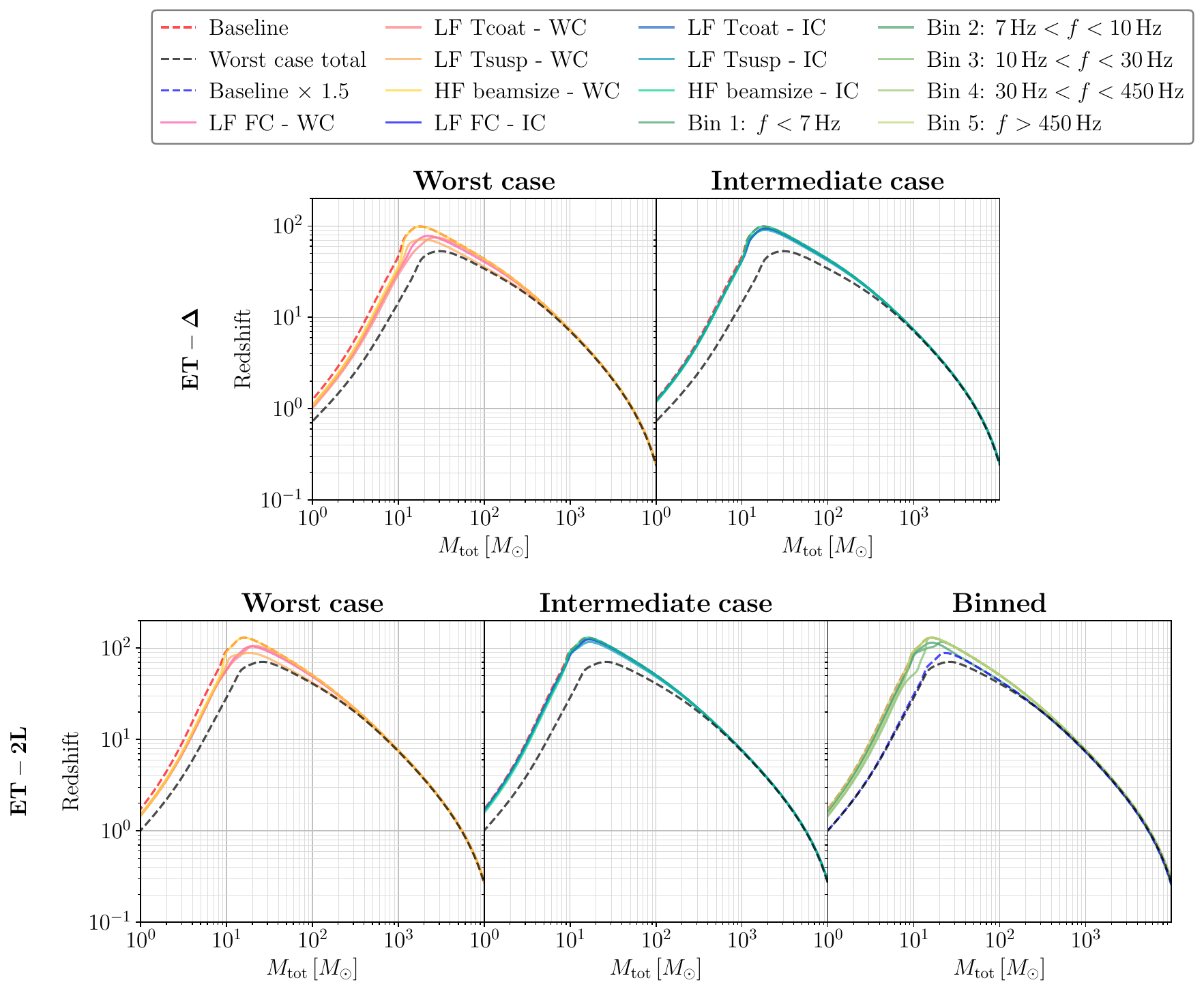}
    \caption{Redshift horizon as a function of total source-frame binary mass $M_{\rm tot}$. The top row corresponds to the ET-$\Delta$ configuration, while the bottom row shows the ET-2L configuration.}
\label{fig:horizon}
\end{figure}

In this section, we report the detection horizons for the various considered detector configurations and sensitivities. We always show results for equal-mass non-spinning binaries as a function of the total source-frame mass. Lowering the mass ratio would result in a lower signal amplitude, and thus a lower horizon. The results are obtained using the \texttt{IMRPhenomD} waveform approximant for the dominant emission mode~\cite{Husa:2015iqa, Khan:2015jqa}. Higher-order harmonics vanish for face-on systems, as we consider in this metric, so their inclusion would not affect our results. Given that all detectors in each configuration share the same sensitivity, the optimal sky location for the network can be computed by maximizing the network's antenna power pattern, which we do using the \texttt{basinhopping} algorithm implemented in \texttt{scipy}~\cite{2020SciPy-NMeth}. Once the optimal location is determined, for a grid of source-frame total masses we minimize as a function of $z$ the difference between the chosen threshold for detection, ${\rm SNR}_{\rm th} = 8$, and the optimal network \ac{snr} 
\begin{equation}\label{eq:snr_net}
    {\rm SNR}^2(\bm{\theta}) = \sum_{d\in {\rm detectors}} 4 {\rm Re} \int_{f_{\rm min}}^{f_{\rm max}} \dfrac{|\tilde{h}^{(d)}(f;\bm{\theta})|^2}{S_{n}^{(d)}(f)}\,{\rm d} f\;,
\end{equation}
where the minimum frequency is $f_{\rm min} = 2\,{\rm Hz}$ in the analyses in the main text (see the discussion in Appendix~\ref{app:minimum_freq}), the maximum frequency is set to 2048\,Hz, 
$S_{n}^{(d)}(f)$ denotes the one-sided noise \ac{psd} of the detector $d$ and $\tilde{h}^{(d)}(f;\bm{\theta})$ is the frequency domain signal described by the set of parameters $\bm{\theta}$ projected onto the detector $d$. The minimization is carried out using the \texttt{Nelder-Mead} algorithm as implemented in \texttt{scipy}~\cite{2020SciPy-NMeth}.

\begin{figure}[tbp]
    \centering
    \includegraphics[width=1.\linewidth]{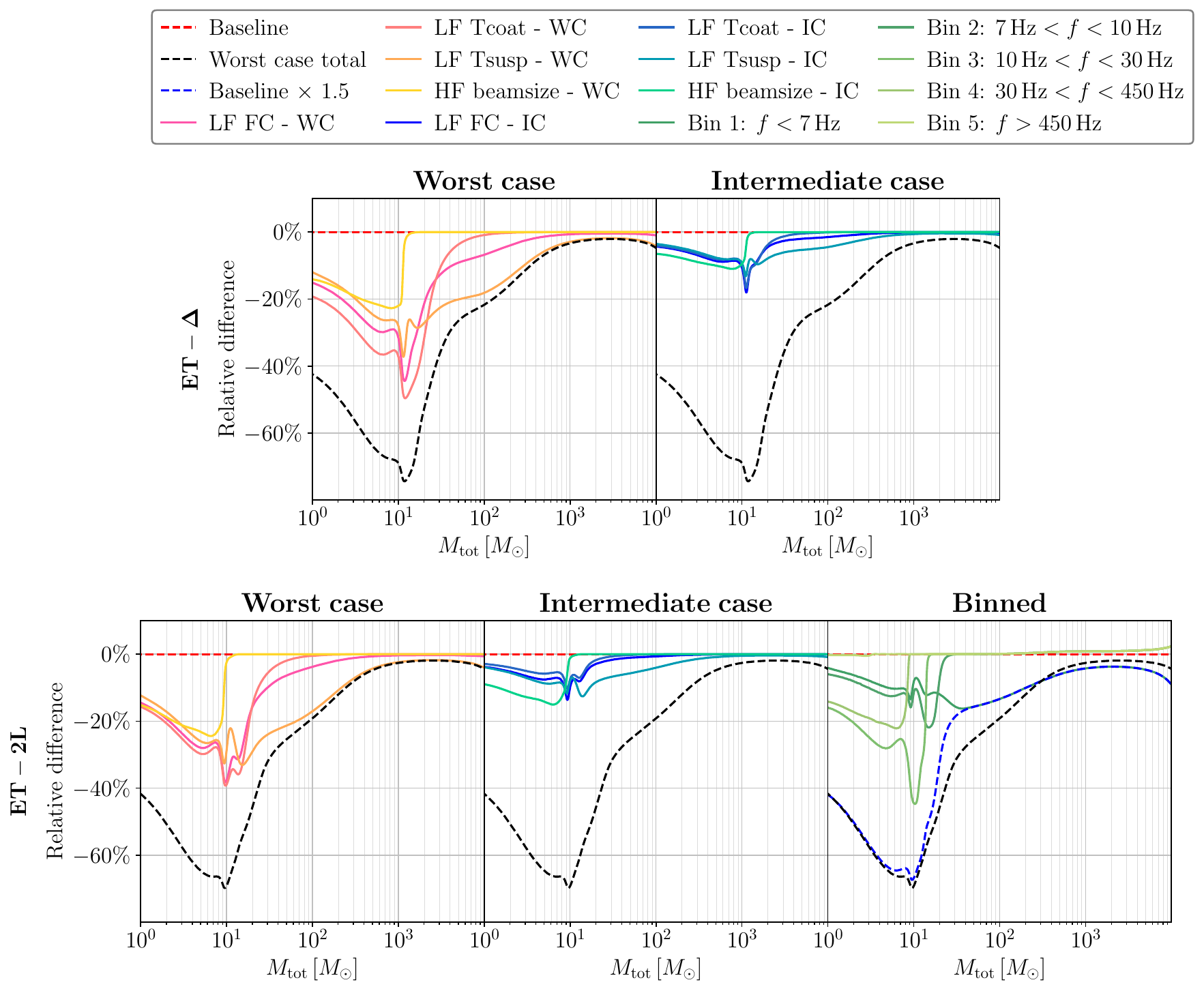}
    \caption{Relative differences in redshift with respect to the Baseline as a function of total source-frame binary mass $M_{\rm tot}$. The horizontal line at 0\% indicates the baseline. The top row corresponds to the ET-$\Delta$ configuration, while the bottom row shows the ET-2L configuration.}
    \label{fig:horizon_relative}
\end{figure}

The horizons for the two different configurations and the different noise curves are reported in \Cref{fig:horizon}. From these results, we observe that the detector performance is only mildly affected by the degradations in the intermediate case, with losses in reach of $\lesssim10\%$. When considering larger degradations, we find that sensitivity losses in the LF instrument affect the horizon across the whole mass range, whereas degradations at high frequency affect the low-mass end only. This is expected, as more massive systems merge at lower frequencies. The single most impactful parameter is the suspension temperature, which can produce variations of $\gtrsim20\%$ across all the mass range in the worst case, followed by the coating temperature. Varying all parameters in the worst case results in losses of $\gtrsim50\%$ in a large fraction of the mass range, even worse than an overall degradation by a factor of $1.5$. This holds for both configurations, which show similar percentage degradations. %
The quoted percentage degradations can be directly inferred from the relative-difference plot shown in \Cref{fig:horizon_relative}. 

\subsection{CBC metrics: detections and parameter estimation}\label{sec:detection_and_fisher}

\begin{table}[tbp]
\caption{We report the number of detections of \ac{cbc} sources with \ac{snr}$\geq 12$ for \ac{bbh} sources and with \ac{snr}$\geq 8$ for \ac{bns} sources in the top row, and \ac{snr}$\geq 50$ for \acp{bns} and \ac{snr}$\geq 100$ for \acp{bbh}  in the bottom rows for each considered sensitivity. We give the results for both the triangular and the 2L configurations. The first row shows the absolute numbers for the Baseline sensitivity curve. For all other scenarios, we report the ratio between the detections for a given degraded sensitivity and the detections obtained with the Baseline sensitivity.}

{\setlength{\tabcolsep}{4.88pt}
\setlength{\aboverulesep}{-1pt}
\setlength{\belowrulesep}{-3pt}
\begin{tabularx}{\textwidth}{lcccccccc}
	\toprule
	  & \multicolumn{2}{c}{\bf BNS} & \multicolumn{2}{c}{\bf BBH} & \multicolumn{2}{c}{\bf Pop.~III} & \multicolumn{2}{c}{\bf PBH} \\
    \cmidrule(r){2-3} \cmidrule(r){4-5} \cmidrule(r){6-7} \cmidrule{8-9}
    &\multicolumn{2}{c}{\makecell{\scriptsize SNR$\geq8$ \scriptsize ($\geq$50)}} &\multicolumn{2}{c}{\makecell{\scriptsize SNR$\geq12$  \scriptsize ($\geq100$)}} &\multicolumn{2}{c}{\makecell{\scriptsize SNR$\geq12$ \scriptsize ($\geq100$)}} &\multicolumn{2}{c}{\makecell{\scriptsize SNR$\geq12$ ($\geq100$)}} \\ 

	 \bf Sensitivity curve & $\Delta$ & 2L & $\Delta$ & 2L & $\Delta$ & 2L & $\Delta$ & 2L \\\midrule
    \multirow{2}{*}{Baseline} &{44836} &{78590} &{77928} &{91393} &{1797} &{2062} &{22806} &{38492}\\ 
&{({\footnotesize 146})} &{({\footnotesize 296})} &{({\footnotesize 1654})} &{({\footnotesize 3286})} &{({\footnotesize 54})} &{({\footnotesize 97})} &{({\footnotesize 107})} &{({\footnotesize 198})}\\ 
\rowcolor{MidnightBlueLight}&{\multirow{2}{*}{--}} &{37\%} &{\multirow{2}{*}{--}} &{73\%} &{\multirow{2}{*}{--}} &{77\%} &{\multirow{2}{*}{--}} &{39\%}\\ 
\rowcolor{MidnightBlueLight}\multirow{-2}{*}{Baseline $\times 1.5$}&{\multirow{-2}{*}{--}} &{({\footnotesize 32\%)}} &{\multirow{-2}{*}{--}} &{({\footnotesize 31\%)}} &{\multirow{-2}{*}{--}} &{({\footnotesize 52\%)}} &{\multirow{-2}{*}{--}} &{({\footnotesize 28\%)}}
\\&{\multirow{2}{*}{--}} &{93\%} &{\multirow{2}{*}{--}} &{98\%} &{\multirow{2}{*}{--}} &{92\%} &{\multirow{2}{*}{--}} &{91\%}\\ 
\multirow{-2}{*}{Bin 1: $f<7\,\mathrm{Hz}$}&{\multirow{-2}{*}{--}} &{({\footnotesize 91\%)}} &{\multirow{-2}{*}{--}} &{({\footnotesize 89\%)}} &{\multirow{-2}{*}{--}} &{({\footnotesize 78\%)}} &{\multirow{-2}{*}{--}} &{({\footnotesize 91\%)}}
\\\rowcolor{MidnightBlueLight}&{\multirow{2}{*}{--}} &{91\%} &{\multirow{2}{*}{--}} &{97\%} &{\multirow{2}{*}{--}} &{96\%} &{\multirow{2}{*}{--}} &{89\%}\\ 
\rowcolor{MidnightBlueLight}\multirow{-2}{*}{Bin 2: $7\,\mathrm{Hz}<f<10\,\mathrm{Hz}$}&{\multirow{-2}{*}{--}} &{({\footnotesize 88\%)}} &{\multirow{-2}{*}{--}} &{({\footnotesize 86\%)}} &{\multirow{-2}{*}{--}} &{({\footnotesize 91\%)}} &{\multirow{-2}{*}{--}} &{({\footnotesize 89\%)}}
\\&{\multirow{2}{*}{--}} &{76\%} &{\multirow{2}{*}{--}} &{92\%} &{\multirow{2}{*}{--}} &{95\%} &{\multirow{2}{*}{--}} &{75\%}\\ 
\multirow{-2}{*}{Bin 3: $10\,\mathrm{Hz}<f<30\,\mathrm{Hz}$}&{\multirow{-2}{*}{--}} &{({\footnotesize 70\%)}} &{\multirow{-2}{*}{--}} &{({\footnotesize 69\%)}} &{\multirow{-2}{*}{--}} &{({\footnotesize 86\%)}} &{\multirow{-2}{*}{--}} &{({\footnotesize 69\%)}}
\\\rowcolor{MidnightBlueLight}&{\multirow{2}{*}{--}} &{80\%} &{\multirow{2}{*}{--}} &{94\%} &{\multirow{2}{*}{--}} &{100\%} &{\multirow{2}{*}{--}} &{86\%}\\ 
\rowcolor{MidnightBlueLight}\multirow{-2}{*}{Bin 4: $30\,\mathrm{Hz}<f<450\,\mathrm{Hz}$}&{\multirow{-2}{*}{--}} &{({\footnotesize 74\%)}} &{\multirow{-2}{*}{--}} &{({\footnotesize 79\%)}} &{\multirow{-2}{*}{--}} &{({\footnotesize 97\%)}} &{\multirow{-2}{*}{--}} &{({\footnotesize 82\%)}}
\\&{\multirow{2}{*}{--}} &{100\%} &{\multirow{2}{*}{--}} &{100\%} &{\multirow{2}{*}{--}} &{100\%} &{\multirow{2}{*}{--}} &{100\%}\\ 
\multirow{-2}{*}{Bin 5: $f>450\,\mathrm{Hz}$}&{\multirow{-2}{*}{--}} &{({\footnotesize 100\%)}} &{\multirow{-2}{*}{--}} &{({\footnotesize 100\%)}} &{\multirow{-2}{*}{--}} &{({\footnotesize 100\%)}} &{\multirow{-2}{*}{--}} &{({\footnotesize 99\%)}}
\\\rowcolor{MidnightBlueLight}&{92\%} &{94\%} &{97\%} &{98\%} &{96\%} &{98\%} &{91\%} &{93\%}\\ 
\rowcolor{MidnightBlueLight}\multirow{-2}{*}{LF FC - IC}&{({\footnotesize 92\%)}} &{({\footnotesize 92\%)}} &{({\footnotesize 90\%)}} &{({\footnotesize 91\%)}} &{({\footnotesize 93\%)}} &{({\footnotesize 96\%)}} &{({\footnotesize 88\%)}} &{({\footnotesize 92\%)}}
\\&{92\%} &{96\%} &{97\%} &{99\%} &{97\%} &{98\%} &{92\%} &{95\%}\\ 
\multirow{-2}{*}{LF Tcoat - IC}&{({\footnotesize 92\%)}} &{({\footnotesize 94\%)}} &{({\footnotesize 90\%)}} &{({\footnotesize 93\%)}} &{({\footnotesize 93\%)}} &{({\footnotesize 97\%)}} &{({\footnotesize 90\%)}} &{({\footnotesize 94\%)}}
\\\rowcolor{MidnightBlueLight}&{93\%} &{94\%} &{97\%} &{98\%} &{95\%} &{96\%} &{92\%} &{92\%}\\ 
\rowcolor{MidnightBlueLight}\multirow{-2}{*}{LF Tsusp - IC}&{({\footnotesize 92\%)}} &{({\footnotesize 92\%)}} &{({\footnotesize 91\%)}} &{({\footnotesize 90\%)}} &{({\footnotesize 91\%)}} &{({\footnotesize 89\%)}} &{({\footnotesize 90\%)}} &{({\footnotesize 92\%)}}
\\&{88\%} &{88\%} &{96\%} &{96\%} &{99\%} &{100\%} &{92\%} &{90\%}\\ 
\multirow{-2}{*}{HF beamsize - IC}&{({\footnotesize 87\%)}} &{({\footnotesize 81\%)}} &{({\footnotesize 88\%)}} &{({\footnotesize 85\%)}} &{({\footnotesize 98\%)}} &{({\footnotesize 97\%)}} &{({\footnotesize 88\%)}} &{({\footnotesize 88\%)}}
\\\rowcolor{MidnightBlueLight}&{70\%} &{78\%} &{88\%} &{92\%} &{84\%} &{90\%} &{71\%} &{74\%}\\ 
\rowcolor{MidnightBlueLight}\multirow{-2}{*}{LF FC - WC}&{({\footnotesize 67\%)}} &{({\footnotesize 72\%)}} &{({\footnotesize 66\%)}} &{({\footnotesize 70\%)}} &{({\footnotesize 74\%)}} &{({\footnotesize 72\%)}} &{({\footnotesize 61\%)}} &{({\footnotesize 71\%)}}
\\&{63\%} &{76\%} &{85\%} &{91\%} &{80\%} &{90\%} &{64\%} &{73\%}\\ 
\multirow{-2}{*}{LF Tcoat - WC}&{({\footnotesize 60\%)}} &{({\footnotesize 71\%)}} &{({\footnotesize 59\%)}} &{({\footnotesize 68\%)}} &{({\footnotesize 70\%)}} &{({\footnotesize 78\%)}} &{({\footnotesize 52\%)}} &{({\footnotesize 68\%)}}
\\\rowcolor{MidnightBlueLight}&{75\%} &{80\%} &{90\%} &{93\%} &{80\%} &{85\%} &{75\%} &{76\%}\\ 
\rowcolor{MidnightBlueLight}\multirow{-2}{*}{LF Tsusp - WC}&{({\footnotesize 75\%)}} &{({\footnotesize 76\%)}} &{({\footnotesize 70\%)}} &{({\footnotesize 72\%)}} &{({\footnotesize 69\%)}} &{({\footnotesize 63\%)}} &{({\footnotesize 67\%)}} &{({\footnotesize 73\%)}}
\\&{74\%} &{78\%} &{92\%} &{93\%} &{98\%} &{99\%} &{83\%} &{83\%}\\ 
\multirow{-2}{*}{HF beamsize - WC}&{({\footnotesize 71\%)}} &{({\footnotesize 73\%)}} &{({\footnotesize 76\%)}} &{({\footnotesize 76\%)}} &{({\footnotesize 91\%)}} &{({\footnotesize 96\%)}} &{({\footnotesize 72\%)}} &{({\footnotesize 79\%)}}
\\\rowcolor{MidnightBlueLight}&{26\%} &{36\%} &{62\%} &{72\%} &{57\%} &{70\%} &{33\%} &{38\%}\\ 
\rowcolor{MidnightBlueLight}\multirow{-2}{*}{Worst case total}&{({\footnotesize 27\%)}} &{({\footnotesize 31\%)}} &{({\footnotesize 24\%)}} &{({\footnotesize 29\%)}} &{({\footnotesize 39\%)}} &{({\footnotesize 37\%)}} &{({\footnotesize 18\%)}} &{({\footnotesize 31\%)}}
\\ \bottomrule

\end{tabularx}
}
\label{tab:snr_table}
\end{table}

\begin{figure}[tbp]
    \centering
    \includegraphics[width=1.\linewidth]{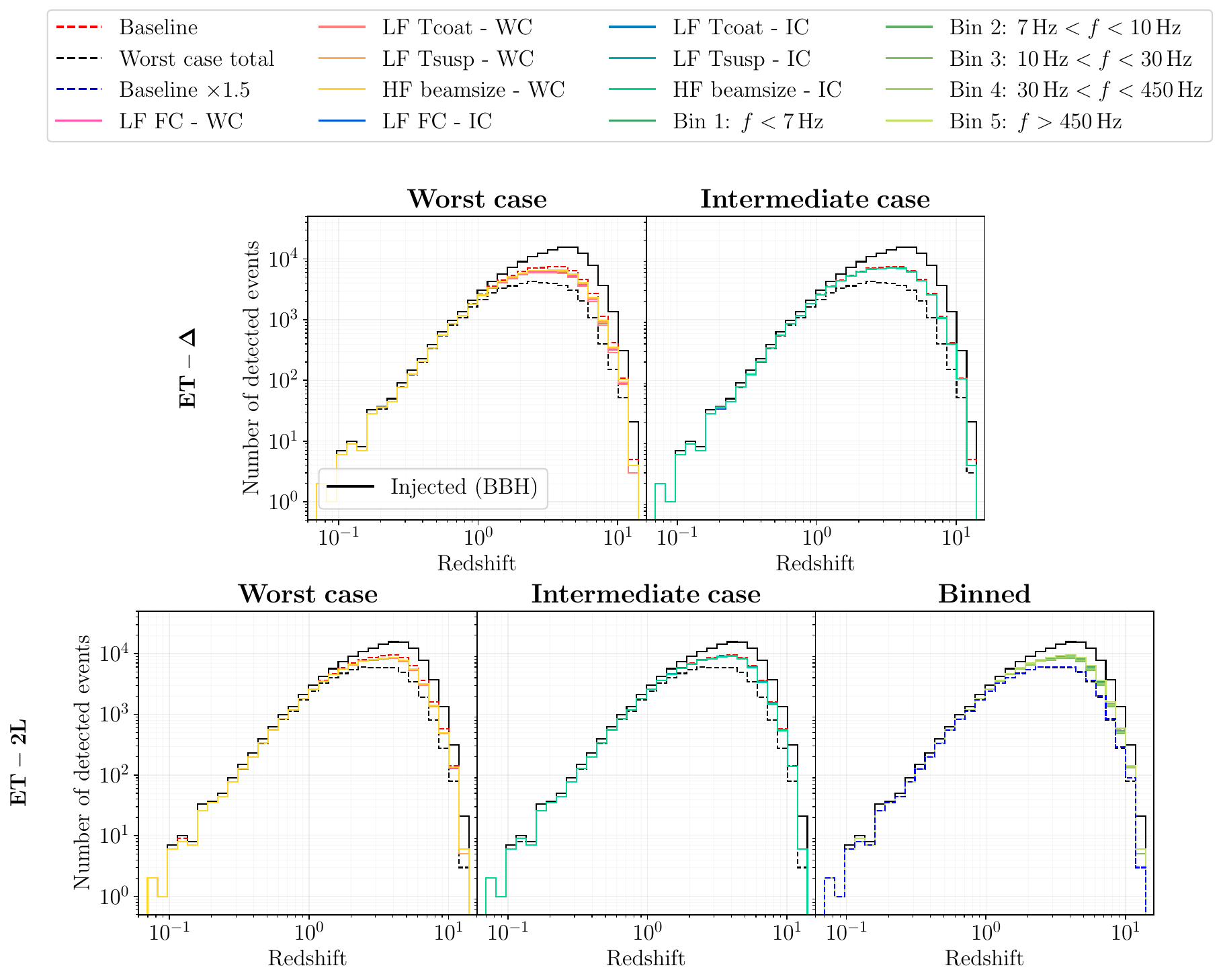}
    \caption{Redshift distribution of detected \ac{bbh} events with $\rm SNR \geq 12$. 
    The top (bottom) row concerns the ET triangular (2L) configuration, while columns correspond to cases as labeled in the panel titles.
    }
    \label{fig:detections_snr12_BBH}
\end{figure}

In the first part of this analysis, we compare the detector sensitivities by the number of detections for the different populations of \ac{cbc} sources described in Section~\ref{subsec:cbcpops}. We consider an event detectable if its optimal network \ac{snr} defined as in Eq.~\eqref{eq:snr_net} is higher than 12 for \acp{bbh} and 8 for \acp{bns}.\footnote{This is motivated by the fact that at least for a fraction of the \ac{bns} detections the \ac{gw} signal might be detected in coincidence with an EM counterpart, increasing the significance of a detection even with a lower \ac{snr}.} For \acp{bbh} we model the signal through the \texttt{IMRPhenomXPHM} waveform model, which includes higher-order emission modes and the effect of spin precession~\cite{Pratten:2020ceb, Garcia-Quiros:2020qpx}. For \ac{bns} systems, instead, we use the waveform model \texttt{IMRPhenomD\_NRTidalv2}, which models only the dominant emission harmonic but incorporates the impact on the signal of tidal deformations of the objects during the inspiral~\cite{Husa:2015iqa, Khan:2015jqa,Dietrich:2019kaq}. In all cases, our results include an uncorrelated 85\% duty cycle for each detector (two detectors in the case of the 2L geometry and three in the case of the $\Delta$ configuration), consistent between the \ac{snr} and \ac{fim} computation, and include the effect of Earth's motion during the observation time as implemented in \texttt{gwfast}~\cite{Iacovelli:2022mbg} and \texttt{GWFish}~\cite{Dupletsa:2022wke}.

The main results are summarized in \Cref{tab:snr_table}, which shows absolute detection numbers for the Baseline configuration and relative variations for all the other scenarios. In parenthesis, we also show the results for a higher \ac{snr} threshold, corresponding to 50 for the \ac{bns} case and 100 for the \ac{bbh} one.
Overall, the intermediate-case degradations have only a minor impact on the detection rate across all populations and leave the \ac{snr} distribution largely unchanged. The HF degradation has the biggest effects only on \acp{bns} and \acp{pbh}, lowering the numbers by more than $10\%$. In contrast, worst-case degradations, particularly those affecting the LF instrument, have a more pronounced effect. For instance, the number of detections for \acp{bbh} from Pop.~III stars and \acp{pbh} can decrease by a factor of 1.5. Despite this, detection rates remain exceptionally high, with always $\sim\!10^4$ \ac{bns} and $\sim\!4.8\times10^4$ \ac{bbh} detections per year. A more significant impact is observed in the high-\ac{snr} tail of the \ac{bbh} distribution. In the worst-case LF degradation scenario, the number of events with \ac{snr}$\ge 500$ is reduced by approximately a factor of $\sim\!5$ for both detector configurations. This reduction may affect science cases that rely on very high-\ac{snr} events, such as precision tests of \ac{gr}. The same trends are confirmed by the binned sensitivity analysis, in which the most impactful bins are always the 2, 3, and 4. In the worst-case scenario and the case of overall degradation of the sensitivity of 1.5, we find comparable results across all kinds of sources. In this case, the number of \ac{bns} detections drops by about a factor of 4, and by a factor of $\gtrsim2$ for high-$z$ \acp{bbh}. While the overall number of \ac{bbh} detections is less impacted, their \ac{snr} distribution is, with 4 times fewer sources having ${\rm SNR}\gtrsim100$.
For completeness, we show in Appendix~\ref{app:additional_plots} the \ac{snr} distributions for \acp{bns} in \Cref{fig:bns_snr}, and for \acp{bbh} in \Cref{fig:bbh_SNR_all}. We show in the same plot \acp{bbh} from Population I and II stars, including \acp{imbh}, \acp{bbh} from Pop.~III stars, and \acp{pbh}. The different statistics -- from highest to lowest, we have \acp{pbh}, \acp{bbh} from Population I and II stars, \acp{bbh} from Pop.~III stars, respectively, as reported in \Cref{tab:pop_properties} -- allow us to distinguish the different \ac{bbh} populations. 

\begin{figure}[tbp]
    \centering
    \includegraphics[width=1.\linewidth]{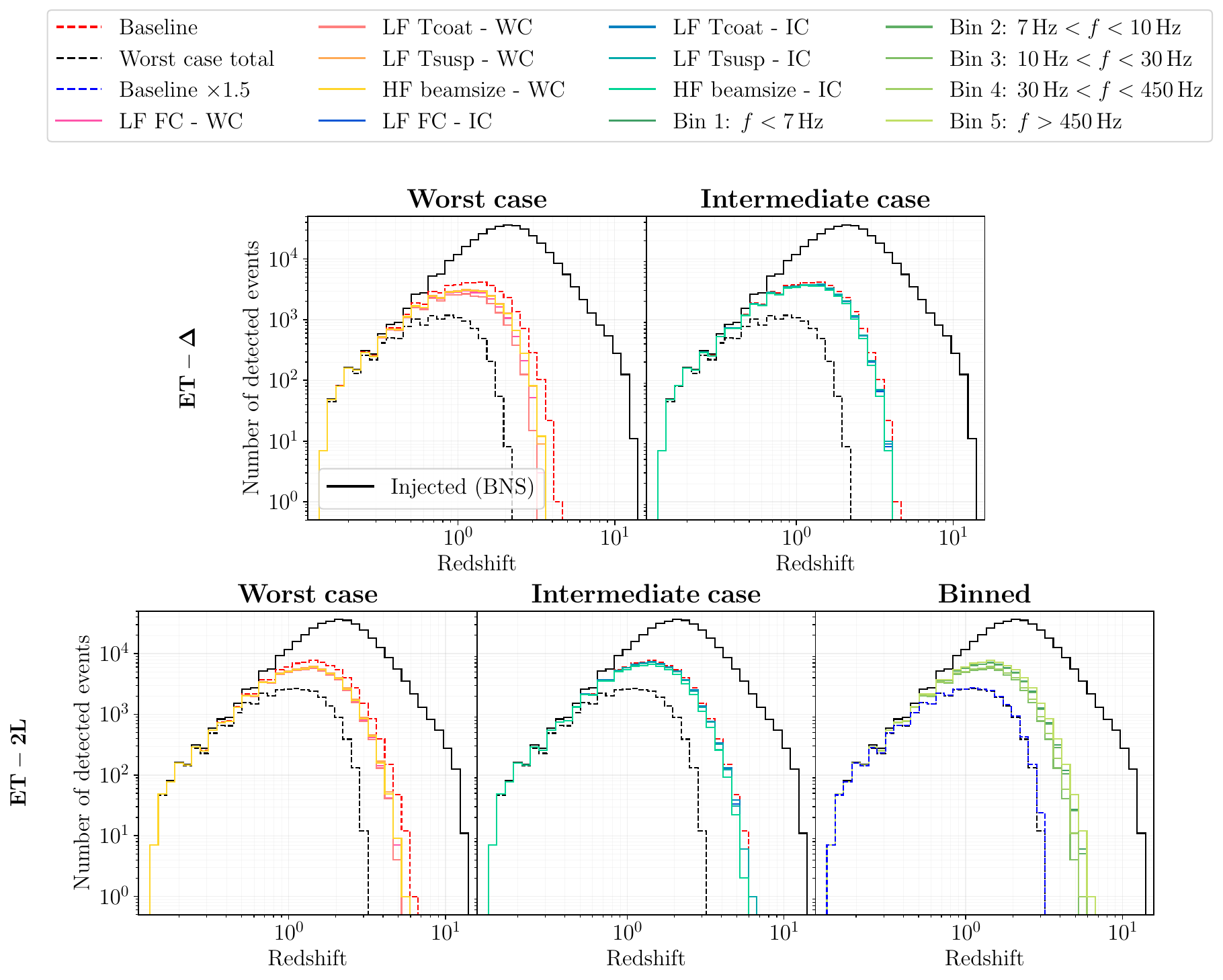}
    \caption{
    Same as \Cref{fig:detections_snr12_BBH}, but for the \ac{bns} case. The detection threshold \ac{snr} for \acp{bns} is 8.} 
    \label{fig:detections_snr12_BNS}
\end{figure}

To further understand these effects, we examine the redshift distribution of detectable events. For \acp{bbh}, we report the redshift distribution of the events with ${\rm SNR}\geq8$ in \Cref{fig:detections_snr12_BBH}. Overall, we find only minor differences in the number of detectable events below $z\sim2$, with all configurations able to detect all the \ac{bbh} events below $z\sim1$. The number of detections drops instead more rapidly with increasing redshift for degradations in the LF instrument or the lower frequency bins. A similar plot for \ac{bns} sources is shown in \Cref{fig:detections_snr12_BNS}. In this case, we find larger differences in the distributions of detectable sources with respect to the injected population even at relatively low redshift. With the most pessimistic degradations, only a limited number of sources beyond the peak of the star formation rate at $z\sim2$ would be observed. Degradations corresponding to the intermediate-case scenario would have a minor impact compared to the baseline results.

As noted previously, even for the most dramatic reductions in the sensitivity we study, the performance remains exquisite. However, the results clearly highlight the critical role of LF sensitivity, both for maximizing detection rates at high redshift and for preserving the population of high-\ac{snr} events. Considering the triangle and 2L performance, while the difference for \acp{bbh} is modest, with $\sim\!15\%$ less detections overall, for the \ac{bns} sources, the 2L configuration offers a better compromise in terms of robustness against noise-budget degradations, with the worst-case performance of the 2L delivering similar results to the intermediate-case scenario of the $\Delta$ configuration in terms of redshift reach as shown in \Cref{fig:detections_snr12_BNS}.

\begin{table}[tbp]
\caption{Summary table for the statistical uncertainties on the source parameters for \ac{bbh} sources. We report relative errors for selected parameters: chirp mass (under $1$\,\textperthousand), luminosity distance (under $10\%$), spin of the primary component (under $10\%$), and sky localization at $90\%$\, confidence level (c.l.) (under $1000$\,deg$^2$). The first row shows the absolute numbers for the Baseline sensitivity curve. For all the other sensitivities, we report the ratio between the measurements for a given sensitivity and those obtained with the Baseline sensitivity. We compare the two configurations side by side.}
{\setlength{\tabcolsep}{5.pt}
\begin{tabularx}{\textwidth}{lcccccccc}
	\toprule
	\multicolumn{9}{c}{\bf BBH}\\ \midrule
    &\multicolumn{2}{c}{\makecell{$\frac{\Delta\mathcal{M}_c} {\mathcal{M}_c}$ \scriptsize{$\le 1$\,\textperthousand}}} &\multicolumn{2}{c}{\makecell{$\frac{\Delta d_L} {d_L}$ \scriptsize{$ \le 10\%$}}} &\multicolumn{2}{c}{\makecell{$\frac{\Delta\chi_1} {\chi_1}$ \scriptsize{$ \le 10\%$}}} &\multicolumn{2}{c}{\makecell{$\Delta\Omega_{90\%}$ \scriptsize{$ \le 1000\,\text{deg}^2$}}} \\ %
    \cmidrule(r){2-3} \cmidrule(r){4-5} \cmidrule(r){6-7} \cmidrule{8-9}
   
	 \bf Sensitivity curve  &$\Delta$ &2L &$\Delta$ &2L &$\Delta$ &2L &$\Delta$ &2L \\\midrule
    Baseline &{50428} &{62968} &{7000} &{16750} &{7808} &{10175} &{20213} &{36569}\\ 
\rowcolor{MidnightBlueLight}Baseline $\times 1.5$&{--} &{64\%} &{--} &{56\%} &{--} &{60\%} &{--} &{59\%}
\\Bin 1: $f<7\,\mathrm{Hz}$&{--} &{95\%} &{--} &{94\%} &{--} &{95\%} &{--} &{96\%}
\\\rowcolor{MidnightBlueLight}Bin 2: $7\,\mathrm{Hz}<f<10\,\mathrm{Hz}$&{--} &{96\%} &{--} &{94\%} &{--} &{96\%} &{--} &{96\%}
\\Bin 3: $10\,\mathrm{Hz}<f<30\,\mathrm{Hz}$&{--} &{90\%} &{--} &{86\%} &{--} &{89\%} &{--} &{89\%}
\\\rowcolor{MidnightBlueLight}Bin 4: $30\,\mathrm{Hz}<f<450\,\mathrm{Hz}$&{--} &{89\%} &{--} &{85\%} &{--} &{84\%} &{--} &{84\%}
\\Bin 5: $f>450\,\mathrm{Hz}$&{--} &{100\%} &{--} &{100\%} &{--} &{99\%} &{--} &{100\%}
\\\rowcolor{MidnightBlueLight}LF FC - IC&{96\%} &{98\%} &{94\%} &{96\%} &{97\%} &{97\%} &{95\%} &{97\%}
\\LF Tcoat - IC&{97\%} &{98\%} &{95\%} &{97\%} &{97\%} &{98\%} &{96\%} &{98\%}
\\\rowcolor{MidnightBlueLight}LF Tsusp - IC&{96\%} &{97\%} &{94\%} &{95\%} &{96\%} &{97\%} &{96\%} &{97\%}
\\HF beamsize - IC&{94\%} &{94\%} &{91\%} &{92\%} &{92\%} &{92\%} &{93\%} &{92\%}
\\\rowcolor{MidnightBlueLight}LF FC - WC&{86\%} &{90\%} &{79\%} &{85\%} &{86\%} &{89\%} &{82\%} &{89\%}
\\LF Tcoat - WC&{83\%} &{90\%} &{75\%} &{85\%} &{84\%} &{89\%} &{78\%} &{89\%}
\\\rowcolor{MidnightBlueLight}LF Tsusp - WC&{86\%} &{89\%} &{81\%} &{85\%} &{87\%} &{89\%} &{85\%} &{90\%}
\\HF beamsize - WC&{87\%} &{89\%} &{81\%} &{85\%} &{84\%} &{85\%} &{84\%} &{86\%}
\\\rowcolor{MidnightBlueLight}Worst case total&{54\%} &{64\%} &{41\%} &{55\%} &{54\%} &{61\%} &{48\%} &{60\%}
\\ \bottomrule

\end{tabularx}
}
\label{tab:fisher_bbh}
\end{table}

\begin{table}[tbp]
\caption{Summary table for the statistical uncertainties on the source parameters for \ac{bns} sources. We report relative errors for selected parameters: chirp mass (under $1$\,\textperthousand), luminosity distance (under $10\%$), tidal deformability (under $20\%$), and sky localization at $90\%$\,c.l. (under $1000$\,deg$^2$). The first row shows the absolute numbers for the Baseline sensitivity curve. For all the other sensitivities, we report the ratio of the measurements for a given sensitivity to those obtained with the Baseline sensitivity. We compare the two configurations side by side.}
{\setlength{\tabcolsep}{6.1pt}
\begin{tabularx}{\textwidth}{lcccccccc}
	\toprule
	\multicolumn{9}{c}{\bf BNS}\\
    \midrule
    &\multicolumn{2}{c}{\makecell{$\frac{\Delta\mathcal{M}_c} {\mathcal{M}_c}$ \scriptsize{$\le 1$\,\textperthousand}}} &\multicolumn{2}{c}{\makecell{$\frac{\Delta d_L} {d_L}$ \scriptsize{$ \le 10\%$}}} &\multicolumn{2}{c}{\makecell{$\frac{\Delta\tilde{\Lambda}} {\tilde{\Lambda}}$ \scriptsize{$ \le 20\%$}}} &\multicolumn{2}{c}{\makecell{$\Delta\Omega_{90\%}$ \scriptsize{$ \le 1000\,\text{deg}^2$}}} \\ %
    \cmidrule(r){2-3} \cmidrule(r){4-5} \cmidrule(r){6-7} \cmidrule{8-9}
   
	 \bf Sensitivity curve &$\Delta$ &2L &$\Delta$ &2L &$\Delta$ &2L &$\Delta$ &2L \\\midrule
    Baseline &{44834} &{78518} &{62} &{315} &{1003} &{1990} &{1353} &{4545}\\ 
\rowcolor{MidnightBlueLight}Baseline $\times 1.5$&{--} &{37\%} &{--} &{29\%} &{--} &{13\%} &{--} &{34\%}
\\Bin 1: $f<7\,\mathrm{Hz}$&{--} &{93\%} &{--} &{68\%} &{--} &{89\%} &{--} &{71\%}
\\\rowcolor{MidnightBlueLight}Bin 2: $7\,\mathrm{Hz}<f<10\,\mathrm{Hz}$&{--} &{91\%} &{--} &{94\%} &{--} &{92\%} &{--} &{96\%}
\\Bin 3: $10\,\mathrm{Hz}<f<30\,\mathrm{Hz}$&{--} &{76\%} &{--} &{84\%} &{--} &{95\%} &{--} &{88\%}
\\\rowcolor{MidnightBlueLight}Bin 4: $30\,\mathrm{Hz}<f<450\,\mathrm{Hz}$&{--} &{80\%} &{--} &{84\%} &{--} &{57\%} &{--} &{79\%}
\\Bin 5: $f>450\,\mathrm{Hz}$&{--} &{100\%} &{--} &{97\%} &{--} &{46\%} &{--} &{94\%}
\\\rowcolor{MidnightBlueLight}LF FC - IC&{92\%} &{94\%} &{92\%} &{95\%} &{95\%} &{97\%} &{91\%} &{96\%}
\\LF Tcoat - IC&{92\%} &{96\%} &{95\%} &{96\%} &{96\%} &{98\%} &{94\%} &{97\%}
\\\rowcolor{MidnightBlueLight}LF Tsusp - IC&{93\%} &{94\%} &{81\%} &{86\%} &{92\%} &{93\%} &{86\%} &{87\%}
\\HF beamsize - IC&{88\%} &{88\%} &{94\%} &{91\%} &{78\%} &{77\%} &{92\%} &{89\%}
\\\rowcolor{MidnightBlueLight}LF FC - WC&{70\%} &{78\%} &{58\%} &{72\%} &{81\%} &{84\%} &{68\%} &{78\%}
\\LF Tcoat - WC&{63\%} &{76\%} &{66\%} &{80\%} &{78\%} &{85\%} &{71\%} &{84\%}
\\\rowcolor{MidnightBlueLight}LF Tsusp - WC&{75\%} &{80\%} &{39\%} &{57\%} &{74\%} &{74\%} &{55\%} &{60\%}
\\HF beamsize - WC&{74\%} &{78\%} &{76\%} &{84\%} &{54\%} &{59\%} &{81\%} &{80\%}
\\\rowcolor{MidnightBlueLight}Worst case total&{26\%} &{36\%} &{23\%} &{28\%} &{33\%} &{35\%} &{24\%} &{35\%}
\\ \bottomrule

\end{tabularx}
}
\label{tab:fisher_bns}
\end{table}

In addition to the number of detectable sources, we provide forecasts for the statistical accuracy attainable in the reconstruction of their parameters. These are computed using the \ac{fim} formalism, in which the likelihood on the event parameters is approximated as a multivariate Gaussian distribution, with covariance matrix given by
\begin{equation}\label{eq:fim_def}
    {\cal C}_{ij}^{-1}(\bm{\theta_0}) = \Gamma_{ij}(\bm{\theta_0}) = \sum_{d\in {\rm detectors}} \left(\dfrac{\partial \tilde{h}^{(d)}(\bm{\theta})}{\partial \theta_i} \Bigg| \dfrac{\partial \tilde{h}^{(d)}(\bm{\theta})}{\partial \theta_j}\right)\Bigg|_{\bm{\theta} = \bm{\theta_0}}\;,
\end{equation}
where $\Gamma_{ij}$ is the \ac{fim}, $\bm{\theta_0}$ denotes the injected event parameters, and the noise-weighted inner product $(\cdot|\cdot)$ is defined as 
\begin{equation}
    (a|b) = 4 {\rm Re} \int_{f_{\rm min}}^{f_{\rm max}} \dfrac{\tilde{a}^*(f) \tilde{b}(f)}{S_{n}(f)}\,{\rm d} f\;,
\end{equation}
where the frequency integration extrema are the same reported below Eq.~\eqref{eq:snr_net}, $a$ and $b$ are two generic time-domain signals, a tilde denotes the Fourier transform in the frequency domain, and the asterisk denotes complex conjugation.
Given the higher computational cost of this analysis, we focus here only on the \ac{bns} and \ac{bbh} events originating from Population I and II stars. %

\begin{figure}[tbp]
\centering
\includegraphics[width=1.\linewidth]{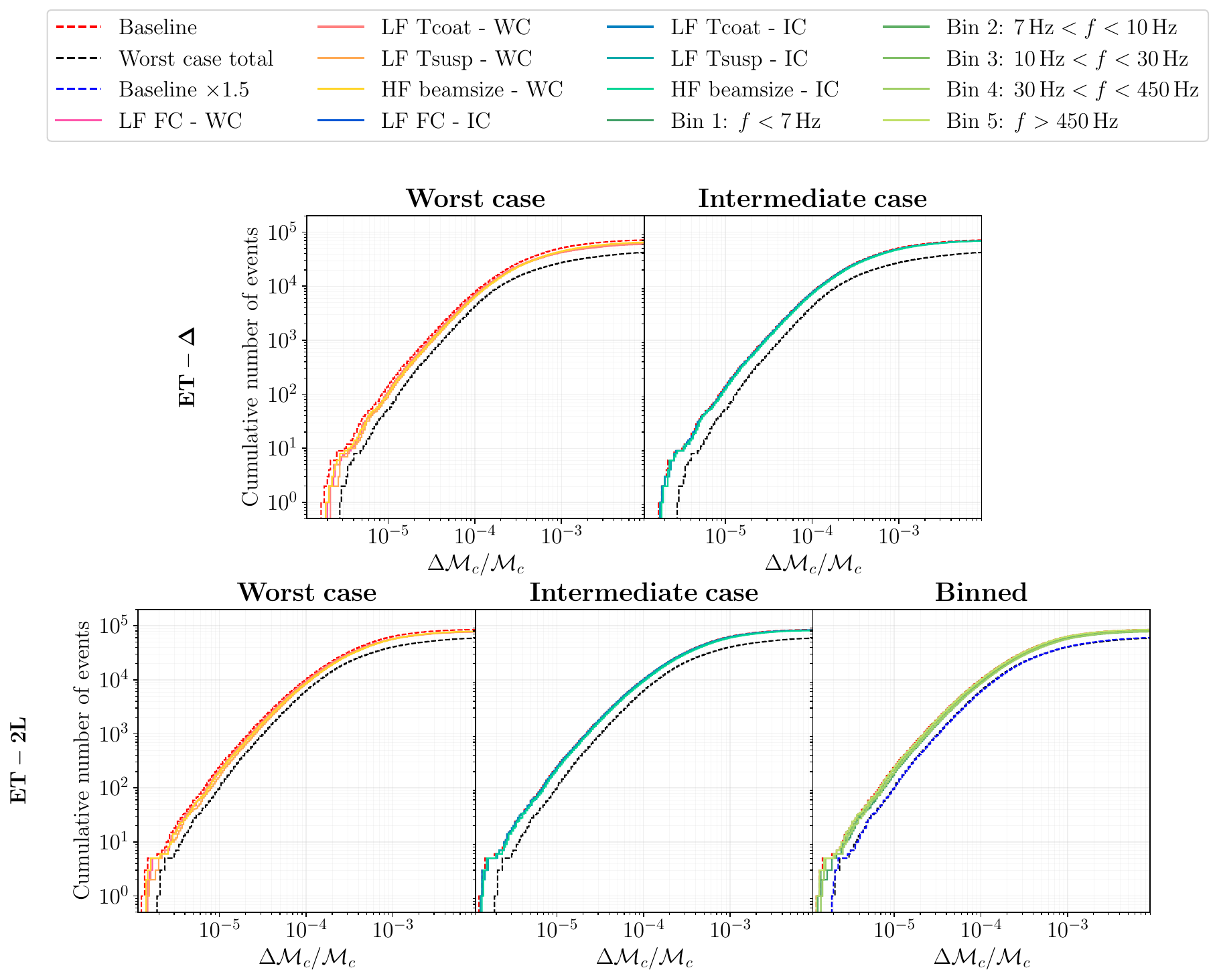}
\caption{Cumulative $\Delta \mathcal{M}_c/\mathcal{M}_c$ distribution for the detected \ac{bbh} events. The top (bottom) row concerns the ET triangular (2L) configuration, while columns correspond to cases as labeled in the panel titles. 
}
\label{fig:chirpmass_BBH}
\end{figure}

\begin{figure}[tbp]
\centering
\includegraphics[width=1.\linewidth]{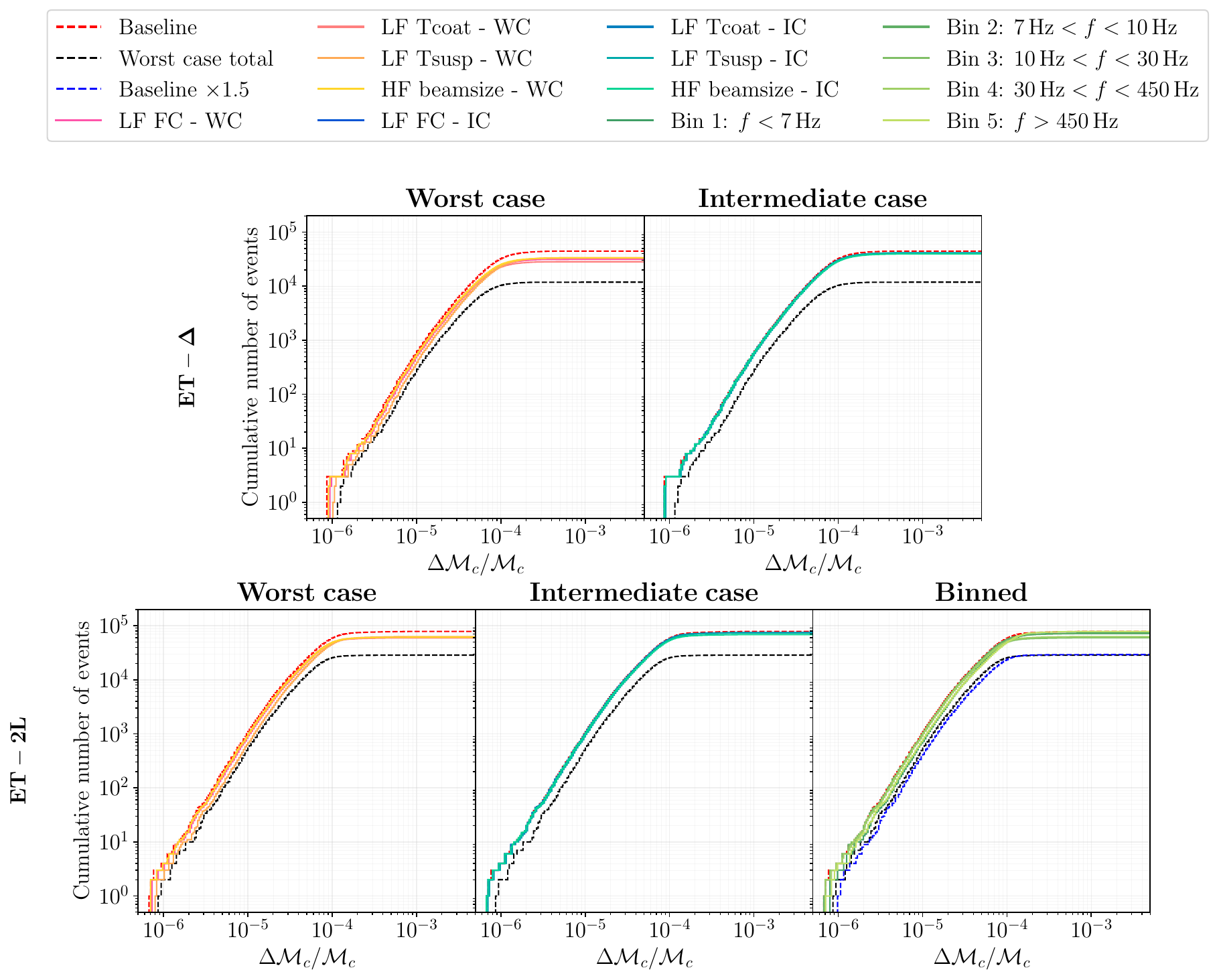}
\caption{Same as \Cref{fig:chirpmass_BBH}, but for the \ac{bns} case. 
}
\label{fig:chirpmass_BNS}
\end{figure}

\begin{figure}[tbp]
\centering
\includegraphics[width=1.\linewidth]{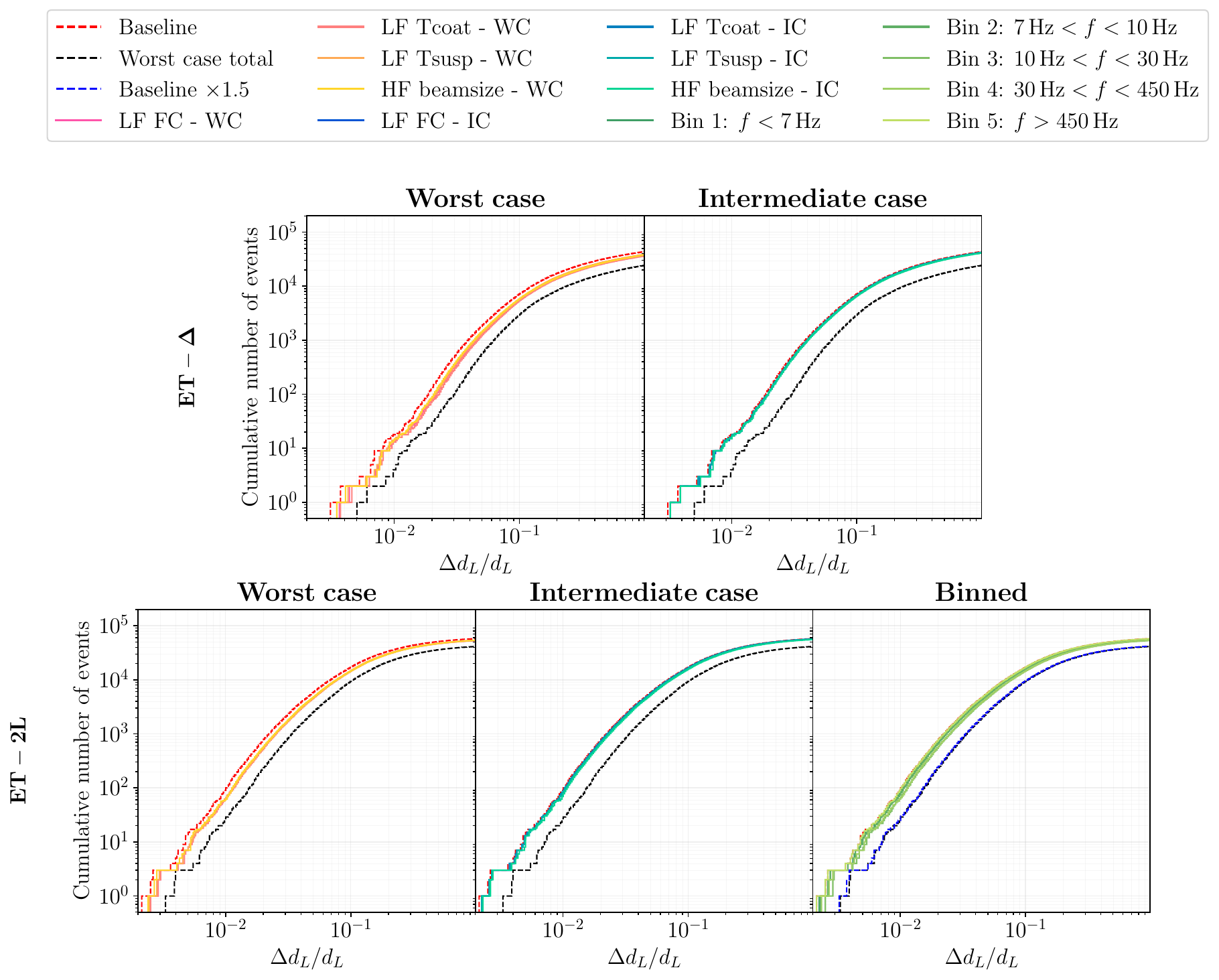}
\caption{Cumulative $\Delta d_L/d_L$ distribution for the detected \ac{bbh} events. The top (bottom) row concerns the ET triangular (2L) configuration, while columns correspond to cases as labeled in the panel titles. 
}
\label{fig:distance_BBH}
\end{figure}

\begin{figure}[tbp]
\centering
\includegraphics[width=1.\linewidth]{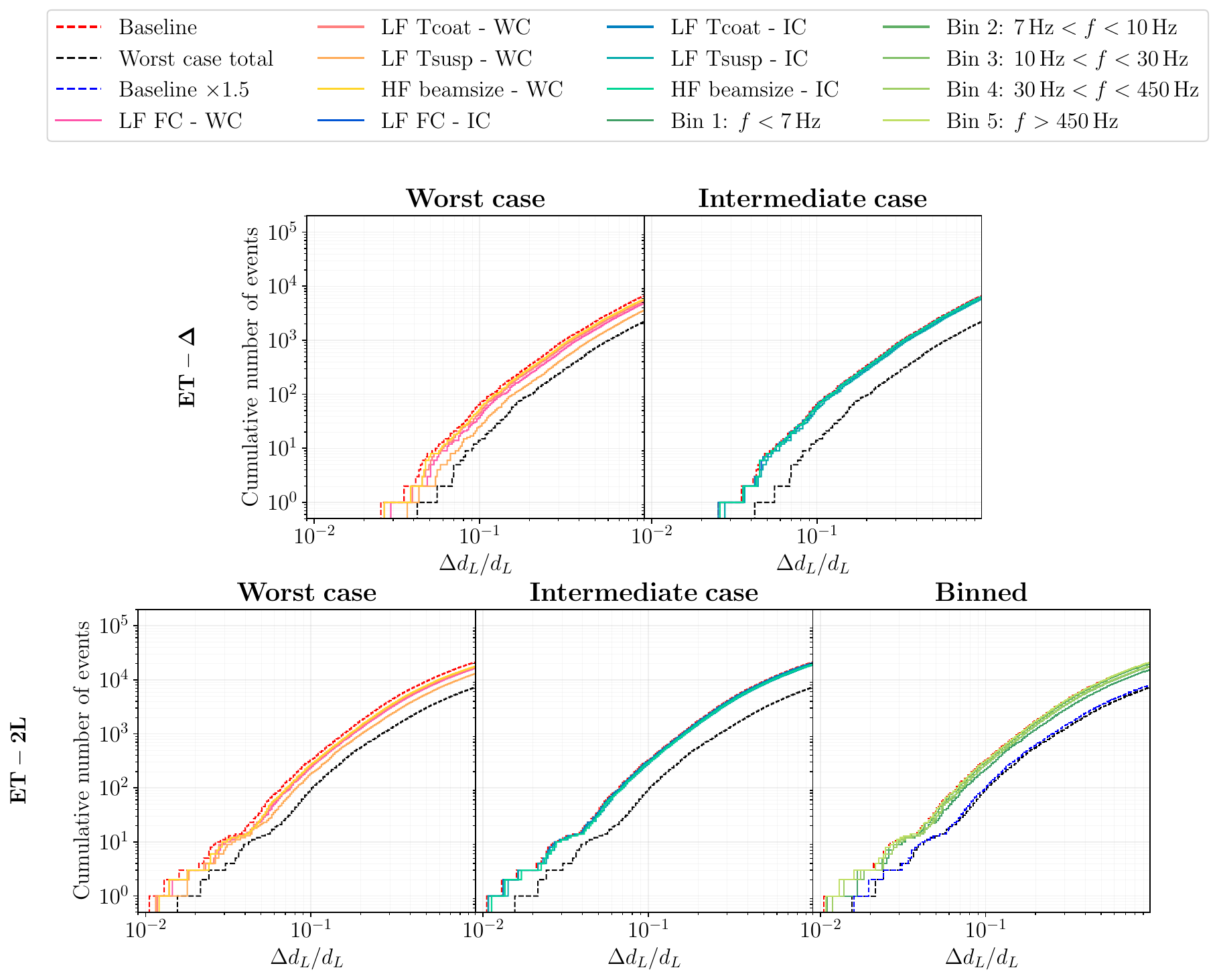}
\caption{Same as \Cref{fig:distance_BBH}, but for the \ac{bns} case. 
}
\label{fig:distance_BNS}
\end{figure}

\begin{figure}[tbp]
\centering
\includegraphics[width=1.\linewidth]{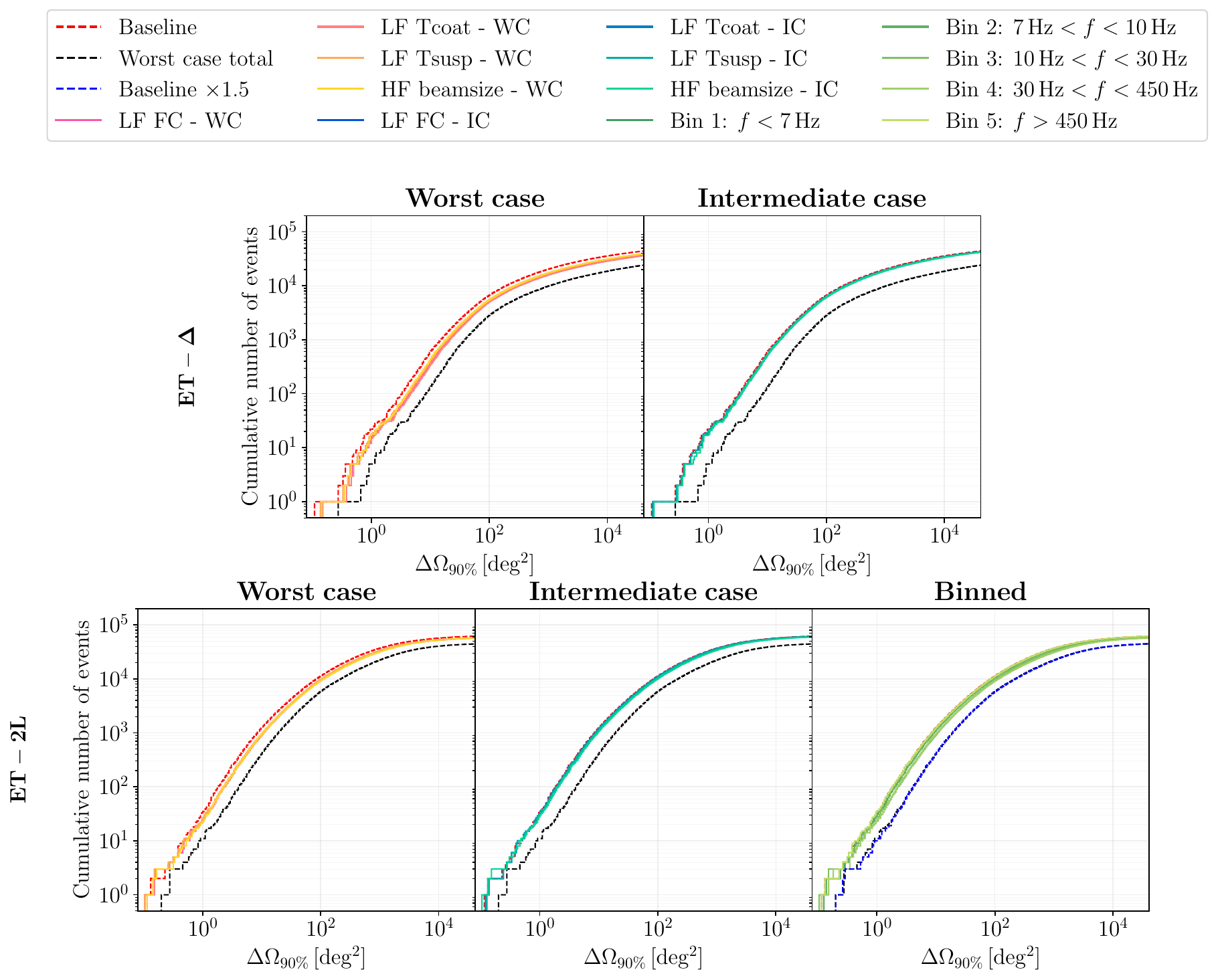}
\caption{Cumulative $\Delta\Omega_{90\%}$ distribution for the detected \ac{bbh} events. The top (bottom) row concerns the ET triangular (2L) configuration, while columns correspond to cases as labeled in the panel titles. 
}
\label{fig:skyloc_BBH}
\end{figure}

\begin{figure}[tbp]
\centering
\includegraphics[width=1.\linewidth]{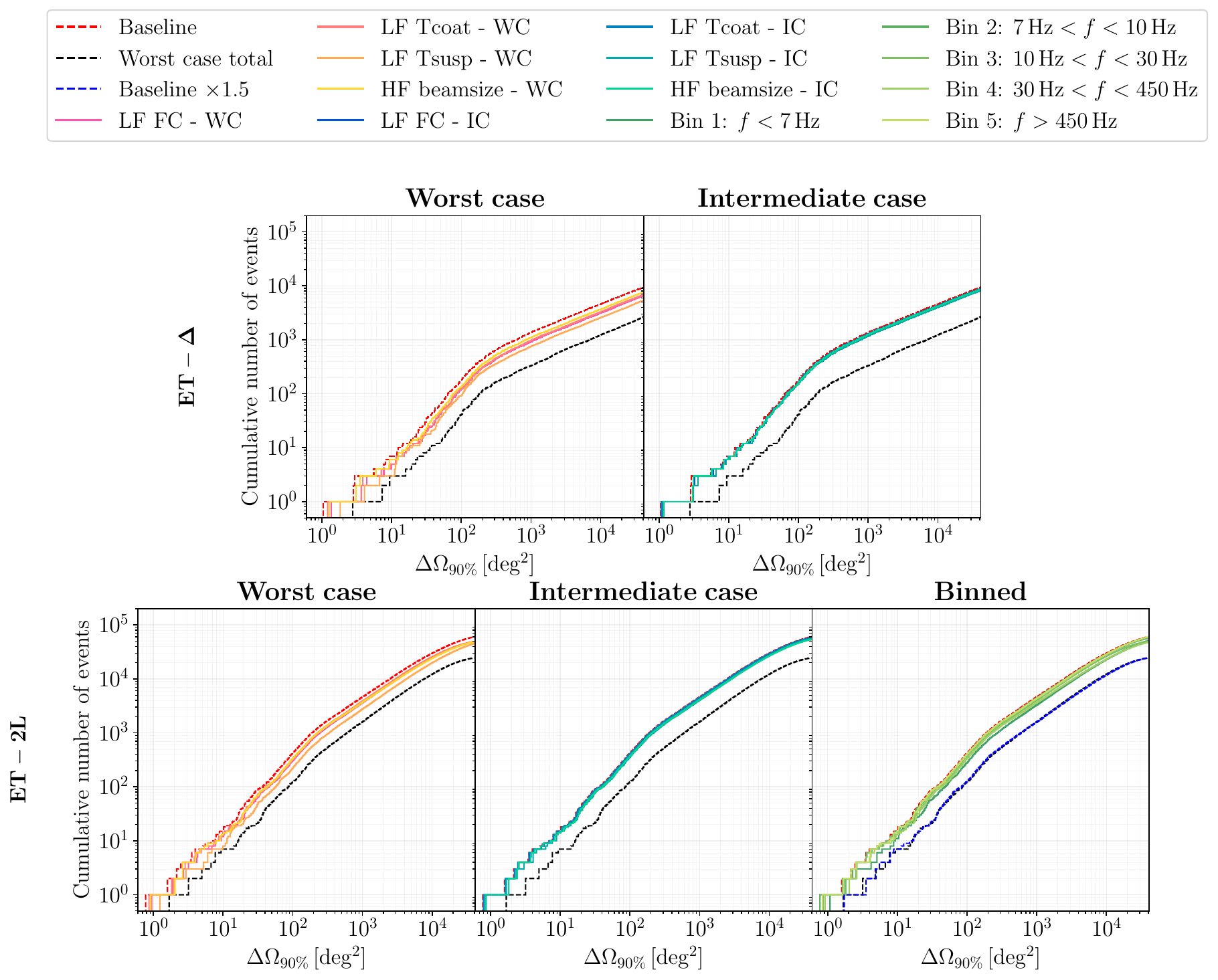}
\caption{Same as \Cref{fig:skyloc_BBH}, but for the \ac{bns} case.  
}
\label{fig:skyloc_BNS}
\end{figure}

We show the results for \acp{bbh} and \acp{bns}, respectively, in Figures~\ref{fig:chirpmass_BBH} and~\ref{fig:chirpmass_BNS} for the detector-frame chirp mass relative error distribution, in Figures~\ref{fig:distance_BBH} and~\ref{fig:distance_BNS} for the luminosity distance relative errors, and in Figures~\ref{fig:skyloc_BBH} and~\ref{fig:skyloc_BNS} for the sky localization error. Additionally, we provide quantitative comparisons across configurations in \Cref{tab:fisher_bbh} for \ac{bbh} systems and in \Cref{tab:fisher_bns} for \ac{bns} systems. In the latter, we include the error in the spin of the primary component for \acp{bbh} and in the tidal deformability for \acp{bns}. Furthermore, these two tables restrict the comparison to well-measured sources. The quantitative results reported in Tables~\ref{tab:fisher_bbh} and~\ref{tab:fisher_bns} reveal that the impact of sensitivity degradations is strongly parameter- and source-dependent. Additionally, the reconstruction of the source properties is more sensitive to degradations in sensitivity than detection rates.

For \acp{bbh}, the chirp mass remains sufficiently well constrained across all configurations. In the worst-case total scenario, the fraction of events with sub-permille accuracy remains above 50\%, confirming that this parameter is less sensitive to moderate losses in sensitivity. In contrast, parameters that depend more directly on the overall \ac{snr} and detector geometry show a more pronounced degradation. The fraction of events with luminosity distance measured to better than $10\%$ drops to $\sim\!41\%-55\%$ in the worst-case total configuration, and the number of well-localized events ($\Delta\Omega_{90\%} \leq 1000\,\mathrm{deg}^2$) is reduced to a similar level. The impact on spin measurements is comparable, with the fraction of events satisfying $\Delta\chi_1/\chi_1 \leq 10\%$ decreasing to $\sim\!54\%-61\%$. Notably, among individual noise contributions, degradations in the LF instrument (in particular suspension and coating thermal noise) produce the largest losses, whereas HF degradations have a comparatively milder effect in the worst-case scenarios.

The situation is more severe for \acp{bns}. While the chirp mass remains robust, other parameters degrade significantly. The fraction of events with luminosity distance measured within $10\%$ drops to $\sim\!23\%-28\%$, indicating a loss by almost a factor of 4. A similar or even stronger reduction is observed for the sky localization. The tidal deformability is particularly affected: even in individual worst-case HF degradations, the fraction of well-measured events can be reduced to $\sim\!54\%-59\%$, and to $\sim\!33\%$ in the total worst-case scenario. This highlights the strong dependence of tidal measurements on HF sensitivity, where tidal effects become more pronounced~\citep{Flanagan:2007ix}. For tidal deformability in particular, the binned analysis shows that the highest-frequency bin above 450\,Hz has the strongest impact, consistent with the fact that tidal effects become increasingly important in the late inspiral. 

Comparing the two source classes, a clear pattern emerges: statistical uncertainties on the source parameters for \acp{bbh} are relatively robust to sensitivity degradations, with most metrics retaining at least $\sim\!55\%-80\%$ of their baseline performance even in pessimistic scenarios. In contrast, \acp{bns} are significantly more vulnerable, particularly for parameters beyond the chirp mass. This difference can be traced to the lower intrinsic \ac{snr} of \acp{bns} and to their stronger reliance on both the early inspiral (for localization and distance) and the late inspiral (for tidal effects). Configuration-wise, when looking at \ac{bns} results especially (see Figures~\ref{fig:detections_snr12_BNS}, \ref{fig:distance_BNS} and~\ref{fig:skyloc_BNS}), we again find that the 2L configuration, even in the worst-case scenario, delivers better results than the triangular configuration. The improved performance of the 2L configuration is mainly due to both longer arms and a longer baseline.

Finally, the intermediate-case scenarios show only marginal deviations from the baseline for both \acp{bbh} and \acp{bns}. This suggests that moderate degradations in individual noise sources can be tolerated without substantially compromising parameter-estimation capabilities.

\subsection{CBC metrics: BNS pre-merger alert}
The low-frequency sensitivity of \ac{et} allows early warnings of \ac{bns} mergers~\cite{Banerjee:2022gkv,Branchesi:2023mws,ET:2025xjr}, by detecting the inspiral signal well before merger and issuing pre-merger alerts with localization information. These early alerts are crucial for multi-messenger astronomy, as they enable rapid follow-up observations and significantly enhance the probability of detecting associated \ac{em} signals. Moreover, they allow for a possible detection of \ac{em} precursors~\cite{Tsang:2011ad,Sridhar:2020uez, Beloborodov:2020ylo, Cooper:2022slk, Dichiara:2023goh, Suvorov:2024cff, Parsons:2026hyt}. Observations of prompt and early-time multi-wavelength emission from \ac{bns} mergers are particularly important for probing the central engine of \acp{GRB}, including the jet composition, particle acceleration processes, and the mechanisms of radiation and energy dissipation. As the emission (both prompt and afterglow) from the merger-driven \acp{GRB} is usually fainter, a rapid localization and early follow-up observations can result in a detection. Moreover, space-based instruments (such as Swift/BAT) can potentially localize gamma-ray emission using early warning alerts~\cite{Tohuvavohu:2024flg}. In the case of kilonovae, early emission is especially sensitive to the properties of the outer, sub-relativistic ejecta and may produce ultraviolet signals detectable by wide-field missions (such as \textit{ULTRASAT}, \textit{QUVIK}), which can strongly benefit from pre-merger alerts.

The detection of \acp{gw} from \ac{bns} systems before the merger strongly depends on the detector design, specifically on the sensitivity at lower frequencies~\cite{Sachdev:2020lfd}. 
In this work, we report the expected number of detections with $\mathrm{SNR}\geq8$ at different times before merger, together with the corresponding sky-localization uncertainty. We consider three pre-merger alert times: 30, 10, and 1 minute before coalescence. Starting from the \ac{bns} population described in Section~\ref{sec:bns}, we use the \texttt{GWFish} framework to assess the localization performance of the 2L 15~km \ac{et} configuration under different scenarios (see Section~\ref{sec:noise_budget}). \Cref{tab:premerger} presents the expected annual detection rates for events with 90\% sky-localization areas smaller than 10, 100, and 1000~deg$^{2}$, normalized to the Baseline sensitivity. In addition, \Cref{tab:premerger} reports the corresponding detection rates for randomly oriented systems and for the subset of \acp{bns} with viewing angle $\Theta_v={\rm min}(\iota,\, 180^\circ-\iota)\leq15^{\circ}$. This subset corresponds to what we consider on-axis systems, for which detecting the jet’s beamed emission is possible. Note that for this science case, given the low statistics, we did not include the duty cycle in the analyses as in \Cref{sec:detection_and_fisher}.

\Cref{tab:premerger} shows that the pre-merger performance of the 2L 15~km \ac{et} configuration is driven mainly by the low-frequency portion of the sensitivity curve. This is evident from the binned analysis: degradations in the first two frequency bins, $f<7\,\mathrm{Hz}$ and $7\,\mathrm{Hz}<f<10\,\mathrm{Hz}$, lead to reductions of more than $50\%$ in the number of pre-merger detections and well-localized events, while the impact is milder in the third bin, $10\,\mathrm{Hz}<f<30\,\mathrm{Hz}$, and negligible at higher frequencies. This behavior is expected, since signals 30, 10, and 1 minute before coalescence lie below 30, 15, and 8\,Hz, respectively, within the detector band (see, e.g., Eq.~(11a) of Ref.~\cite{Marsat:2018oam}). As a consequence, the pre-merger alert and localization capabilities are determined almost entirely by the detector sensitivity in the low-frequency inspiral regime. In Appendix~\ref{app:pre_merger_frequency}, we briefly discuss the relation between pre-merger time and \ac{gw} frequency, and show the distribution for our \ac{bns} population.

This behavior is confirmed when comparing individual degradations for both intermediate and worst-case scenarios. The LF cases (LF FC, LF Tcoat, and LF Tsusp) have a major impact, with the degradation in the suspension temperature leading to the worst performance among the four, down to $\sim\!25\%$ of the baseline. %
The HF beam size degradation, on the other hand, is essentially neutral. \Cref{fig:noise_scenarios}, in fact, shows that varying the HF beam size changes the ET-HF sensitivity curve only at higher frequencies, while below $\sim\!20$\,Hz the degraded and baseline curves are nearly identical, so pre-merger counts and localizations are almost unchanged. For on-axis events, the absolute numbers are much lower, but the overall trend confirms the results from randomly oriented systems.

{

\newcolumntype{L}{>{\raggedright\arraybackslash\normalsize\hsize=2.4\hsize}X}

\newcolumntype{M}{>{\centering\arraybackslash\normalsize\hsize=0.95\hsize}X}

\newcolumntype{N}{>{\centering\arraybackslash\normalsize\hsize=0.68\hsize}X}

\setlength{\tabcolsep}{2pt}

\begin{xltabular}{\textwidth}{L M N N N N N N}
\caption{Pre-merger \ac{bns} counts for 2L 15\,km \ac{et} configuration for different cuts on the sky localization accuracy at 30\,min, 10\,min, and 1\,min prior to merger. We report the results for binaries with all orientations and for those with a viewing angle $\Theta_v\leq15^\circ$. The first row shows the absolute numbers for the Baseline sensitivity curve. For all the other sensitivities, we report the ratio of counts for a given sensitivity relative to the Baseline sensitivity.} 
\label{tab:premerger}\\
\toprule
\multicolumn{1}{c}{} &
\multicolumn{1}{c}{} &
\multicolumn{3}{c}{All orientation BNSs} &
\multicolumn{3}{c}{BNSs with $\Theta_v \leq 15^\circ$} \\ 
\cmidrule(r){3-5} \cmidrule{6-8}
\textbf{Sensitivity curve} & $\leq\Delta\Omega_{90\%}$ & 30 min & 10 min & 1 min & 30 min & 10 min & 1 min  \\ \hline
\endfirsthead
\toprule
\multicolumn{1}{c}{} &
\multicolumn{1}{c}{} &
\multicolumn{3}{c}{All orientation BNSs} &
\multicolumn{3}{c}{BNSs with $\Theta_v \leq 15^\circ$} \\
\cmidrule(r){3-5} \cmidrule{6-8}
\textbf{Sensitivity curve} & $\leq\Delta\Omega_{90\%}$ & 30 min & 10 min & 1 min & 30 min & 10 min & 1 min  \\
\midrule
\endhead
{Baseline} & 10 & 2 & 5 & 9 & 0 & 2 & 2 \\
{} & 100 & 33 & 84 & 201 & 4 & 8 & 20 \\
{} & 1000 & 350 & 839 & 1949 & 32 & 91 & 244 \\
{} & Total & 2590 & 10430 & 39586 & 235 & 942 & 3776 \\
\rowcolor{MidnightBlueLight}{Baseline $\times 1.5$} & 10 & 0\% & 40\% & 44\% & - & 0\% & 50\% \\
\rowcolor{MidnightBlueLight}{} & 100 & 42\% & 31\% & 34\% & 75\% & 38\% & 30\% \\
\rowcolor{MidnightBlueLight}{} & 1000 & 32\% & 35\% & 35\% & 31\% & 34\% & 35\% \\
\rowcolor{MidnightBlueLight}{} & Total & 37\% & 36\% & 34\% & 39\% & 41\% & 37\% \\
{Bin 1: $f<7\,\mathrm{Hz}$} & 10 & 0\% & 60\% & 56\% & - & 0\% & 50\% \\
{} & 100 & 48\% & 52\% & 65\% & 75\% & 62\% & 55\% \\
{} & 1000 & 37\% & 56\% & 66\% & 31\% & 57\% & 74\% \\
{} & Total & 39\% & 65\% & 84\% & 39\% & 64\% & 85\% \\
\rowcolor{MidnightBlueLight}{Bin 2: $7\,\mathrm{Hz}<f<10\,\mathrm{Hz}$} & 10 & 100\% & 80\% & 78\% & - & 50\% & 100\% \\
\rowcolor{MidnightBlueLight}{} & 100 & 91\% & 74\% & 86\% & 100\% & 62\% & 80\% \\
\rowcolor{MidnightBlueLight}{} & 1000 & 95\% & 77\% & 89\% & 100\% & 67\% & 90\% \\
\rowcolor{MidnightBlueLight}{} & Total & 99\% & 70\% & 80\% & 100\% & 78\% & 81\% \\
{Bin 3: $10\,\mathrm{Hz}<f<30\,\mathrm{Hz}$} & 10 & 100\% & 80\% & 78\% & - & 50\% & 100\% \\
{} & 100 & 100\% & 94\% & 68\% & 100\% & 100\% & 60\% \\
{} & 1000 & 100\% & 98\% & 72\% & 100\% & 100\% & 73\% \\
{} & Total & 100\% & 100\% & 70\% & 100\% & 100\% & 77\% \\
\rowcolor{MidnightBlueLight}{Bin 4: $30\,\mathrm{Hz}<f<450\,\mathrm{Hz}$} & 10 & 100\% & 100\% & 100\% & - & 100\% & 100\% \\
\rowcolor{MidnightBlueLight}{} & 100 & 100\% & 100\% & 100\% & 100\% & 100\% & 100\% \\
\rowcolor{MidnightBlueLight}{} & 1000 & 100\% & 100\% & 100\% & 100\% & 100\% & 100\% \\
\rowcolor{MidnightBlueLight}{} & Total & 100\% & 100\% & 100\% & 100\% & 100\% & 100\% \\
{Bin 5: $f>450\,\mathrm{Hz}$} & 10 & 100\% & 100\% & 100\% & - & 100\% & 100\% \\
{} & 100 & 100\% & 100\% & 100\% & 100\% & 100\% & 100\% \\
{} & 1000 & 100\% & 100\% & 100\% & 100\% & 100\% & 100\% \\
{} & Total & 100\% & 100\% & 100\% & 100\% & 100\% & 100\% \\
\rowcolor{MidnightBlueLight}{LF FC - IC} & 10 & 100\% & 100\% & 78\% & - & 100\% & 100\% \\
\rowcolor{MidnightBlueLight}{} & 100 & 97\% & 90\% & 89\% & 100\% & 88\% & 85\% \\
\rowcolor{MidnightBlueLight}{} & 1000 & 91\% & 92\% & 91\% & 84\% & 88\% & 93\% \\
\rowcolor{MidnightBlueLight}{} & Total & 91\% & 90\% & 90\% & 91\% & 91\% & 91\% \\
{LF Tcoat - IC} & 10 & 100\% & 100\% & 89\% & - & 100\% & 100\% \\
{} & 100 & 97\% & 93\% & 93\% & 100\% & 88\% & 85\% \\
{} & 1000 & 95\% & 95\% & 94\% & 91\% & 91\% & 95\% \\
{} & Total & 96\% & 93\% & 91\% & 95\% & 94\% & 93\% \\
\rowcolor{MidnightBlueLight}{LF Tsusp - IC} & 10 & 100\% & 80\% & 89\% & - & 50\% & 100\% \\
\rowcolor{MidnightBlueLight}{} & 100 & 82\% & 80\% & 81\% & 100\% & 75\% & 70\% \\
\rowcolor{MidnightBlueLight}{} & 1000 & 72\% & 77\% & 82\% & 72\% & 71\% & 85\% \\
\rowcolor{MidnightBlueLight}{} & Total & 69\% & 74\% & 86\% & 68\% & 75\% & 87\% \\
{HF beamsize - IC} & 10 & 100\% & 100\% & 100\% & - & 100\% & 100\% \\
{} & 100 & 100\% & 100\% & 98\% & 100\% & 100\% & 90\% \\
{} & 1000 & 100\% & 100\% & 99\% & 100\% & 100\% & 100\% \\
{} & Total & 100\% & 100\% & 99\% & 100\% & 100\% & 100\% \\
\rowcolor{MidnightBlueLight}{LF FC - WC} & 10 & 100\% & 60\% & 78\% & - & 0\% & 100\% \\
\rowcolor{MidnightBlueLight}{} & 100 & 73\% & 55\% & 61\% & 100\% & 62\% & 45\% \\
\rowcolor{MidnightBlueLight}{} & 1000 & 63\% & 65\% & 63\% & 66\% & 59\% & 66\% \\
\rowcolor{MidnightBlueLight}{} & Total & 59\% & 57\% & 59\% & 60\% & 60\% & 63\% \\
{LF Tcoat - WC} & 10 & 100\% & 60\% & 78\% & - & 0\% & 100\% \\
{} & 100 & 85\% & 61\% & 62\% & 100\% & 62\% & 50\% \\
{} & 1000 & 76\% & 69\% & 66\% & 75\% & 64\% & 66\% \\
{} & Total & 73\% & 59\% & 55\% & 75\% & 63\% & 59\% \\
\rowcolor{MidnightBlueLight}{LF Tsusp - WC} & 10 & 0\% & 20\% & 44\% & - & 0\% & 50\% \\
\rowcolor{MidnightBlueLight}{} & 100 & 42\% & 32\% & 47\% & 75\% & 38\% & 35\% \\
\rowcolor{MidnightBlueLight}{} & 1000 & 22\% & 33\% & 49\% & 31\% & 29\% & 54\% \\
\rowcolor{MidnightBlueLight}{} & Total & 26\% & 33\% & 58\% & 27\% & 36\% & 59\% \\
{HF beamsize - WC} & 10 & 100\% & 100\% & 100\% & - & 100\% & 100\% \\
{} & 100 & 100\% & 100\% & 97\% & 100\% & 100\% & 90\% \\
{} & 1000 & 100\% & 100\% & 98\% & 100\% & 100\% & 100\% \\
{} & Total & 100\% & 100\% & 99\% & 100\% & 100\% & 99\% \\
\rowcolor{MidnightBlueLight}{Worst case total} & 10 & 0\% & 20\% & 22\% & - & 0\% & 0\% \\
\rowcolor{MidnightBlueLight}{} & 100 & 30\% & 19\% & 24\% & 50\% & 38\% & 25\% \\
\rowcolor{MidnightBlueLight}{} & 1000 & 17\% & 19\% & 26\% & 25\% & 16\% & 25\% \\
\rowcolor{MidnightBlueLight}{} & Total & 19\% & 19\% & 26\% & 22\% & 22\% & 28\% \\

\bottomrule

\end{xltabular}

}

\subsection{CBC metrics: BNS post-merger}\label{sec:bns_postmerg}

The emission in the post-merger phase of \ac{bns} mergers is a goldmine for understanding the \ac{ns} structure~\cite{Baiotti:2016qnr,Sarin:2020gxb}. Despite being extremely energetic, potentially more than the merger itself, this type of signal is short-lived and sits on the high-frequency part of the spectrum accessible to ground-based instruments, where the detector sensitivity begins to decrease~\cite{Torres-Rivas:2018svp,Wijngaarden:2022sah}. For this reason, its detectability is challenging. While in practice the detection of the post-merger signal will follow the detection of the inspiral, we focus here on the \ac{snr} in the post-merger only as a proxy for how well we will be able to characterize it with future instruments. 

\begin{figure}[tbp]
    \centering
    \includegraphics[width=1.\linewidth]{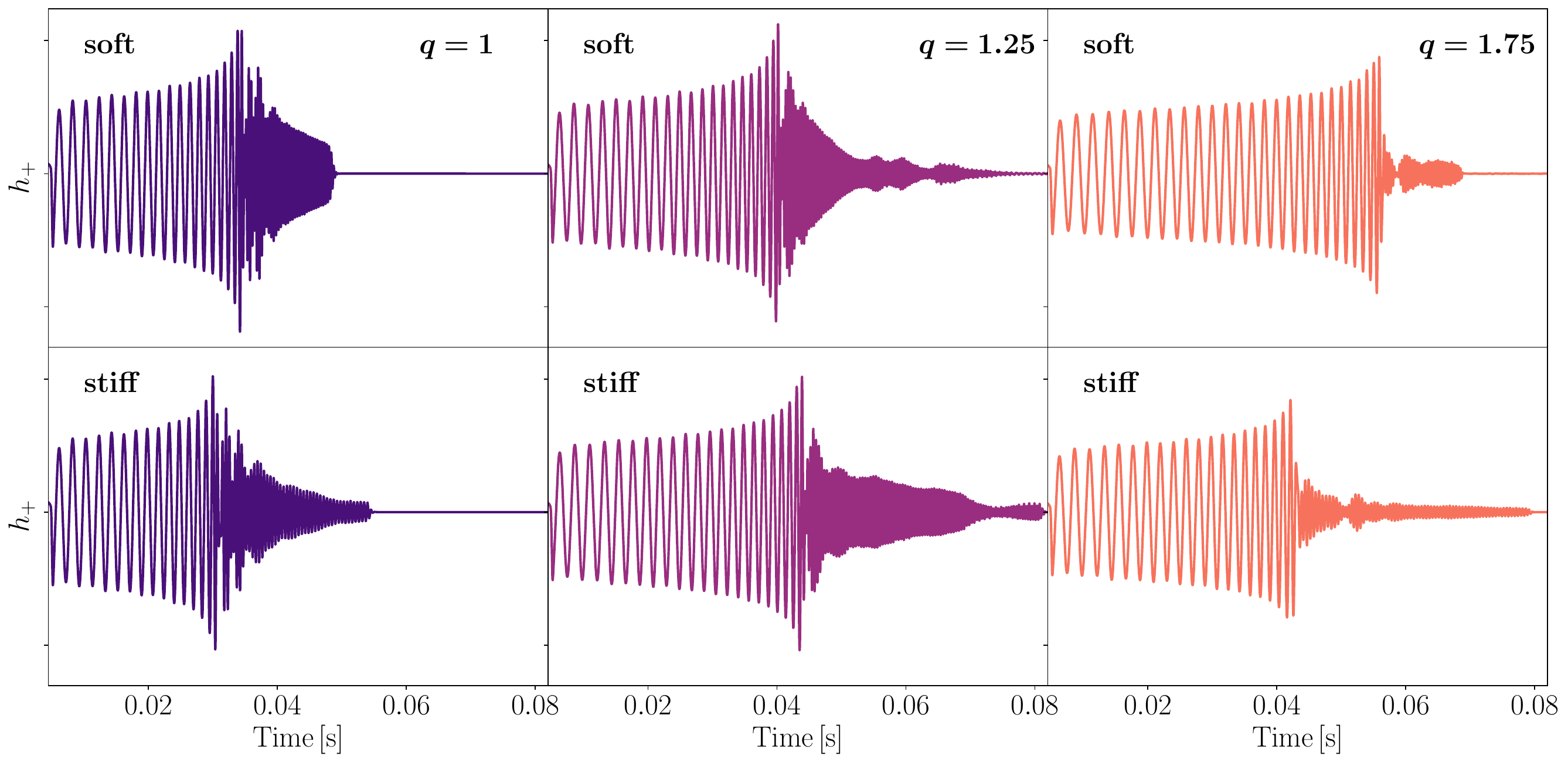}
    \caption{Dimensionless strain for the $+$ polarization of the chosen \ac{bns} \ac{nr} simulations as a function of time. Each column corresponds to a different mass ratio, with the top and bottom panels showing the signal for a soft and stiff \ac{eos}, respectively. Arbitrary units are used for the $y$ axis for ease of visualization.}
    \label{fig:bns_pm_nr_wfs}
\end{figure}

The post-merger morphology strongly depends on the binary properties and the assumed \ac{eos}. For this reason, we select a sample of six \ac{bns} \ac{nr} simulations from the \href{http://www.computational-relativity.org}{CoRe} database~\cite{Gonzalez:2022mgo} with different masses and \acp{eos}, and restrict to non-spinning systems. The chosen simulations have varying mass ratios, equal-mass ($m_1=m_2=1.35\,M_\odot$), mild ($m_1=1.528\,M_\odot,\ m_2=1.222\,M_\odot$), and moderate ($m_1=1.75\,M_\odot,\ m_2=1\,M_\odot$), and for each mass ratio, we further select two different \acp{eos}, one soft and one stiff, associated with lower and higher tidal deformabilities, respectively. The simulations span a range of tidal deformabilities from $\Lambda\sim350$ to $\Lambda\sim7000$. For $q=1$, the two selected simulations are \texttt{BAM:0003} and \texttt{BAM:0060}~\cite{Bernuzzi:2014kca}, for $q=1.25$ \texttt{BAM:0012}~\cite{Dietrich:2016hky} and \texttt{BAM:0048}, and for $q=1.75$ \texttt{BAM:0021} and \texttt{BAM:0093}~\cite{Dietrich:2016hky}, respectively, for the lower (soft \ac{eos}) and higher (stiff \ac{eos}) value of $\Lambda$. The chosen waveforms are shown in \Cref{fig:bns_pm_nr_wfs}.
After computing the Fourier transform for each simulation, we compute the \ac{snr} as in Eq.~\eqref{eq:snr_net} for a system with optimal sky position and orientation at the chosen configuration (obtained by numerical maximization, and different for 2L and $\Delta$), starting the analysis from the merger frequency determined as in Eq.~(11) of Ref.~\cite{Dietrich:2018uni} and extending it to $f_{\rm max}=8192\,{\rm Hz}$. Concerning the extrinsic parameters, we select the optimal sky position and orientation for the sources for the two detector configurations considered at a reference time, and fix the distance to 100\,Mpc.

\begin{table}[tb]
\caption{\acp{snr} for the different \ac{bns} post-merger waveforms, evaluated using a triangular detector configuration and the sensitivity curves considered in this work. In the first row, we report the SNR value for the Baseline sensitivity curve, and in all subsequent rows, the results relative to this value. The values in parentheses refer to the case in which the \ac{lwa} is used.}
    \centering
    {\setlength{\tabcolsep}{9.6pt}
    \begin{tabularx}{\textwidth}{lcccccc}
    	\toprule
    	\multicolumn{7}{c}{\textbf{Triangle}}\\\midrule
	 & \multicolumn{6}{c}{SNR}\\ \cmidrule{2-7}
	 & $q=1$ & $q=1.25$ & $q=1.75$ & $q=1$ & $q=1.25$ & $q=1.75$ \\
	\textbf{Sensitivity curve} & soft & soft & soft & stiff & stiff & stiff \\ \midrule
        \multirow{2}{*}{Baseline} & 8.21 & 7.85 & 4.35 & 9.97 & 8.26 & 5.39 \\
	 & {\footnotesize (8.61)} & {\footnotesize (8.21)} & {\footnotesize (4.45)} & {\footnotesize (10.23)} & {\footnotesize (8.58)} & {\footnotesize (5.48)} \\
	\rowcolor{MidnightBlueLight} & 99\% & 99\% & 99\% & 99\% & 99\% & 99\% \\
	\rowcolor{MidnightBlueLight} \multirow{-2}{*}{LF FC - IC} & {\footnotesize (99\%)} & {\footnotesize (99\%)} & {\footnotesize (99\%)} & {\footnotesize (99\%)} & {\footnotesize (99\%)} & {\footnotesize (99\%)} \\
	 \multirow{2}{*}{LF Tcoat - IC} & 99\% & 99\% & 99\% & 99\% & 99\% & 99\% \\
	 & {\footnotesize (99\%)} & {\footnotesize (99\%)} & {\footnotesize (99\%)} & {\footnotesize (99\%)} & {\footnotesize (99\%)} & {\footnotesize (99\%)} \\
	\rowcolor{MidnightBlueLight} & 99\% & 99\% & 99\% & 99\% & 99\% & 99\% \\
	\rowcolor{MidnightBlueLight} \multirow{-2}{*}{LF Tsusp - IC} & {\footnotesize (99\%)} & {\footnotesize (99\%)} & {\footnotesize (99\%)} & {\footnotesize (99\%)} & {\footnotesize (99\%)} & {\footnotesize (99\%)} \\
	 \multirow{2}{*}{HF beam size - IC} & 99\% & 99\% & 99\% & 99\% & 99\% & 99\% \\
	 & {\footnotesize (99\%)} & {\footnotesize (99\%)} & {\footnotesize (99\%)} & {\footnotesize (99\%)} & {\footnotesize (99\%)} & {\footnotesize (99\%)} \\
	\rowcolor{MidnightBlueLight} & 99\% & 99\% & 99\% & 99\% & 99\% & 99\% \\
	\rowcolor{MidnightBlueLight} \multirow{-2}{*}{LF FC - WC} & {\footnotesize (99\%)} & {\footnotesize (99\%)} & {\footnotesize (99\%)} & {\footnotesize (99\%)} & {\footnotesize (99\%)} & {\footnotesize (99\%)} \\
	 \multirow{2}{*}{LF Tcoat - WC} & 99\% & 99\% & 99\% & 99\% & 99\% & 99\% \\
	 & {\footnotesize (99\%)} & {\footnotesize (99\%)} & {\footnotesize (99\%)} & {\footnotesize (99\%)} & {\footnotesize (99\%)} & {\footnotesize (99\%)} \\
	\rowcolor{MidnightBlueLight} & 99\% & 99\% & 99\% & 99\% & 99\% & 99\% \\
	\rowcolor{MidnightBlueLight} \multirow{-2}{*}{LF Tsusp - WC} & {\footnotesize (99\%)} & {\footnotesize (99\%)} & {\footnotesize (99\%)} & {\footnotesize (99\%)} & {\footnotesize (99\%)} & {\footnotesize (99\%)} \\
	 \multirow{2}{*}{HF beam size - WC} & 99\% & 99\% & 97\% & 98\% & 99\% & 97\% \\
	 & {\footnotesize (99\%)} & {\footnotesize (99\%)} & {\footnotesize (97\%)} & {\footnotesize (98\%)} & {\footnotesize (99\%)} & {\footnotesize (97\%)} \\
	\rowcolor{MidnightBlueLight} & 99\% & 99\% & 98\% & 98\% & 99\% & 97\% \\
	\rowcolor{MidnightBlueLight} \multirow{-2}{*}{Worst case total} & {\footnotesize (99\%)} & {\footnotesize (99\%)} & {\footnotesize (98\%)} & {\footnotesize (98\%)} & {\footnotesize (99\%)} & {\footnotesize (97\%)} \\
	\bottomrule

    \end{tabularx}
    }
    \label{tab:bns_pm_table_delta}
\end{table}

\begin{table}[tbp]
\caption{Same as \Cref{tab:bns_pm_table_delta} for the 2L configuration.}
    \centering
    {\setlength{\tabcolsep}{7pt}
    \begin{tabularx}{\textwidth}{lcccccc}
    	\toprule
    	\multicolumn{7}{c}{\textbf{2L misaligned}}\\\midrule
	 & \multicolumn{6}{c}{SNR}\\ \cmidrule{2-7}
	 & $q=1$ & $q=1.25$ & $q=1.75$ & $q=1$ & $q=1.25$ & $q=1.75$ \\
	\textbf{Sensitivity curve} & soft & soft & soft & stiff & stiff & stiff \\ \midrule
        \multirow{2}{*}{Baseline} & 7.85 & 7.57 & 4.47 & 10.06 & 8.10 & 5.60 \\
	 & {\footnotesize (8.76)} & {\footnotesize (8.37)} & {\footnotesize (4.69)} & {\footnotesize (10.66)} & {\footnotesize (8.80)} & {\footnotesize (5.82)} \\
	\rowcolor{MidnightBlueLight} & 66\% & 66\% & 66\% & 66\% & 66\% & 66\% \\
	\rowcolor{MidnightBlueLight} \multirow{-2}{*}{Baseline $\times 1.5$} & {\footnotesize (66\%)} & {\footnotesize (66\%)} & {\footnotesize (66\%)} & {\footnotesize (66\%)} & {\footnotesize (66\%)} & {\footnotesize (66\%)} \\
	 \multirow{2}{*}{Bin 1: $f<7\,{\rm Hz}$} & 100\% & 100\% & 100\% & 100\% & 100\% & 100\% \\
	 & {\footnotesize (100\%)} & {\footnotesize (100\%)} & {\footnotesize (100\%)} & {\footnotesize (100\%)} & {\footnotesize (99\%)} & {\footnotesize (100\%)} \\
	\rowcolor{MidnightBlueLight} & 100\% & 100\% & 100\% & 100\% & 100\% & 100\% \\
	\rowcolor{MidnightBlueLight} \multirow{-2}{*}{Bin 2: $7\,{\rm Hz}<f<10\,{\rm Hz}$} & {\footnotesize (100\%)} & {\footnotesize (100\%)} & {\footnotesize (100\%)} & {\footnotesize (100\%)} & {\footnotesize (99\%)} & {\footnotesize (100\%)} \\
	 \multirow{2}{*}{Bin 3: $10\,{\rm Hz}<f<30\,{\rm Hz}$} & 100\% & 100\% & 100\% & 100\% & 100\% & 100\% \\
	 & {\footnotesize (100\%)} & {\footnotesize (100\%)} & {\footnotesize (100\%)} & {\footnotesize (100\%)} & {\footnotesize (99\%)} & {\footnotesize (100\%)} \\
	\rowcolor{MidnightBlueLight} & 100\% & 100\% & 100\% & 100\% & 100\% & 100\% \\
	\rowcolor{MidnightBlueLight} \multirow{-2}{*}{Bin 4: $30\,{\rm Hz}<f<450\,{\rm Hz}$} & {\footnotesize (100\%)} & {\footnotesize (100\%)} & {\footnotesize (100\%)} & {\footnotesize (100\%)} & {\footnotesize (99\%)} & {\footnotesize (100\%)} \\
	 \multirow{2}{*}{Bin 5: $f>450\,{\rm Hz}$} & 66\% & 66\% & 66\% & 66\% & 66\% & 66\% \\
	 & {\footnotesize (66\%)} & {\footnotesize (66\%)} & {\footnotesize (66\%)} & {\footnotesize (66\%)} & {\footnotesize (66\%)} & {\footnotesize (66\%)} \\
	\rowcolor{MidnightBlueLight} & 99\% & 99\% & 99\% & 99\% & 99\% & 99\% \\
	\rowcolor{MidnightBlueLight} \multirow{-2}{*}{LF FC - IC} & {\footnotesize (99\%)} & {\footnotesize (99\%)} & {\footnotesize (99\%)} & {\footnotesize (99\%)} & {\footnotesize (99\%)} & {\footnotesize (99\%)} \\
	 \multirow{2}{*}{LF Tcoat - IC} & 99\% & 99\% & 99\% & 99\% & 99\% & 99\% \\
	 & {\footnotesize (99\%)} & {\footnotesize (99\%)} & {\footnotesize (99\%)} & {\footnotesize (99\%)} & {\footnotesize (99\%)} & {\footnotesize (99\%)} \\
	\rowcolor{MidnightBlueLight} & 99\% & 99\% & 99\% & 99\% & 99\% & 99\% \\
	\rowcolor{MidnightBlueLight} \multirow{-2}{*}{LF Tsusp - IC} & {\footnotesize (99\%)} & {\footnotesize (99\%)} & {\footnotesize (99\%)} & {\footnotesize (99\%)} & {\footnotesize (99\%)} & {\footnotesize (99\%)} \\
	 \multirow{2}{*}{HF beam size - IC} & 99\% & 99\% & 99\% & 99\% & 99\% & 99\% \\
	 & {\footnotesize (99\%)} & {\footnotesize (99\%)} & {\footnotesize (99\%)} & {\footnotesize (99\%)} & {\footnotesize (99\%)} & {\footnotesize (99\%)} \\
	\rowcolor{MidnightBlueLight} & 99\% & 99\% & 99\% & 99\% & 99\% & 99\% \\
	\rowcolor{MidnightBlueLight} \multirow{-2}{*}{LF FC - WC} & {\footnotesize (99\%)} & {\footnotesize (99\%)} & {\footnotesize (99\%)} & {\footnotesize (99\%)} & {\footnotesize (99\%)} & {\footnotesize (99\%)} \\
	 \multirow{2}{*}{LF Tcoat - WC} & 99\% & 99\% & 99\% & 99\% & 99\% & 99\% \\
	 & {\footnotesize (99\%)} & {\footnotesize (99\%)} & {\footnotesize (99\%)} & {\footnotesize (99\%)} & {\footnotesize (99\%)} & {\footnotesize (99\%)} \\
	\rowcolor{MidnightBlueLight} & 99\% & 99\% & 99\% & 99\% & 99\% & 99\% \\
	\rowcolor{MidnightBlueLight} \multirow{-2}{*}{LF Tsusp - WC} & {\footnotesize (99\%)} & {\footnotesize (99\%)} & {\footnotesize (99\%)} & {\footnotesize (99\%)} & {\footnotesize (99\%)} & {\footnotesize (99\%)} \\
	 \multirow{2}{*}{HF beam size - WC} & 99\% & 99\% & 98\% & 99\% & 99\% & 98\% \\
	 & {\footnotesize (99\%)} & {\footnotesize (99\%)} & {\footnotesize (98\%)} & {\footnotesize (99\%)} & {\footnotesize (99\%)} & {\footnotesize (98\%)} \\
	\rowcolor{MidnightBlueLight} & 99\% & 99\% & 98\% & 99\% & 99\% & 98\% \\
	\rowcolor{MidnightBlueLight} \multirow{-2}{*}{Worst case total} & {\footnotesize (99\%)} & {\footnotesize (99\%)} & {\footnotesize (98\%)} & {\footnotesize (99\%)} & {\footnotesize (99\%)} & {\footnotesize (98\%)} \\
	\bottomrule

    \end{tabularx}
    }
    \label{tab:bns_pm_table_2l}
\end{table}

A relevant aspect of this analysis concerns the \ac{lwa}. This is widely used in \ac{gw} data analysis and assumes that the \ac{gw} frequency of the signal satisfies $f \ll f_*$, where
$f_* = c / (2\pi L_{\rm arm})$
is the characteristic frequency of an interferometric detector with arm length $L_{\rm arm}$. For a detector with 10\,km arms, such as the $\Delta$ configuration, this corresponds to
$f_* \simeq 4.8\,{\rm kHz}$,
while for a detector with 15\,km arms, such as the 2L configuration, one finds
$f_* \simeq 3.2\,{\rm kHz}$.
However, due to the high-frequency content of post-merger \ac{gw} signals, the \ac{lwa} is no longer valid when computing the \acp{snr}. In this regime, it is necessary to account for the full frequency dependence of the detector response by including the complete transfer function in the antenna pattern functions, as derived in Refs.~\cite{Estabrook:1975jtn, Romano:2016dpx}. This treatment is implemented in \texttt{GWFish}, which adopts the so-called two-arm Michelson transfer function, thereby making the detector tensor explicitly frequency-dependent. As a consequence, the detector response is suppressed at high frequencies, with damping becoming significant for $f \gtrsim f_*$.

Thus, in general, we expect higher \acp{snr} for signals analyzed under the \ac{lwa}, since neglecting the frequency-dependent transfer function leads to an overestimation of the detector response at high frequencies. Moreover, detectors with longer arm lengths are expected to be affected more strongly by the damping induced by the transfer function than shorter-arm configurations. This is because, at the frequencies corresponding to the peak of the post-merger \ac{gw} amplitude, the damping factor increases with arm length, resulting in a larger suppression of the detector response for longer-arm interferometers (see Ref.~\cite{Dhani:2025} and Appendix~\ref{app:psd_cfr}).

In Tables~\ref{tab:bns_pm_table_delta} and~\ref{tab:bns_pm_table_2l} we report the \acp{snr} for the chosen configurations and sensitivity curves evaluated at the optimal sky position for the chosen configurations. The main values we report correspond to the case in which we do not resort to the \ac{lwa}, while the smaller values in parentheses correspond to the case in which the \ac{lwa} is used. As expected, we find that the \acp{snr} obtained without the \ac{lwa} are smaller than the corresponding ones computed within this approximation. Concerning the detector configuration, we find \ac{snr} values for the 2L configuration to be higher compared to the triangular configuration when (incorrectly) resorting to the \ac{lwa}. The difference for the simulations with a stiff \ac{eos} reduces when the \ac{lwa} is dropped. For the simulations with soft \ac{eos} (with mergers at higher frequencies compared to the stiff case, due to the higher compactness) the triangle performs better. This can also partly be traced to the fact that, at high frequencies, the noise curves for arm lengths of 10\,km and 15\,km are closer to each other than in the rest of the frequency band, resulting in an improved \ac{snr} for a configuration consisting of three detectors (see Appendix~\ref{app:psd_cfr}). 

The \acp{snr} for the injections with smaller mass ratios are never found to be less than $\sim\!7$, close to the detection threshold used in \Cref{sec:detection_and_fisher} for full inspiral-merger \ac{bns} signals. The only exception is a degradation in the highest frequency bin, for which, however, the \ac{snr} remains above 5. As expected, however, the \acp{snr} for the more asymmetric binaries are consistently lower, pointing to the fact that their characterization will be more challenging. Concerning the comparison of different noise curves, this metric is largely impacted by degradations in the high-frequency sensitivity. We do not find any appreciable variation in the \acp{snr} for degradations in parameters related to the low-frequency instrument. The variation of the beam size at high frequencies is found to result only in a minor degradation. When lowering instead the sensitivity in the highest frequency bin of the binned analysis, we find \acp{snr} to decrease by about 34\%, and similarly when scaling the full noise curve by a factor of 1.5.

\subsection{Sensitivity to stochastic backgrounds}

A key figure of merit for evaluating \ac{gw} detector performance in searches for \acp{sgwb} is the \ac{pls}, a dimensionless, frequency-dependent curve that quantifies the detectability of power-law-like \acp{sgwb} by a detector network through cross-correlation~\cite{Thrane:2013oya}. Denoting by $\beta$ the background spectral index, the \ac{pls} is defined as

\begin{equation}
\label{eq:PLS}
\Omega_{\rm PLS}(f) = \max_{\beta \in (-\infty,+\infty)} \, \Omega_{\rm GW}(f; \beta) = \max_{\beta \in (-\infty,+\infty)}  \, \left\{\dfrac{\rho\,\left(f/f_{\rm ref} \right)^{\beta}}{\left[2T\,\mathcal{I}_1(\beta)
\right]^{1/2}}
\right\} \; ,
\end{equation}
where
\begin{equation}
{\cal I}_1(\beta) = \int_{0}^{\infty} \Omega_{\mathrm{eff}}^{-2}(f) 
\left( \frac{f}{f_{\mathrm{ref}}} \right)^{2\beta}\,{\rm d}f\,, \quad\; 
\Omega_{\mathrm{eff}}(f) = 
\dfrac{4\pi^2}{3H_0^2} f^3
\sqrt{\sum_{a=1}^{N_{\mathrm{det}}}
\sum_{b>a}^{N_{\mathrm{det}}}\dfrac{\Gamma_{ab}^2(f)}
{S_n^{(a)}(f) S_n^{(b)}(f)}}\;.
\end{equation}
Note that $\Omega_{\rm PLS}(f)$ is a function of the observation time $T$, the \ac{snr} $\rho$ one wishes to achieve, the detector \acp{psd} $S_n(f)$, the geometry of the network parametrized by the overlap reduction function $\Gamma_{ab}(f)$, and an arbitrary positive reference frequency $f_{\rm ref}$.%

\begin{figure}[tbp]
\centering
\includegraphics[width=1.\linewidth]{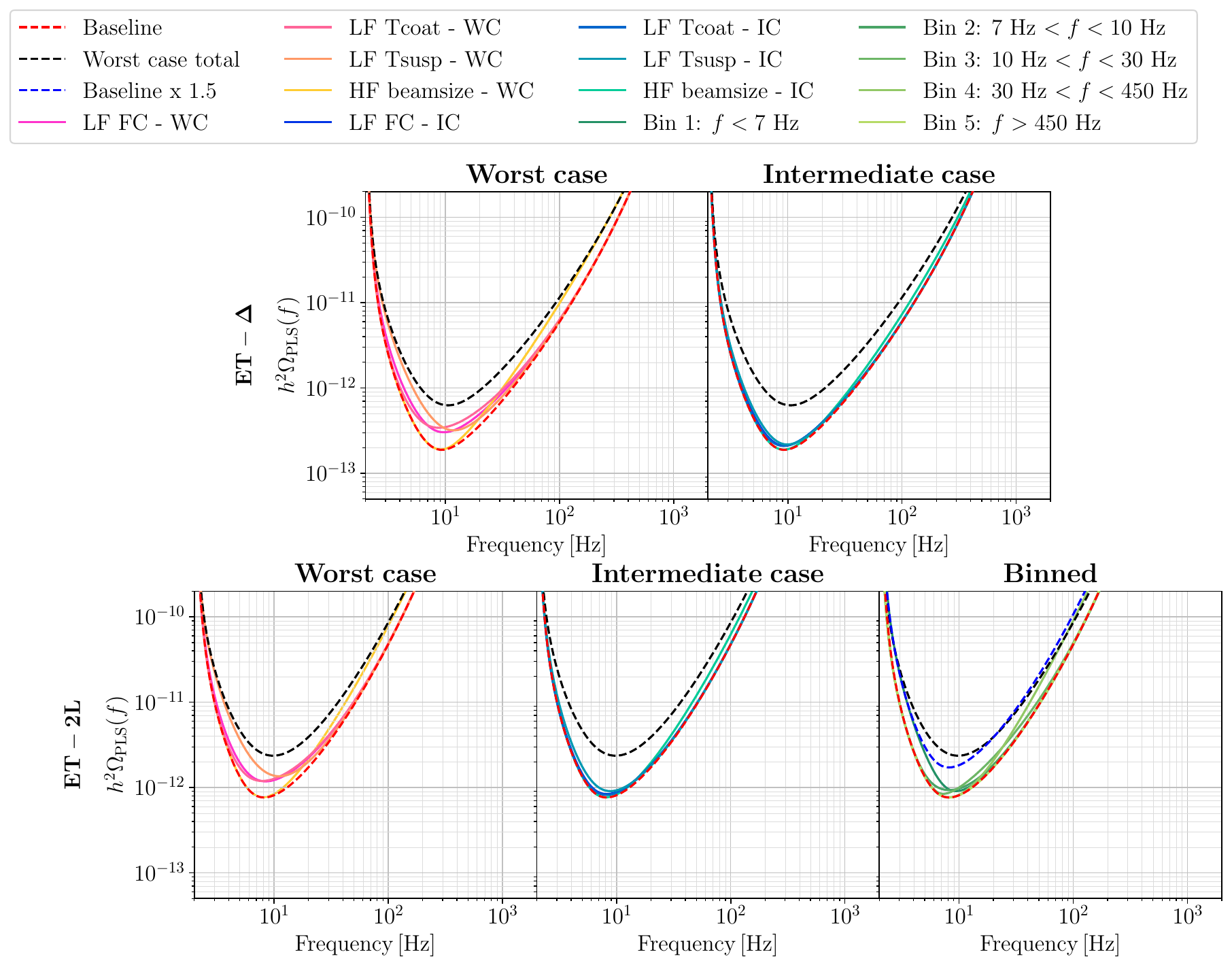}
\caption{Cosmology-independent \ac{pls} comparison under different instrumental configurations, colors as in legend. The top (bottom) row concerns the ET triangular (2L) configuration, while columns correspond to cases as labeled in the panel titles.}
\label{fig:pls}
\end{figure}

Alternatively, as shown in Ref.~\cite{Belgacem:2025oom}, the frequency itself can be defined as a function of $\beta$:
\begin{equation}
f(\beta) = f_{\mathrm{ref}} 
\exp\left\{\frac{\mathcal{I}_2(\beta)}{\mathcal{I}_1(\beta)}\right\} \; ,
\end{equation} 
where
\begin{equation}
\mathcal{I}_{2}(\beta) = \int_{0}^{\infty} \Omega_{\mathrm{eff}}^{-2}(f) 
\left( \frac{f}{f_{\mathrm{ref}}} \right)^{2\beta} 
\ln\left( \frac{f}{f_{\mathrm{ref}}} \right) \, {\rm d} f \; .
\end{equation}
In this framework, the \ac{pls} is given by:
\begin{equation}
\Omega_{\mathrm{PLS}}(f) \equiv \Omega_{\mathrm{PLS}}(f(\beta)) =
\dfrac{\rho}{\left[2T\,\mathcal{I}_1(\beta)
\right]^{1/2}}\, \exp\left\{\beta\dfrac{\mathcal{I}_2(\beta)}{\mathcal{I}_1(\beta)}\right\} \; .
\end{equation}
Note that in these two forms, the sensitivity is cosmology-dependent. One can define an alternative version through

\begin{equation}
    h^2 \Omega_{\rm PLS}(f) = \left(\dfrac{H_0}{100 \, \rm{km \, s^{-1} \, Mpc^{-1}}}\right)^2 \Omega_{\rm PLS}(f) \; ,
\end{equation}
so that the overall quantity does not depend on the adopted cosmology.

\begin{figure}[tbp]
\centering
\includegraphics[width=0.7\linewidth]{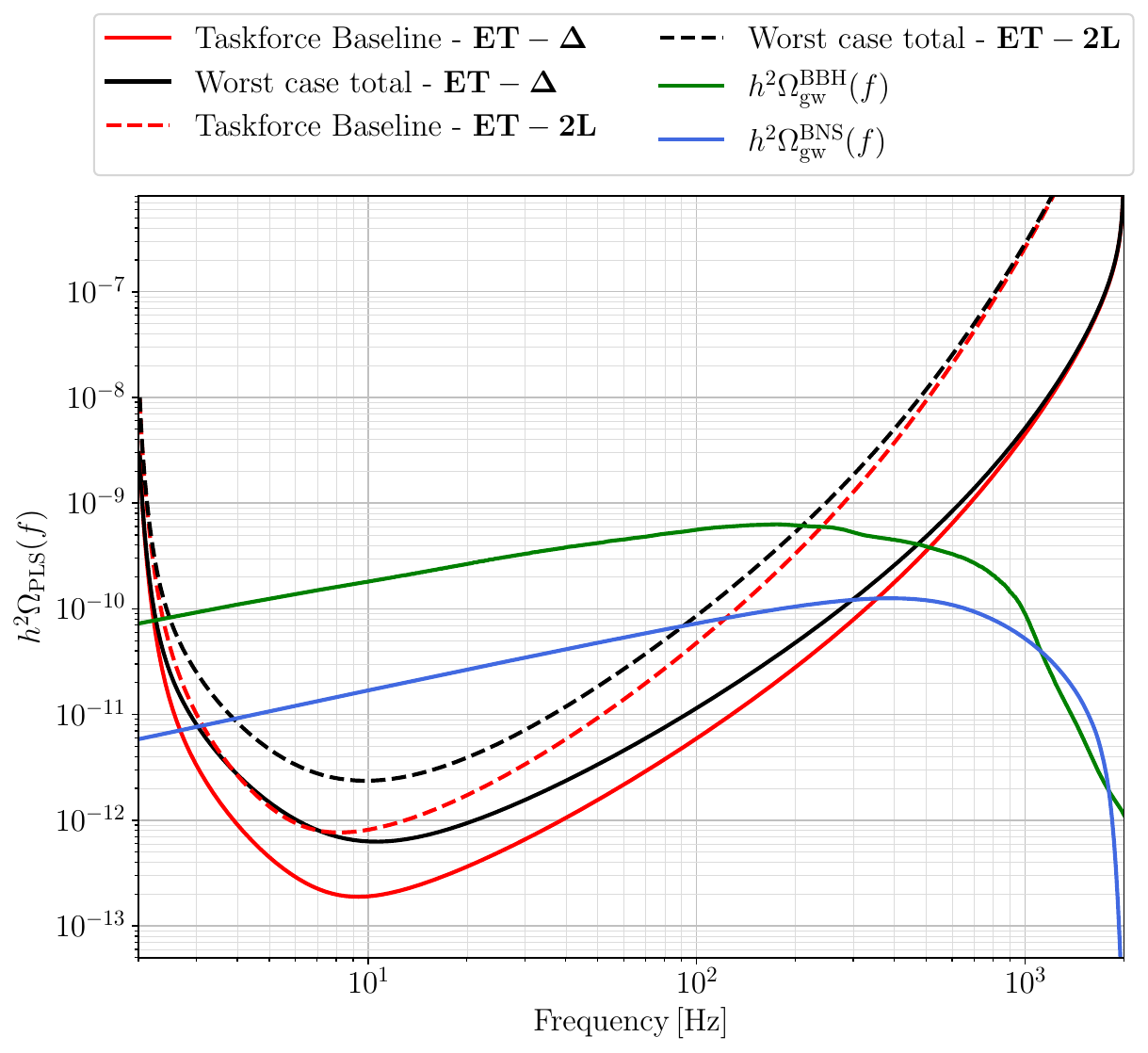}
\caption{Comparison of selected cosmology-independent \ac{pls} curves, together with the astrophysical background produced by the superposition of signals generated by the \ac{bbh} and \ac{bns} populations adopted in this study, colors as in legend. %
}
\label{fig:pls_with_bg}
\end{figure}

Intuitively, the \ac{pls} represents the boundary below which a common stochastic process is undistinguishable from the noise: the lower the bucket, the higher the chances of detection. In \Cref{fig:pls}, we show how the sensitivity changes under different technical layouts of the detector --- both in its triangular and 2L configurations --- and we compare the curves with standard benchmarks plotted with dashed lines.\footnote{Note that the sensitivity for the 2L configuration is not vanishing since, as in Ref.~\cite{Branchesi:2023mws}, the detectors are misaligned with respect to local East, and their relative angle with respect to the great circle joining the two locations is of $42.49^\circ$.} These results are obtained with \texttt{gwfast.stochastic}~\cite{Belgacem:2024ntv}.

All panels show that the Baseline provides the best sensitivity. Nevertheless, particularly around $10\,\mathrm{Hz}$, both the HF beam size \acp{psd} and the Bin 5 configuration yield comparable performance. This can be understood by noting that, given the damping of $\Gamma_{ab}(f)$ with increasing frequency for a fixed detector configuration, the \ac{pls} is mostly affected by variations in the low frequency sensitivity. We find the degradations in the worst-case scenarios when changing the parameters of the low-frequency instruments to be particularly severe, while in the intermediate case, the results remain similar to the baseline design. We also observe that the triangular configuration is able to deliver improved results compared to the 2L design. At frequencies above $\sim\!800\,{\rm Hz}$, this is mainly driven by the geometry itself, with the proximity of the instruments composing the triangle resulting in larger $\Gamma_{ab}(f)$ in this range. At lower frequencies, the results are driven by the adopted alignment of the 2Ls, which has a strong impact on $\Gamma_{ab}(f)$ and thus the \ac{pls} (see, e.g., Figure~71 of Ref.~\cite{Branchesi:2023mws}). For comparison, in \Cref{app:additional_plots} we report the results for a configuration with aligned instruments, which are optimal for \ac{sgwb} detection.\footnote{The \ac{pls} for the two configurations have the same minimum (albeit at different frequencies) for an alignment of the 2Ls with respect to the great circle joining them of $\sim\!35^\circ$.} A crucial point when interpreting these results is the assumption of uncorrelated noise in all instruments in both configurations. Correlated magnetic noise in both configurations~\cite{Janssens:2021cta,Janssens:2022tdj,Branchesi:2023mws} and correlated seismic and Newtonian noise in the triangular configuration can severely impact \ac{sgwb} searches at frequencies below $\sim\!40\,$Hz, potentially reducing the sensitivity to the level of second generation detectors~\cite{Janssens:2024jln}.
A faithful model of the correlated noise components will be needed in order to reconstruct in an unbiased way the \ac{sgwb} parameters~\cite{Caporali:2025mum}, and the reconstruction would depend on the specific spectral shape and amplitude of the signal~\cite{ET:2025xjr}.

Finally, as an example of a realistic signal, in \Cref{fig:pls_with_bg} we report the amplitude of the total \ac{sgwb} generated by our \ac{bbh} and \ac{bns} populations (see, e.g., Refs.~\cite{Phinney:2001di,Belgacem:2024ohp}) without filtering for detectability, and compare it to the \ac{pls} for the Baseline sensitivity and the one corresponding to the worst case scenario for both the considered configurations. While the low-frequency behavior of the \ac{sgwb} generated by \acp{bns} follows the $f^{2/3}$ behavior expected for the inspiral of compact binaries~\cite{Phinney:2001di}, the one generated by \acp{bbh} starts slightly deviating already at 3\,Hz. We verified that this is due to the presence of several high-mass \acp{bbh} in our population, and when computing the background at frequencies below $f\lesssim1\,{\rm Hz}$ the inspiral behavior is recovered.

\subsection{Pulsar detectability}

\begin{figure}[tbp]
    \centering
    \includegraphics[width=1.\linewidth]{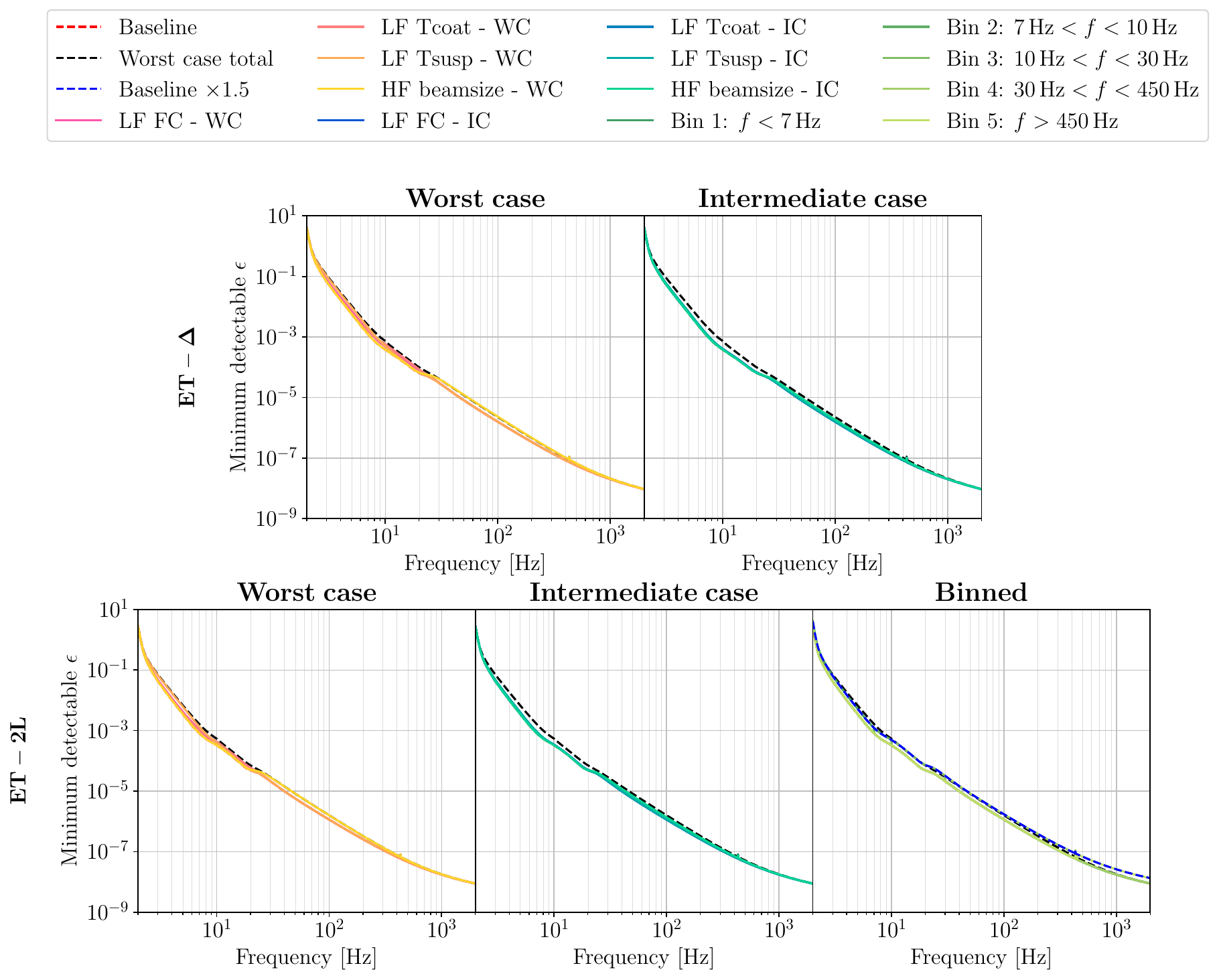}
    \caption{Minimum detectable ellipticity comparison under different instrumental configurations, colors as in legend. The top (bottom) row concerns the ET triangular (2L) configuration, while columns correspond to cases as labeled in the panel titles.}
    \label{fig:min_ellipticity}
\end{figure}

Another interesting yet-undetected source of \acp{gw} are non-axisymmetric, rapidly spinning \acp{ns} (i.e., compact stars with some degree of ellipticity). These indeed possess a quadrupole moment capable of generating a persistent, almost monochromatic \ac{gw} signal at twice the rotational frequency of the object for objects rotating about one of their principal axes of inertia.

Searches for these systems can be either fully coherent and targeted, in which one exploits knowledge of the sky position, distance, orbital period and orbital period variation from electromagnetic observations (plus Keplerian parameters for binaries), narrow-band if limited knowledge of the rotational parameters is available from the electromagnetic data, broadband searches in specific sky regions, and finally, all-sky broadband searches, which probe the largest parameter space at a considerable computational cost. Similar to Refs.~\cite{Branchesi:2023mws,ET:2025xjr}, we consider figures of merit representative of the first and last classes of
searches. 

For targeted searches, we compute the signal emitted by the objects present in the ATNF Pulsar Catalogue\footnote{\url{https://www.atnf.csiro.au/research/pulsar/psrcat/}.} assuming \ac{gw} emission is the dominant source of rotational speed loss, and compute the number of events with detectable amplitude for different choices of the ellipticity of the object. In particular, we take from the catalog the rotational frequency of each object, $f_{\rm gw}/2$, its first derivative, $\dot{f}_{\rm gw}/2$, and the distance to the object. We compute the strain corresponding to the spin-down limit and the corresponding limit on the star ellipticity as
\begin{equation}\label{eq:psr_spindown_strain_eps}
    h_0^{({\rm sd})} = \dfrac{1}{d_L}\left(\dfrac{5G I_{3}}{2c^3}\dfrac{|\dot{f}_{\rm gw}|}{f_{\rm gw}}\right)^{1/2}\,, \qquad\quad \epsilon^{({\rm sd})} = \dfrac{h_0^{({\rm sd})}}{I_{3}} \,\dfrac{c^4 d_L}{4\pi^2 G f_{\rm gw}^2}\;,
\end{equation}
where $I_{3}$ is the \ac{ns} moment of inertia with respect to the principal axis aligned with the rotation axis. For this parameter, we consider a reference value $I_{3}\approx(2/5)m_{\rm NS}R_{\rm NS}^2 \sim 10^{38}\,{\rm kg\,m}^2$ obtained for a typical  $1.33\,M_\odot$ \ac{ns} with a radius of 10\,km. We find that 2048 sources out of the 2698 in the catalog fall in the detector frequency band. The SNR in each detector is obtained as~\cite{Maggiore:2007ulw}
\begin{equation}
    {\rm SNR}_{\rm targeted}^{{\rm PSR}, (d)} = \dfrac{4\pi^2 G}{c^4} \,\dfrac{I_3 f_{\rm gw}^2}{d_L}\,\epsilon\,\sqrt{\dfrac{4}{5}\,\dfrac{F_{\rm ang}}{2}\,\dfrac{T_{\rm obs}}{S_n^{(d)}(f_{\rm gw})}}\;,
\end{equation}
where $\epsilon$ is the \ac{ns} ellipticity, $T_{\rm obs}$ is the observation time, the factor of $4/5$ comes from averaging over the \ac{ns} inclination, and the angular efficiency factor $F_{\rm ang} = 2/5$ for an L-shaped detector and $F_{\rm ang} = 3/10$ for a triangular detector. In this targeted analysis we neglect the impact of Doppler shifts, as they can be corrected for using knowledge about the source position from the chosen catalog~\cite{Maggiore:2007ulw}.
In \Cref{tab:psr_targeted}, we report the number of detected objects in the catalog assuming an observation time of 1\,yr and an \ac{snr} threshold of 8. The results are given for three different choices for the ellipticity: 
\begin{itemize}[label=--]
    \item an ellipticity equal to the spin-down limit $\epsilon^{({\rm sd})}$ which should be regarded as optimistic;
    \item an ellipticity equal to the minimum of $\epsilon^{({\rm sd})}$ and $4\times10^{-6}$, which represents the maximum allowed ellipticity in the ideal scenario in which the whole crust for a $1.4\,M_\odot$ \ac{ns} reaches the breaking strain~\cite{Horowitz:2009ya,Morales:2022wxs}. This should still be regarded as optimistic.
    \item an ellipticity equal to the minimum of $\epsilon^{({\rm sd})}$ and $10^{-9}$, which has been found to be a long-term residual value sustainable by an \ac{ns} for billions of years~\cite{Morales:2024hwl}. This can be regarded as a more conservative estimate.
\end{itemize}

\begin{table}[tbp]
\caption{Number of detectable pulsars in the ATNF Pulsar Catalogue through their \ac{cgw} emission for the different networks and sensitivities considered assuming an observation time of 1\,yr and an SNR threshold of 8. We report the numbers for different choices of the \ac{ns} ellipticity; see text.}
{\setlength{\tabcolsep}{9.7pt}
\begin{tabularx}{\textwidth}{lcccccc}
	\toprule
	& \multicolumn{2}{c}{$\epsilon^{({\rm sd})}$} & \multicolumn{2}{c}{${\rm min}\{\epsilon^{(\rm sd)},\, 4\times10^{-6}\}$} & \multicolumn{2}{c}{${\rm min}\{\epsilon^{(\rm sd)},\, 10^{-9}\}$} \\
    \cmidrule(r){2-3} \cmidrule(r){4-5} \cmidrule{6-7}
	\textbf{Sensitivity curve} & $\Delta$ & 2L & $\Delta$ & 2L & $\Delta$ & 2L \\\midrule
    Baseline & 870 & 1006 & 187 & 223 & 10 & 12 \\
\rowcolor{MidnightBlueLight} Baseline $\times 1.5$ & -- & 81\% & -- & 60\% & -- & 50\% \\
Bin 1: $f<7\,{\rm Hz}$ & -- & 90\% & -- & 100\% & -- & 100\% \\
\rowcolor{MidnightBlueLight} Bin 2: $7\,{\rm Hz}<f<10\,{\rm Hz}$ & -- & 98\% & -- & 99\% & -- & 100\% \\
Bin 3: $10\,{\rm Hz}<f<30\,{\rm Hz}$ & -- & 99\% & -- & 98\% & -- & 100\% \\
\rowcolor{MidnightBlueLight} Bin 4: $30\,{\rm Hz}<f<450\,{\rm Hz}$ & -- & 97\% & -- & 85\% & -- & 100\% \\
Bin 5: $f>450\,{\rm Hz}$ & -- & 95\% & -- & 78\% & -- & 58\% \\
\rowcolor{MidnightBlueLight} LF FC - IC & 99\% & 99\% & 100\% & 99\% & 100\% & 100\% \\
LF Tcoat - IC & 99\% & 99\% & 100\% & 99\% & 100\% & 100\% \\
\rowcolor{MidnightBlueLight} LF Tsusp - IC & 97\% & 96\% & 100\% & 99\% & 100\% & 100\% \\
HF beam size - IC & 98\% & 98\% & 92\% & 92\% & 90\% & 83\% \\
\rowcolor{MidnightBlueLight} LF FC - WC & 94\% & 95\% & 99\% & 97\% & 100\% & 100\% \\
LF Tcoat - WC & 98\% & 97\% & 100\% & 98\% & 100\% & 100\% \\
\rowcolor{MidnightBlueLight} LF Tsusp - WC & 87\% & 87\% & 100\% & 99\% & 100\% & 100\% \\
HF beam size - WC & 96\% & 97\% & 81\% & 87\% & 80\% & 83\% \\
\rowcolor{MidnightBlueLight} Worst case total & 80\% & 81\% & 82\% & 85\% & 80\% & 83\% \\
\bottomrule

\end{tabularx}
}
\label{tab:psr_targeted}
\end{table}

From the results in \Cref{tab:psr_targeted}, we can observe the large number of detectable objects in the considered catalog for the most optimistic case in which the ellipticity equals the spin-down limit: $\sim\!43\%$ for the triangular configuration and $\sim\!49\%$ for the 2L. For the conservative estimate, these numbers are considerably reduced but never vanish. Regarding the comparison of variations in the sensitivity curve, we find only minor degradations when varying one parameter at a time, but a loss of $\sim\!20\%$ in the detections when all of them are varied simultaneously in the worst case scenario. A similar result also holds for the binned analysis, albeit with a smaller number of detectable signals when more conservative choices for the ellipticity are considered. From this analysis, we find that the bins mostly impacting the results are those at the two extrema of the frequency range. 

For broadband searches, we compute the minimum detectable ellipticity at a distance of 8\,kpc as a function of frequency with values representative of the \texttt{FrequencyHough} pipeline~\cite{2014PhRvD..90d2002A}, as~\cite{Branchesi:2023mws,ET:2025xjr}
\begin{equation}
    \epsilon_{95\%}^{\rm min}(f; d_L) \approx \dfrac{4.97 c^4 d_L}{4\pi^2 G I_3 f^2} \left(\dfrac{T_{\rm FFT}}{T_{\rm obs}}\right)^{1/4} \sqrt{\dfrac{S_n(f)[{\rm CR}_{\rm th} + 1.6449]}{T_{\rm FFT}}}\;,
\end{equation}
where $T_{\rm FFT} = 10$ days is the duration of the considered data segments, $T_{\rm obs}=1$\,yr is the observation time, and ${\rm CR}_{\rm th} = 4$ is the critical ratio threshold used to select outliers. The results are reported in \Cref{fig:min_ellipticity}, while in \Cref{fig:ellipticity_relative} we report the relative variation with respect to the Baseline noise curve. We find that results for this metric are only mildly affected by variations in the detectors' sensitivity, even in scenarios corresponding to the larger degradations, and similar for the two configurations.

\subsection{Core-Collapse Supernova  Detectability}

\begin{figure}[tbp]
    \centering
    \includegraphics[width=\linewidth]{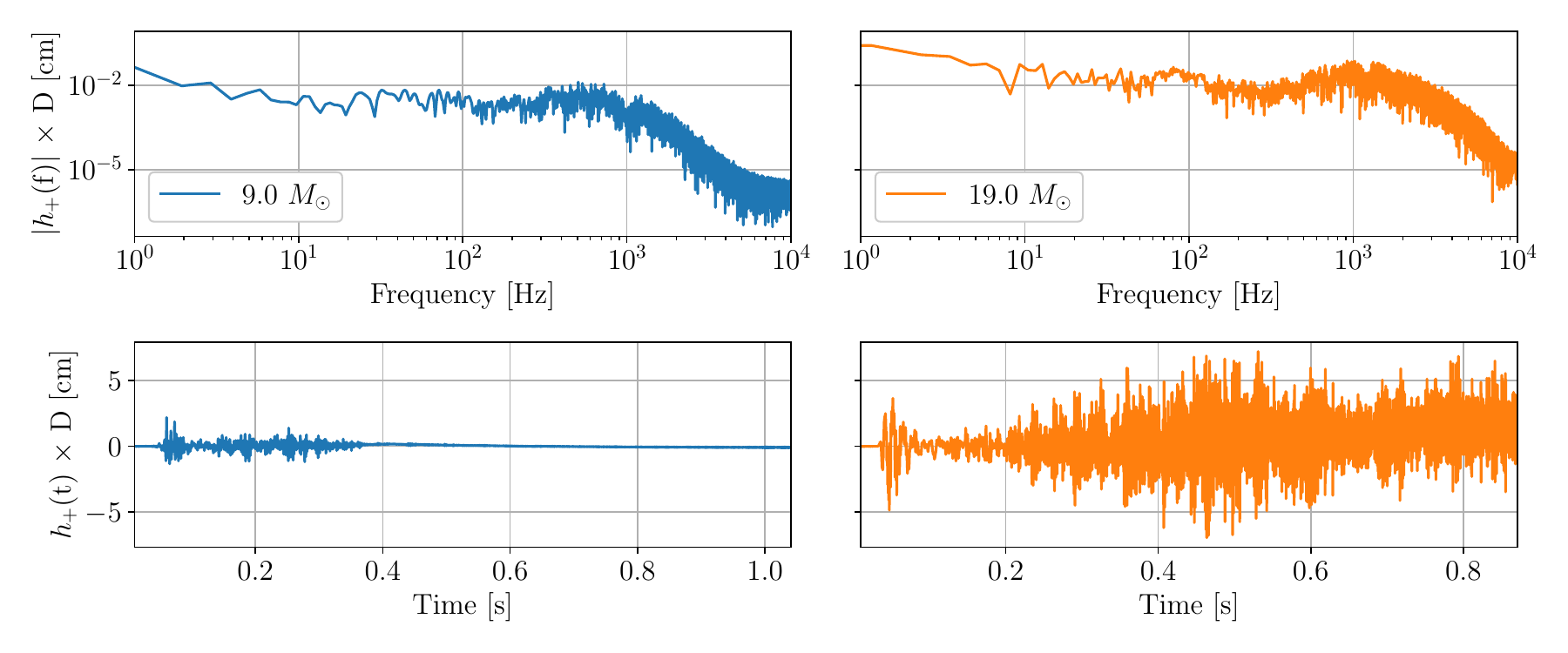}
    \caption{\ac{ccsn} \ac{gw} strain multiplied by source distance (in cm), for the chosen simulations corresponding to our best ($19\,M_\odot$, orange) and worst case ($9\,M_\odot$, blue) scenario. The strain is shown in both the time and frequency domains.}
    \label{fig:waveform_sn}
\end{figure}

To assess the detectability of \acp{gw} from \acp{ccsn}, we compute the \ac{snr} for a representative set of potential events using three-dimensional \ac{ccsn} simulations from Ref.~\cite{Radice:2018usf}.\footnote{The simulation data is available at \url{https://www.astro.princeton.edu/~burrows/gw.3d/}.} We select models from $19\,M_\odot$ and $9\,M_\odot$ progenitors. These represent the highest and lowest total energy radiated in \acp{gw}, respectively. The strain amplitudes for these two waveforms are shown in \Cref{fig:waveform_sn}. Following Ref.~\cite{Radice:2018usf}, we report the \ac{gw} strain multiplied by source distance, $h \times D$. This is a convenient parametrization used in numerical supernova studies (see, e.g., Ref.~\cite{Andresen:2016pdt}) as detection horizons are too low to consider redshift effects, making the observed strain proportional to the distance. 

We randomly sample sky locations and polarization angles to compute the \ac{snr} using Eq.~\eqref{eq:snr_net} for simulated events. \Cref{tab:sne} presents the resulting detection horizons, defined as the luminosity distance, $d_L$, at which a network detects a source with an \ac{snr} of 8. Crucially, as in \Cref{sec:bns_postmerg}, we do not resort to the \ac{lwa} in this analysis, given the large high-frequency component of \acp{ccsn} waveforms.

{
\setlength{\tabcolsep}{14pt} 
\begin{table}[tbp]
\caption{Detection horizon at ${\rm SNR}=8$ for two different \ac{ccsn} waveforms with different progenitor masses for both the triangular and the 2L configurations for the sensitivity curves analyzed. We report the median $d_L$ value together with the 90\%\,c.l.}
\begin{tabularx}{\textwidth}{l  c c  c c}
	\toprule
    \multicolumn{1}{c}{}& \multicolumn{2}{c}{$d_{L,\,{\rm max}}$ [kpc] -- $19\,M_\odot$} & \multicolumn{2}{c}{$d_{L,\,{\rm max}}$ [kpc] -- $9\,M_\odot$}\\
    \cmidrule(r){2-3} \cmidrule{4-5}
	\textbf{Sensitivity curve}  &$\Delta$ &2L &$\Delta$ &2L \\\midrule  
    {Baseline}  &{$129.4^{+84.5}_{-50.5}$} &{$143.1^{+92.9}_{-54.6}$} &{$14.7^{+9.8}_{-6.5}$} &{$17.5^{+11.6}_{-7.5}$}
\\\rowcolor{MidnightBlueLight}{Baseline $\times 1.5$}  &{--} &{$95.4^{+61.9}_{-36.4}$} &{--} &{$11.7^{+7.7}_{-5.0}$}
\\{Bin 1: $f<7\,\mathrm{Hz}$}  &{--} &{$143.1^{+92.9}_{-54.6}$} &{--} &{$17.5^{+11.6}_{-7.5}$}
\\\rowcolor{MidnightBlueLight}{Bin 2: $7\,\mathrm{Hz}<f<10\,\mathrm{Hz}$}  &{--} &{$143.1^{+92.9}_{-54.6}$} &{--} &{$17.5^{+11.6}_{-7.5}$}
\\{Bin 3: $10\,\mathrm{Hz}<f<30\,\mathrm{Hz}$}  &{--} &{$142.9^{+92.8}_{-54.6}$} &{--} &{$17.5^{+11.5}_{-7.5}$}
\\\rowcolor{MidnightBlueLight}{Bin 4: $30\,\mathrm{Hz}<f<450\,\mathrm{Hz}$}  &{--} &{$139.0^{+90.2}_{-52.4}$} &{--} &{$15.1^{+10.0}_{-6.3}$}
\\{Bin 5: $f>450\,\mathrm{Hz}$}  &{--} &{$101.4^{+66.2}_{-39.6}$} &{--} &{$14.9^{+9.8}_{-6.4}$}
\\\rowcolor{MidnightBlueLight}{LF FC - IC}  &{$129.4^{+84.5}_{-50.5}$} &{$143.1^{+92.9}_{-54.6}$} &{$14.7^{+9.8}_{-6.5}$} &{$17.5^{+11.6}_{-7.5}$}
\\{LF Tcoat - IC}  &{$129.4^{+84.5}_{-50.5}$} &{$143.1^{+92.9}_{-54.6}$} &{$14.7^{+9.8}_{-6.5}$} &{$17.5^{+11.6}_{-7.5}$}
\\\rowcolor{MidnightBlueLight}{LF Tsusp - IC}  &{$129.4^{+84.5}_{-50.5}$} &{$143.1^{+92.9}_{-54.6}$} &{$14.7^{+9.8}_{-6.5}$} &{$17.5^{+11.6}_{-7.5}$}
\\{HF beamsize - IC}  &{$125.8^{+82.1}_{-48.9}$} &{$138.9^{+90.1}_{-52.8}$} &{$13.8^{+9.2}_{-6.1}$} &{$16.3^{+10.8}_{-7.0}$}
\\\rowcolor{MidnightBlueLight}{LF FC - WC}  &{$129.3^{+84.5}_{-50.5}$} &{$143.0^{+92.8}_{-54.6}$} &{$14.7^{+9.8}_{-6.5}$} &{$17.5^{+11.5}_{-7.5}$}
\\{LF Tcoat - WC}  &{$129.3^{+84.5}_{-50.5}$} &{$143.0^{+92.8}_{-54.6}$} &{$14.7^{+9.8}_{-6.5}$} &{$17.5^{+11.5}_{-7.5}$}
\\\rowcolor{MidnightBlueLight}{LF Tsusp - WC}  &{$129.3^{+84.5}_{-50.5}$} &{$143.0^{+92.9}_{-54.6}$} &{$14.7^{+9.8}_{-6.5}$} &{$17.5^{+11.5}_{-7.5}$}
\\{HF beamsize - WC}  &{$119.5^{+77.9}_{-46.2}$} &{$135.0^{+87.6}_{-51.0}$} &{$12.3^{+8.2}_{-5.4}$} &{$15.3^{+10.1}_{-6.5}$}
\\\rowcolor{MidnightBlueLight}{Worst case total}  &{$120.0^{+78.3}_{-46.4}$} &{$134.7^{+87.4}_{-50.9}$} &{$12.3^{+8.2}_{-5.4}$} &{$15.1^{+10.0}_{-6.4}$}
\\ \bottomrule
 
\end{tabularx}
\label{tab:sne}
\end{table}
}

A key caveat is that selecting waveforms from a finite set of simulations entails assuming our template bank functions as a complete representation of a possible signal. Variations in progenitor mass, rotation, and explosion dynamics yield varied \ac{gw} signatures~\cite{Andresen:2016pdt,Schnauck:2025xxs}. Methods exist to correlate \ac{gw} and neutrino emissions for more robust detection strategies~\cite{Drago:2023cve}. Further research models signal features, including low-frequency components~\cite{Richardson:2025ldi} and the intermediate frequency ($\sim\!100$\,Hz) standing accretion shock instability (SASI)~\cite{Foglizzo:2024oqf}. We therefore assume the selected waveforms provide realistic proxies, especially as multi-messenger triggers could inform targeted \ac{gw} searches for fainter signals. 

We note that the $19\,M_\odot$ model yields higher \ac{gw} energy emission and consequently greater detection ranges than the $15\,M_\odot$ progenitor results reported in Ref.~\cite{Branchesi:2023mws}. For all strains we find that the 2L configuration provides larger detection horizons for both considered models. 
Concerning the different sensitivity curves, we notice that the impact of changing specific design parameters is modest, with the HF beam size being the most impactful. This is expected, as it results in changes in the high-frequency region of the noise curve, where \ac{ccsn} waveforms accumulate most of the \ac{snr}. For the same reason, the largest loss in detection horizon is found when degrading the noise curve in the highest frequency bin.
The horizons for the $9\,M_\odot$ model are generally confined to the Galactic neighborhood. This underscores that for lower-mass progenitors with weak \ac{gw} emission, detection will rely heavily on a rare Galactic event. Conversely, the $19\,M_\odot$ model extends the detection horizon beyond the Magellanic Clouds, offering a higher event rate probability. A note of caution is needed here: as remarked in \Cref{sec:metrics_and_pops}, template-based searches for \acp{ccsn} will be challenging, hence our results should be regarded as optimistic in absolute terms.

These results highlight the necessity of a multi-messenger approach. The coincident detection of low-energy neutrinos by facilities such as Hyper-Kamiokande or DUNE would provide a crucial time-stamp for the core bounce~\cite{Kara:2024xug, Gossan:2015xda}. This external trigger lowers the \ac{snr} threshold required for confident \ac{gw} detection, expanding the horizons reported here and allowing for the extraction of physics from deep within the \ac{ccsn} core~\cite{Halim:2021yqa}.

\section{Discussion}\label{sec:discussion}

In the previous sections, we presented the results for the metrics described in Section~\ref{sec:metrics_and_pops} for the two considered detector configurations for \ac{et}, a triangle with 10\,km arms and 2 Ls with 15\,km arms for the set of noise curves outlined in Section~\ref{sec:noise_budget}. We summarize here schematically the main findings, with the aim of associating losses in science output with degradations in the sensitivity. In particular, we look at how different science cases respond to sensitivity losses for different parts of the frequency spectrum. The idea is to clarify which regions of the sensitivity curve are critical for which types of science, and where performance losses become particularly costly. When the complete set of metrics presented in this paper is considered, the following picture emerges: different science tasks depend on different frequency ranges, and therefore, their sensitivity to instrumental degradations is nonuniform. This motivates the need to consider a broad set of metrics to evaluate the detector performance, as we do in this work. This pattern appears throughout, both when sensitivity is altered through specific technical parameters, e.g., the coating temperature or beam size, and when it is degraded in a more agnostic way within selected frequency bands. Considering specific frequency ranges, we find that:
\begin{description}[align=left]
    \item[Low frequencies:] Sensitivity at low frequencies (below $\sim\!30$\,Hz) is essential for science cases that rely on long signals. In particular, it dominates the performance of early-warning detections of \ac{bns} mergers and their localization before coalescence. When sensitivity in this frequency band is degraded, the number of systems that can be detected tens of minutes before merger drops significantly, and sky-localization accuracy worsens accordingly. This has direct consequences for the prospects of prompt electromagnetic follow-up. This frequency range is also of pivotal importance to detect massive, high-redshift \ac{bbh} sources, with the horizon distance dropping appreciably across all the mass spectrum when sensitivity drops below 10\,Hz, and a lower number of \ac{pbh} detections. Indeed, for the worst cases considered, the horizon can decrease by as much as 40\% over a significant fraction of the mass range, including the peak. Concerning the reconstruction of the \ac{bbh} signals properties, those most impacted in this frequency range are the localization and distance of the objects, followed by the spins. The number of \ac{bbh} detections with relative uncertainties in $d_L$ at sub-percent level can be reduced by as much as a factor of 2 in the worst scenarios considered, and a factor of $\sim\!1.5$ fewer sources would be localized to better than $10\,{\rm deg}^2$. This has a direct impact on science cases relying on localization, in particular the reconstruction of the local expansion rate of the Universe through the statistical host identification technique. This is also the frequency region where a good fraction of the \ac{bbh} sources build most of the \ac{snr}, necessary for precision studies, such as tests of the nature of the objects and analyses exploring deviations from GR. We lose almost half of \ac{bbh} sources when considering very high-SNR signals (above 500), resulting in significant losses and stressing once again the paramount importance of the low-frequency sensitivity.
    Another science case for which sensitivity at $f\lesssim30$\,Hz is relevant is \acp{sgwb}, in particular for the 2L design, whose sensitivity degrades faster with increasing frequency compared to the triangular geometry: at low frequencies, the overlap reduction functions are maximal and the \ac{pls} is lowest, hence, a higher noise level in this region can significantly impact searches for an \ac{sgwb} signal. Indeed, the \ac{pls} can increase by as much as a factor of 1.7 in a large fraction of the most sensitive range at $\sim\!10$\,Hz. 
    These conclusions stem from the analysis of changes in the ET-LF instrument, all impactful in this range, and the analysis in the three lowest frequency bins considered.       
    \item[Intermediate frequencies:] The intermediate-frequency band, (between $\sim\!30$\,Hz and $450$\,Hz) is where most of the \acp{bbh} in our simulated catalogs merge and accumulate the last portion of their \ac{snr}. This frequency band corresponds to the bucket of the sensitivity and impacts the detection rates and parameter reconstruction of the signals. The statistical uncertainties attainable on the masses, spins, distances, and sky localizations of the sources increase. For \acp{bbh}, the degradation does not exceed $\sim\!10-15\%$ in all the cases considered, but worsens for \ac{bns} systems with around 20\% degradation for chirp mass and distance errors, and up to 40\% for tidal deformability measurement errors. 
    Loss of sensitivity in this large frequency range also impacts the number of detectable pulsars in direct searches, since a fraction of them emit at those frequencies. These conclusions stem from the analysis of changes in the filter cavity length of the ET-LF instrument, the coating thermal noise of the ET-HF instrument, and the analysis in the dedicated frequency bin.   
    \item[High frequencies:] for $f>450$\,Hz, the impact of changes in the sensitivity on \ac{bbh} sources is marginal. This includes most of the low-mass \ac{pbh} binaries in our population: their high $z$ results in larger detector-frame masses, and thus lower merger frequencies. This range is where \ac{bns} systems merge, hence, it has a pivotal impact on the reconstruction of the tidal deformability of the objects, which mostly affects the evolution of the binaries from the late inspiral onward. Indeed, we find that about 50\% fewer systems would have relative uncertainty of the effective tidal deformability parameter $\tilde{\Lambda}$ below 20\%.
    It also determines the sensitivity to the post-merger signal of \ac{bns} mergers, whose observation can be hampered by sensitivity losses in this region, with \ac{snr} losses as high as 35\%. 
    This frequency band is also relevant for observing a good fraction of the pulsars in the chosen catalog for the more conservative choices of ellipticity. Finally, a large fraction of the signal from the selected \acp{ccsn} simulations lies at these frequencies, thus, the horizon for such events diminishes with a degraded sensitivity. For the worst-case scenario, a degradation in this frequency range is reflected almost entirely on the maximum detectable distance, whose median lowers by a factor of 1.4. These conclusions stem from analysis of changes in the coating thermal noise of the ET-HF instrument and the analysis in the dedicated frequency bin.     
\end{description}

In general, we find that, while impactful for the scientific return, sensitivity degradations even in the worst-case scenarios considered would still allow \ac{et} to be an outstanding instrument, across all considered metrics: (\emph{i}) it will be able to detect binaries at cosmological distances well beyond the peak of the star formation rate; (\emph{ii}) it will be able to observe tens of thousands of \acp{cbc} every year and (\emph{iii}) provide an exquisite reconstruction of the parameters for a large number of them; (\emph{iv}) it will give us the opportunity to detect yet-elusive sources, including \acp{sgwb} of different origins, deliver observations of \acp{cgw} from known pulsars and potentially unknown ones, and give us the opportunity to detect the \ac{gw} signal from \acp{ccsn} out to $\gtrsim100\,$kpc. This confirms the potential of \ac{et} as a precision and discovery machine. Moreover, variations in design parameters at the intermediate level considered have minor impact on several of the metrics considered, pointing to the robustness of its science case.

While the direct comparison of the two configurations, \ac{et}-$\Delta$-10\,km and \ac{et}-2L-15\,km, falls beyond the scope of this study, we highlight two relevant points: (i) when looking at the relative impact of degradations in the noise curve on the same configuration, \ac{et}-$\Delta$-10\,km and \ac{et}-2L-15\,km perform similarly for several metrics; (ii) when looking at absolute numbers, for some metrics, the worst degradation scenarios of the \ac{et}-2L-15\,km configuration perform similarly or even better than the baseline scenario of \ac{et}-$\Delta$-10\,km (see e.g. Figures~\ref{fig:detections_snr12_BNS}, \ref{fig:distance_BNS} and~\ref{fig:skyloc_BNS}, and Tables~\ref{tab:snr_table} and~\ref{tab:sne}), with the exceptions of \ac{bns} post-merger and \acp{sgwb}, depending on the alignment of the 2Ls. The first point proves the overall robustness of both designs against noise-budget degradations, while the second points to the fact that \ac{et}-2L-15\,km could yeld a better scientific return; see also Ref.~\cite{Branchesi:2023mws}. The main driver of the improved performance of the \ac{et}-2L-15\,km configuration for several metrics is the better sensitivity due to the longer arms.

It is important to stress again that, while some of the metrics we used to evaluate the performance of the detector are general, such as detection horizons and the \ac{pls}, others are based on specific astrophysical assumptions regarding the population of sources in the Universe, and thus inherently carry uncertainties. While the absolute values of population-dependent metrics can change under different astrophysical assumptions, we expect the relative impact of changes in detector sensitivity to be largely preserved. Population-based results in this paper should thus be understood as representative forecasts rather than rigid predictions. Similar reasoning applies also to the sensitivity degradations considered: while changes in specific design parameters allow us to understand their impact on the science case, the binned frequency analysis without reference to specific noise sources allows our results to remain general.

\section{Summary and Conclusions}\label{sec:conclusions}

Characterizing the scientific output of an experiment like \ac{et}, which will dominate the scientific landscape for decades, and has the potential to be both a precision and discovery machine, is a complex task.
In this work, we have identified a set of metrics to evaluate the detector performance in a fast yet meaningful way. These include detectability and parameter estimation capabilities of \ac{cbc} systems, pre-merger localization and post-merger observability for \acp{bns}, sensitivity to \acp{sgwb}, detectability of \acp{cgw} from isolated pulsars, and of signals from \acp{ccsn}. Some of the metrics depend on specific choices for the astrophysical distribution of sources, hence, our results carry some inherent uncertainty. The chosen populations are built from state-of-the-art population synthesis codes and are meant to provide a useful benchmark for comparing different detector sensitivities. In particular, we do not expect the relative impact of degradations to be affected by differences in the overall merger rate. Our framework allowed us to study how the science prospects for \ac{et} would change following variations in the detector sensitivity, and, vice versa, to link scientific targets to sensitivity requirements. 

We have studied how the \ac{et} noise curve would degrade in response to variations in relevant design parameters. In particular, we have explored variations in quantum noise, coating thermal noise, and suspension thermal noise of the \ac{et} low-frequency instrument, and the coating thermal noise of the \ac{et} high-frequency instrument. Moreover, we have identified crucial frequency ranges for the detector sensitivity, and conducted an analysis in which the sensitivity is degraded in each one of them. This allowed us to assess their impact on the \ac{et} science case, while remaining agnostic about the cause of a degradation. Our analysis has been conducted considering two \ac{et} configurations: a triangular detector with 10\,km-long arms, and 2 L-shaped interferometers with 15\,km-long arms.

The key result is that even considering large degradations in sensitivity the science case for both the considered detector configurations remains largely intact: \ac{et} will still be able to (\emph{i}) detect a large fraction of \ac{cbc} events in the Universe, from different populations and out to very high redshift; (\emph{ii}) provide exquisite inference of the parameters of the systems giving rise to the signals; (\emph{iii}) have promising sensitivity to \acp{sgwb}, allowing us to detect the astrophysical component generated by unresolved binary mergers, and potentially a cosmological component generated in the very early Universe; (\emph{iv}) detect the \ac{cgw} signals from some known pulsars; (\emph{v}) enlarge the detectability horizon for \acp{ccsn}. This proves the soundness and feasibility of the proposed science case, even if, in a first stage, the target sensitivity should not be achieved.

Concerning the studied design parameters, we find that the most impactful for several science cases to be those related to the low-frequency instrument, in particular, the suspension and coating thermal noise. As also confirmed also by our binned analysis, a higher noise level at frequencies $f\lesssim30\,{\rm Hz}$ can strongly impact the localization and pre-merger alert capabilities of \ac{et}, prevent detection of high-$z$ sources, and lower the sensitivity to \acp{sgwb}. It is also particularly important for high-SNR \ac{bbh} events, thus for precision studies, e.g., tests of GR. This confirms the relevance of the low-frequency instrument to achieving \ac{et} science objectives.
The intermediate frequency range, from 30\,Hz to 450\,Hz, is relevant for parameter estimation in general. The impact of sensitivity at higher frequencies is instead more limited to specific science cases, such as the observability of the post-merger signal from \ac{bns} mergers and the reconstruction of their tidal deformability parameters, all related to the \ac{ns} \ac{eos}, and the observability of yet-elusive sources, such as pulsars and \acp{ccsn}.

Our comparative study evaluates how variations in the noise level affect the scientific performance of \ac{et}, providing a practical reference for directly relating changes in the noise budget to their impact on scientific return.

\section*{Data availability}
The data used in this paper, the injected populations, and all the results are publicly available on \textsc{Zenodo}~\raisebox{-1pt}{\href{https://doi.org/10.5281/zenodo.21682194}{\includegraphics[width=9pt]{zenodo-icon-blue.pdf}}}~\cite{pubdataZenodo}. The corresponding scripts can be found on the \ac{et} \textsc{GitLab}~\raisebox{-1pt}{\href{https://gitlab.et-gw.eu/pub/impact-on-science-case}{\includegraphics[width=9pt]{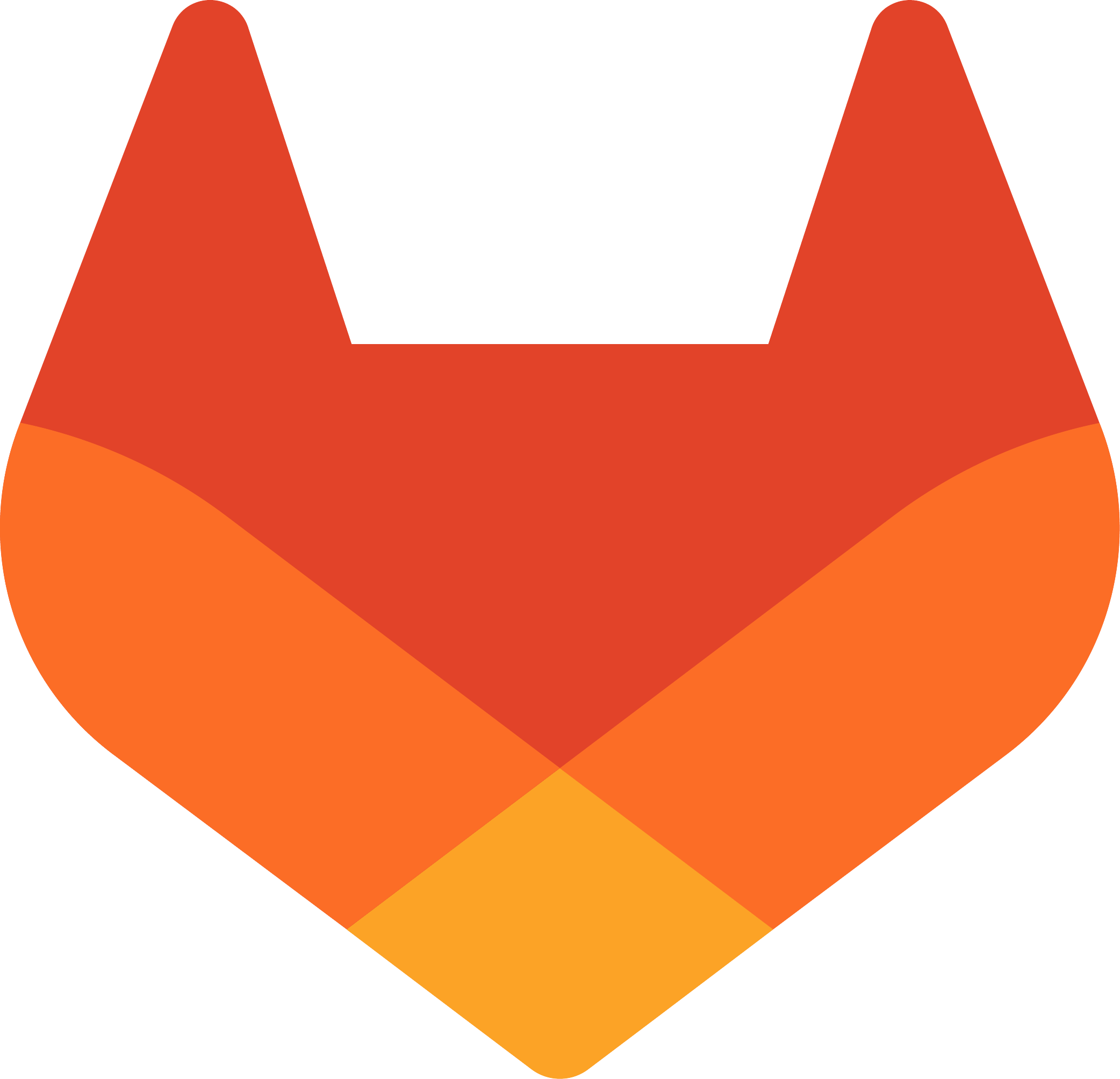}}}~\cite{pubdataGit}. 

\section*{Acknowledgments}
The authors thank INFN-Tier 1 at CNAF for providing the computational resources to carry out most of the analyses presented in this work. N. Muttoni acknowledges support by the SwissMap National Center for Competence in Research. F. Iacovelli is supported by a Miller Postdoctoral Fellowship and by NSF Grants No.~AST-2307146, No.~PHY-2513337, No.~PHY-090003, and No.~PHY-20043, by NASA Grant No.~21-ATP21-0010, by John Templeton Foundation Grant No.~62840, by the Simons Foundation [MPS-SIP-00001698, E.B.], by the Simons Foundation International [SFI-MPS-BH-00012593-02], and by Italian Ministry of Foreign Affairs and International Cooperation Grant No.~PGR01167. G. Franciolini acknowledges support by the Italian MUR Departments of Excellence grant 2023–2027 “Quantum Frontier” and from Istituto Nazionale di Fisica Nucleare (INFN) through the Theoretical Astroparticle Physics (TAsP) project. F. Santoliquido has been funded by the European Union – NextGenerationEU under the Italian Ministry of University and Research (MUR) “Decreto per l’assunzione di ricercatori internazionali post-dottorato PNRR” – Missione 4 “Istruzione e Ricerca” Componente 2 “Dalla Ricerca all’Impresa” del PNRR – Investimento 1.2 “Finanziamento di progetti presentati da giovani ricercatori” – CUP D13C25000700001. Y. Li was supported by the European Research Council through the Consolidator grant BHianca (grant agreement ID 101002761). M. Arca Sedda acknowledges funding from the European Union’s Horizon 2020 research and innovation programme under the Marie Skłodowska-Curie grant agreement No.~101025436 (project GRACE-BH) and from the MERAC Foundation through the 2023 MERAC prize, and acknowledges the ACME project which has received funding from the European Union's Horizon Europe Research and Innovation programme under Grant Agreement No.~101131928. M. Korobko was supported by the Deutsche Forschungsgemeinschaft (DFG) under Germany's Excellence Strategy EXC 2121 ``Quantum Universe''-390833306. F. Crescimbeni is partially supported by the MUR FIS2 Advanced Grant ET-NOW (CUP:~B53C25001080001) and by the INFN TEONGRAV initiative. 

\appendix

\begin{figure}[tb]
    \centering
    \begin{minipage}{0.49\linewidth}
        \includegraphics[width=1\linewidth]{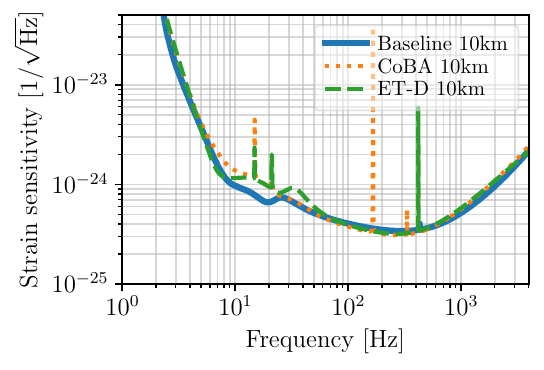}
    \end{minipage}
    \begin{minipage}{0.49\linewidth}
        \includegraphics[width=1\linewidth]{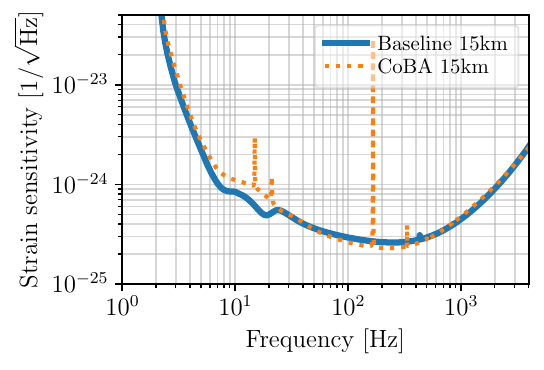}
    \end{minipage}
    \caption{Comparison of different sensitivities for 10\,km (left) and 15\,km (right) detectors: the current Baseline sensitivity adopted in this study (blue), the previous iteration used in Refs.~\cite{Branchesi:2023mws, ET:2025xjr} (orange dotted), and the ET-D sensitivity (green dashed), present only for the 10\,km design~\cite{Hild:2010id}.}
    \label{fig:sens_cfr_coba}
\end{figure}

\section{ET sensitivity curve comparison}\label{app:sens_curves}
The Baseline sensitivity curves used in this paper adopt the sensitivity by 2025 ETO Task Force~\cite{taskforce}. This is an updated sensitivity with respect to the one used in Refs.~\cite{Branchesi:2023mws,ET:2025xjr}, which itself was a major update to one of the initial \ac{et} sensitivity models referred to as ET-D~\cite{Hild:2010id}. A comparison of these curves is provided in \Cref{fig:sens_cfr_coba} for a 10\,km detector (left panel) and a 15\,km detector (right panel).

The main evolution of the sensitivity curve from ET-D to the curves used in Refs.~\cite{Branchesi:2023mws, ET:2025xjr} was in the level of detail for various noise contributions, in particular, optical losses and thermal noise modeling. The current Baseline sensitivity features significant parameter changes to optimize the sensitivity compared to the one used in Refs.~\cite{Branchesi:2023mws, ET:2025xjr}: an improved and more detailed optical-loss budget; longer filter cavities; optimized suspensions for ET-HF; a realistic optical layout leading to longer signal recycling cavities, among other modifications. 

The current Baseline sensitivity curves are publicly available on \textsc{GitLab}~\raisebox{-1pt}{\href{https://gitlab.et-gw.eu/pub/et-sensitivity-curves}{\includegraphics[width=9pt]{GitLab_icon.pdf}}}~\cite{pubcurves}.\footnote{The repository contains 3 different results for the HF and LF interferometers separately. The Baseline sensitivities used in this paper come from the results provided in the second column (labeled ``\texttt{displacement ASD}''), obtained by combining the \ac{asd} for HF and LF as follows, rescaled by the arm length of the considered detector:
\[
{\rm ASD}_{\rm tot} = \sqrt{\left(({\rm ASD}_{\rm LF}/\rm{L})^{-2} + {(\rm ASD}_{\rm HF}/\rm{L})^{-2}\right)^{-1}}\;.
\]
with L the detector arm length in meters.} 

\section{PSD comparison for different configurations}\label{app:psd_cfr}

\begin{figure}[tb]
    \centering
    \begin{minipage}{0.49\linewidth}
        \includegraphics[width=1\linewidth]{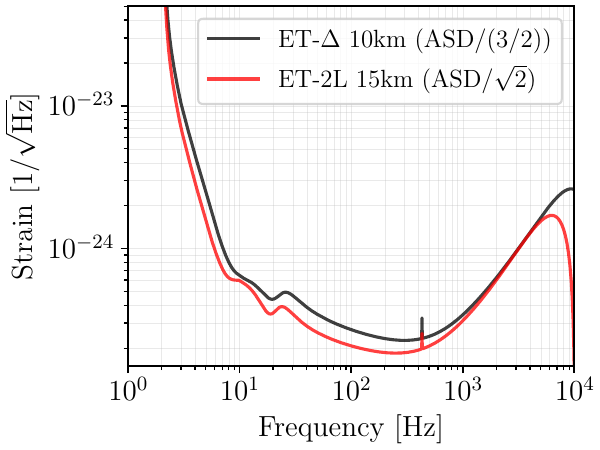}
    \end{minipage}
    \begin{minipage}{0.49\linewidth}
        \includegraphics[width=1\linewidth]{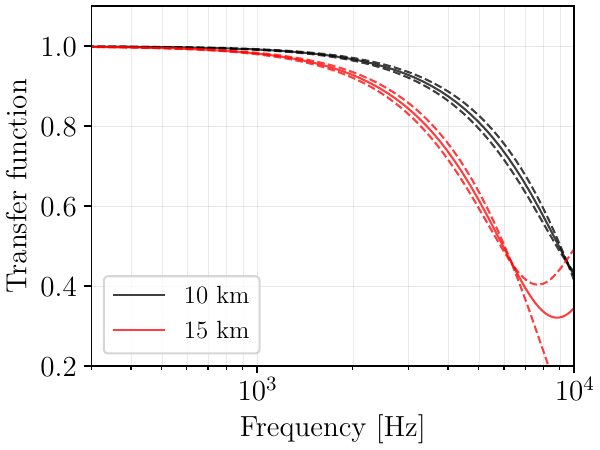}
    \end{minipage}
\caption{\emph{Left panel}: effective network amplitude spectral densities for the ET-$\Delta$ 10\,km and ET-2L 15\,km configurations. To enable a direct comparison between the two configurations, the \acp{asd} are normalized by the corresponding network response factors, namely $3/2$ for the triangular configuration and $\sqrt{2}$ for the 2L configuration. \emph{Right panel}: transfer function comparing the effective sensitivities of the 10\,km and 15\,km ET arm length. The solid line shows the median and the dashed ones the 90\%\,c.l. after averaging over sky position and polarization angle.}
    \label{fig:effective_psd_and_transfer_func}
\end{figure}

In this appendix, we compare the effective strain sensitivities for the ET-$\Delta$ and ET-2L configurations. Since the two configurations differ both in arm length and in the number and orientation of interferometers, a comparison of the raw \acf{asd} is not fully representative of the actual network sensitivity. Therefore, in Figure~\ref{fig:effective_psd_and_transfer_func}, we show an effective \ac{asd}, normalized to account for the number and orientation of the interferometers. For the ET-$\Delta$ configuration, the \ac{asd} is rescaled by the factor $\sqrt{3\sin^2{(60^{\circ})}}=3/2$, where 3 accounts for the number of interferometers and $\sin(60^{\circ})$ for the opening angle. For the 2L geometry, on the other side, the effective \ac{asd} is rescaled by a factor $\sqrt{2}$ accounting for 2 interferometers with $90^{\circ}$ opening angle. Although being an approximation that does not take into account the full antenna pattern response, the left plot in Figure~\ref{fig:effective_psd_and_transfer_func} demonstrates the better low-frequency sensitivity of the 2L configuration. At high frequencies, above $\sim\!2$\,kHz, instead, the two configurations offer a similar return.

Another factor to account for is the transfer response when the \ac{lwa} is dropped. In the low-frequency regime, where the \ac{gw} wavelength is much larger than the interferometer arm length, the detector response is approximately frequency-independent, and the strain response simply scales with the arm length. In this regime, the two configurations differ through the geometrical strain normalization as discussed above. At higher frequencies, however, the \ac{lwa} is no longer valid, because the \ac{gw} wavelength becomes comparable to the detector arm length. The detector response thus acquires a frequency dependence, as shown in the right panel of \Cref{fig:effective_psd_and_transfer_func}. We can see how at frequencies above $\sim\!1$\,kHz the longer-arm configuration gets more penalized.

\section{The role of the minimum frequency}\label{app:minimum_freq}

The sensitivity curves analyzed in this study extend below 3\,Hz, reaching down to frequencies where detector performance is still evolving. To fully exploit the available bandwidth for our comparative analysis of the different \acp{psd}, we adopt a minimum frequency of $f_{\rm min} = 2\,{\rm Hz}$ in our calculations. Below $2\,{\rm Hz}$, the \ac{psd} degrades substantially across all configurations, by more than 10 orders of magnitude, while it decreases by less than three orders of magnitude between $2\,{\rm Hz}$ and $3\,{\rm Hz}$, making $2\,{\rm Hz}$ a meaningful lower cutoff. This choice allows us to assess performance differences across the complete sensitive band of the instrument, which is particularly relevant when evaluating variations in low-frequency sensitivity. This is the same choice that was adopted in Refs.~\cite{Branchesi:2023mws,ET:2025xjr}.

\begin{table}[tbp]
\caption{We report the number of detections of \ac{cbc} sources with ${\rm SNR}\geq 12$ for \ac{bbh} sources and with ${\rm SNR}\geq 8$ for \ac{bns} sources. We select the two limiting-case sensitivity curves and perform the analysis starting from 3\,Hz. The corresponding results for 2\,Hz as shown in Table~\ref{tab:snr_table} are reported for reference in parentheses.}

{\setlength{\tabcolsep}{5.36pt}
\setlength{\aboverulesep}{-1pt}
\setlength{\belowrulesep}{-3pt}
\begin{tabularx}{\textwidth}{lcccccccc}
	\toprule
	  & \multicolumn{2}{c}{\bf BNS} & \multicolumn{2}{c}{\bf BBH} & \multicolumn{2}{c}{\bf Pop.~III} & \multicolumn{2}{c}{\bf PBH} \\
    \cmidrule(r){2-3} \cmidrule(r){4-5} \cmidrule(r){6-7} \cmidrule{8-9}
    &\multicolumn{2}{c}{\makecell{\scriptsize SNR$\geq8$ }} &\multicolumn{2}{c}{\makecell{\scriptsize SNR$\geq12$  \scriptsize }} &\multicolumn{2}{c}{\makecell{\scriptsize SNR$\geq12$ \scriptsize }} &\multicolumn{2}{c}{\makecell{\scriptsize SNR$\geq12$}} \\ 

	 \bf Sensitivity curve & $\Delta$ & 2L & $\Delta$ & 2L & $\Delta$ & 2L & $\Delta$ & 2L \\\midrule
    \multirow{2}{*}{Baseline} &{44951} &{78403} &{77858} &{91332} &{1767} &{1999} &{22761} &{38368}\\ 
&{({\footnotesize 44836})} &{({\footnotesize 78590})} &{({\footnotesize 77928})} &{({\footnotesize 91393})} &{({\footnotesize 1797})} &{({\footnotesize 2062})} &{({\footnotesize 22806})} &{({\footnotesize 38492})}\\ 
\rowcolor{MidnightBlueLight}&{11908} &{28399} &{48172} &{65358} &{1014} &{1403} &{7480} &{14667}\\ 
\rowcolor{MidnightBlueLight}\multirow{-2}{*}{Worst case total}&{({\footnotesize 11874)}} &{({\footnotesize 28604)}} &{({\footnotesize 48251)}} &{({\footnotesize 65454)}} &{({\footnotesize 1032)}} &{({\footnotesize 1443)}} &{({\footnotesize 7509)}} &{({\footnotesize 14705)}}
\\ \bottomrule

\end{tabularx}
}
\label{tab:snr_table_3Hz_low_snr}
\end{table}

\begin{table}[tbp]
\caption{We report the number of detections of \ac{cbc} sources with ${\rm SNR}\geq 100$ for \ac{bbh} sources and with ${\rm SNR}\geq 50$ for \ac{bns} sources. We select the two limiting-case sensitivity curves and perform the analysis starting from 3\,Hz. The corresponding results for 2\,Hz as shown in Table~\ref{tab:snr_table} are reported for reference in parentheses.}

{\setlength{\tabcolsep}{9.2pt}
\setlength{\aboverulesep}{-1pt}
\setlength{\belowrulesep}{-3pt}
\begin{tabularx}{\textwidth}{lcccccccc}
	\toprule
	  & \multicolumn{2}{c}{\bf BNS} & \multicolumn{2}{c}{\bf BBH} & \multicolumn{2}{c}{\bf Pop.~III} & \multicolumn{2}{c}{\bf PBH} \\
    \cmidrule(r){2-3} \cmidrule(r){4-5} \cmidrule(r){6-7} \cmidrule{8-9}
    &\multicolumn{2}{c}{\makecell{\scriptsize SNR$\geq50$ }} &\multicolumn{2}{c}{\makecell{\scriptsize SNR$\geq100$  \scriptsize }} &\multicolumn{2}{c}{\makecell{\scriptsize SNR$\geq100$ \scriptsize }} &\multicolumn{2}{c}{\makecell{\scriptsize SNR$\geq100$}} \\ 

	 \bf Sensitivity curve & $\Delta$ & 2L & $\Delta$ & 2L & $\Delta$ & 2L & $\Delta$ & 2L \\\midrule
    \multirow{2}{*}{Baseline} &{149} &{309} &{1656} &{3267} &{57} &{94} &{107} &{197}\\ 
&{({\footnotesize 146})} &{({\footnotesize 296})} &{({\footnotesize 1654})} &{({\footnotesize 3286})} &{({\footnotesize 54})} &{({\footnotesize 97})} &{({\footnotesize 107})} &{({\footnotesize 198})}\\ 
\rowcolor{MidnightBlueLight}&{35} &{97} &{389} &{956} &{15} &{40} &{19} &{59}\\ 
\rowcolor{MidnightBlueLight}\multirow{-2}{*}{Worst case total}&{({\footnotesize 40)}} &{({\footnotesize 93)}} &{({\footnotesize 398)}} &{({\footnotesize 958)}} &{({\footnotesize 21)}} &{({\footnotesize 36)}} &{({\footnotesize 19)}} &{({\footnotesize 62)}}
\\ \bottomrule

\end{tabularx}
}
\label{tab:snr_table_3Hz_high_snr}
\end{table}

\begin{table}[tbp]
\caption{Summary table for the statistical uncertainties on the source parameters for \ac{bbh} sources for the two limiting-case sensitivity curves as in Table~\ref{tab:fisher_bbh}. Here we report absolute numbers for each criteria, comparing the results obtained starting from 3\,Hz with those obtained starting from 2\,Hz (in parentheses).}
{\setlength{\tabcolsep}{4.87pt}
\begin{tabularx}{\textwidth}{lcccccccc}
	\toprule
	\multicolumn{9}{c}{\bf BBH}\\ \midrule
    &\multicolumn{2}{c}{\makecell{$\frac{\Delta\mathcal{M}_c} {\mathcal{M}_c}$ \scriptsize{$\le 1$\,\textperthousand}}} &\multicolumn{2}{c}{\makecell{$\frac{\Delta d_L} {d_L}$ \scriptsize{$ \le 10\%$}}} &\multicolumn{2}{c}{\makecell{$\frac{\Delta\chi_1} {\chi_1}$ \scriptsize{$ \le 10\%$}}} &\multicolumn{2}{c}{\makecell{$\Delta\Omega_{90\%}$ \scriptsize{$ \le 1000\,\text{deg}^2$}}} \\ %
    \cmidrule(r){2-3} \cmidrule(r){4-5} \cmidrule(r){6-7} \cmidrule{8-9}
   
	 \bf Sensitivity curve  &$\Delta$ &2L &$\Delta$ &2L &$\Delta$ &2L &$\Delta$ &2L \\\midrule
    \multirow{2}{*}{Baseline} &{42046} &{51626} &{7146} &{16931} &{7451} &{9648} &{20348} &{36835}\\ 
&{({\footnotesize 50428})} &{({\footnotesize 62968})} &{({\footnotesize 7000})} &{({\footnotesize 16750})} &{({\footnotesize 7808})} &{({\footnotesize 10175})} &{({\footnotesize 20213})} &{({\footnotesize 36569})}\\ 
\rowcolor{MidnightBlueLight}&{22405} &{32473} &{2995} &{9216} &{3957} &{5739} &{9863} &{22281}\\ 
\rowcolor{MidnightBlueLight}\multirow{-2}{*}{Worst case total}&{({\footnotesize 27101})} &{({\footnotesize 40066})} &{({\footnotesize 2883})} &{({\footnotesize 9157})} &{({\footnotesize 4219})} &{({\footnotesize 6157})} &{({\footnotesize 9765})} &{({\footnotesize 22060})}
\\ \bottomrule

\end{tabularx}
}
\label{tab:fisher_bbh_3Hz}
\end{table}

\begin{table}[tbp]
\caption{Summary table for the statistical uncertainties on the source parameters for \ac{bns} sources for the two limiting-case sensitivity curves as in Table~\ref{tab:fisher_bns}. Here we report absolute numbers for each criteria, comparing the results obtained starting from 3\,Hz with those obtained starting from 2\,Hz (in parentheses).}
{\setlength{\tabcolsep}{6.9pt}
\begin{tabularx}{\textwidth}{lcccccccc}
	\toprule
	\multicolumn{9}{c}{\bf BNS}\\ \midrule
    &\multicolumn{2}{c}{\makecell{$\frac{\Delta\mathcal{M}_c} {\mathcal{M}_c}$ \scriptsize{$\le 1$\,\textperthousand}}} &\multicolumn{2}{c}{\makecell{$\frac{\Delta d_L} {d_L}$ \scriptsize{$ \le 10\%$}}} &\multicolumn{2}{c}{\makecell{$\frac{\Delta\tilde{\Lambda}} {\tilde{\Lambda}}$ \scriptsize{$ \le 20\%$}}} &\multicolumn{2}{c}{\makecell{$\Delta\Omega_{90\%}$ \scriptsize{$ \le 1000\,\text{deg}^2$}}} \\ %
    \cmidrule(r){2-3} \cmidrule(r){4-5} \cmidrule(r){6-7} \cmidrule{8-9}
   
	 \bf Sensitivity curve  &$\Delta$ &2L &$\Delta$ &2L &$\Delta$ &2L &$\Delta$ &2L \\\midrule
    Baseline &{44926} &{77944} &{49} &{284} &{981} &{1997} &{1217} &{4008}\\ 
&{({\footnotesize 44834})} &{({\footnotesize 78518})} &{({\footnotesize 62})} &{({\footnotesize 315})} &{({\footnotesize 1003})} &{({\footnotesize 1990})} &{({\footnotesize 1353})} &{({\footnotesize 4545})}\\ 
\rowcolor{MidnightBlueLight}&{11908} &{28382} &{9} &{71} &{342} &{692} &{295} &{1337}\\ 
\rowcolor{MidnightBlueLight}\multirow{-2}{*}{Worst case total}&{({\footnotesize 11874})} &{({\footnotesize 28603})} &{({\footnotesize 14})} &{({\footnotesize 88})} &{({\footnotesize 336})} &{({\footnotesize 697})} &{({\footnotesize 330})} &{({\footnotesize 1576})}
\\ \bottomrule

\end{tabularx}
}
\label{tab:fisher_bns_3Hz}
\end{table}

Earlier studies, such as ET Conceptual Design Report~\cite{ETconceptual2020}, used a higher minimum frequency $f_{\rm min} = 3\,{\rm Hz}$. To verify that our comparative conclusions remain robust to this choice, we have repeated representative analyses using $f_{\rm min} = 3\,{\rm Hz}$. The results, summarized below and in Tables~\ref{tab:snr_table_3Hz_low_snr}--\ref{tab:premerger_1min_23hz}, demonstrate that the relative performance of different sensitivity curves is preserved. In Tables~\ref{tab:snr_table_3Hz_low_snr}--\ref{tab:fisher_bns_3Hz} we select the Baseline sensitivity and Worst Case Total, the two limiting cases, and perform the analysis only for those. We note that:
\begin{description}[align=left]
	\item[Detection horizons and rates:] Horizon distances for equal-mass systems change by $\lesssim3\%$ across most of the mass range when shifting from $f_{\rm min}=2\,{\rm Hz}$ to $3\,{\rm Hz}$, with relative differences between sensitivity curves remaining nearly identical. \Cref{tab:snr_table_3Hz_low_snr} reports detection numbers for the Baseline and worst-case scenarios at both $f_{\rm min}$ values. There is practically no impact on the detection number for \ac{cbc} sources: the variation in the number of detections is driven more by randomness in the duty cycle than by differences in the starting frequency (we also checked that without including the duty cycle in our analyses we miss only a handful of events starting from 3\,Hz). The fractional change in detection rates is comparable, confirming that our comparative assessment of instrumental degradations is not sensitive to this choice. The same results are confirmed when a higher \ac{snr} threshold is chosen, as shown in~\Cref{tab:snr_table_3Hz_high_snr}.
    \item[Parameter estimation:] Considerations made for the detection rates described in the previous point hold for the statistical uncertainties on the source parameters for \acp{bbh} and \acp{bns}, as shown in Table~\ref{tab:fisher_bbh_3Hz} and Table~\ref{tab:fisher_bns_3Hz}, respectively.
	\item[High-frequency sources:] Metrics dominated by high-frequency signal content, i.e., \ac{bns} post-merger observability and \ac{ccsn} detectability, are entirely unaffected, as expected. 
	\item[Stochastic backgrounds:] The relative difference in \ac{pls} for \ac{sgwb} searches changes negligibly in the $2-3\,{\rm Hz}$ band, where sensitivity is already subdominant to the instrumental bucket at $\sim\!10\,{\rm Hz}$.
	\item[Continuous waves from pulsars:] The number of catalog sources within the detector band decreases from 2048 to 1668 when adopting $f_{\rm min}=3\,{\rm Hz}$. However, the excluded systems predominantly fall below detection thresholds even in optimistic scenarios. Consequently, the number of detectable sources and, crucially, the relative variation between sensitivity curves remain nearly identical.
	\item[Pre-merger BNS localization:] The metric most sensitive to the $f_{\rm min}$ choice is early-warning localization of \ac{bns} mergers, particularly at $\Delta t = 30\,{\rm min}$ before coalescence, where a fraction of the population emits at $f \lesssim 3\,{\rm Hz}$ (see \Cref{fig:pre_merger_freq}). \Cref{tab:premerger_30min_23hz} shows that detection numbers at 30\,min decrease by a few when adopting $f_{\rm min}=3\,{\rm Hz}$, as expected from the loss of low-frequency information. However, the relative degradation between sensitivity curves remains consistent: for instance, the worst-case scenario retains $\sim\!18\%$ of the Baseline performance at both $f_{\rm min}$ values. Importantly, the comparative conclusion that low-frequency sensitivity is critical for early warnings holds regardless of the minimum frequency choice. At shorter warning times, $\Delta t = 10\,{\rm min}$ and $1\,{\rm min}$, where the signal shifts to higher frequencies, the impact diminishes further, as can be seen from Tables~\ref{tab:premerger_10min_23hz} and~\ref{tab:premerger_1min_23hz}, respectively. Moreover, given the limited sample sizes at 30\,min, one should be cautious in over-interpreting absolute numbers. The trends across sensitivity configurations remain robust.
\end{description}

Our choice of $f_{\rm min}=2\,{\rm Hz}$ allows us to completely assess the low-frequency performance differences that are central to this study. While absolute detection numbers for early-warning \ac{bns} localization are modestly overestimated compared to $f_{\rm min}=3\,{\rm Hz}$, the relative comparison between sensitivity curves, which forms the basis of our conclusions, is preserved across all science cases. The primary findings of this work remain valid under either choice.

{\setlength{\tabcolsep}{6.9pt}
\begin{xltabular}{\textwidth}{X c c c c c}
\caption{Pre-merger \ac{bns} counts for 2L 15\,km \ac{et} configuration for different cuts on the sky localization accuracy at 30\,min prior to merger comparing two different starting frequencies of 2\,Hz and 3\,Hz. We report the results for binaries with all orientations and for those with a viewing angle $\Theta_v\leq15^\circ$.}
\label{tab:premerger_30min_23hz} \\ \toprule
\multicolumn{6}{c}{\bf 30\,min} \\ \midrule
\multicolumn{1}{c}{} &
\multicolumn{1}{c}{} &
\multicolumn{2}{c}{All orientation BNSs} &
\multicolumn{2}{c}{BNSs with $\Theta_v \leq 15^\circ$} \\ %
\cmidrule(r){3-4} \cmidrule{5-6}
{\bf Sensitivity curve} & $\leq\Delta\Omega_{90\%}$ &2\,Hz &3\,Hz  &2\,Hz &3\,Hz \\ \midrule
\endfirsthead
\toprule
\multicolumn{6}{c}{\bf 30\,min} \\ \midrule
\multicolumn{1}{c}{} &
\multicolumn{1}{c}{} &
\multicolumn{2}{c}{All orientation BNSs} &
\multicolumn{2}{c}{BNSs with $\Theta_v \leq 15^\circ$} \\ %
\cmidrule(r){3-4} \cmidrule{5-6}
{\bf Sensitivity curve} & $\leq\Delta\Omega_{90\%}$ &2\,Hz &3\,Hz  &2\,Hz &3\,Hz \\
\midrule
\endhead
{Baseline} & 10 & 2 & 2 & 0 & 0 \\
{} & 100 & 33 & 31 & 4 & 4 \\
{} & 1000 & 350 & 289 & 32 & 26 \\
{} & Total & 2590 & 2533 & 235 & 227 \\

\rowcolor{MidnightBlueLight}{Baseline $\times 1.5$} & 10 & 0 & 0 & 0 & 0 \\
\rowcolor{MidnightBlueLight}{} & 100 & 14 & 13 & 3 & 3 \\
\rowcolor{MidnightBlueLight}{} & 1000 & 111 & 85 & 10 & 10 \\
\rowcolor{MidnightBlueLight}{} & Total & 947 & 927 & 91 & 90 \\

{Bin 1: $f<7\,\mathrm{Hz}$} & 10 & 0 & 0 & 0 & 0 \\
{} & 100 & 16 & 15 & 3 & 3 \\
{} & 1000 & 128 & 107 & 10 & 10 \\
{} & Total & 1022 & 1007 & 92 & 91 \\

\rowcolor{MidnightBlueLight}{Bin 2: $7\,\mathrm{Hz}<f<10\,\mathrm{Hz}$} & 10 & 2 & 2 & 0 & 0 \\
\rowcolor{MidnightBlueLight}{} & 100 & 30 & 26 & 4 & 4 \\
\rowcolor{MidnightBlueLight}{} & 1000 & 332 & 277 & 32 & 26 \\
\rowcolor{MidnightBlueLight}{} & Total & 2563 & 2504 & 235 & 227 \\

{Bin 3: $10\,\mathrm{Hz}<f<30\,\mathrm{Hz}$} & 10 & 2 & 2 & 0 & 0 \\
{} & 100 & 33 & 31 & 4 & 4 \\
{} & 1000 & 350 & 289 & 32 & 26 \\
{} & Total & 2590 & 2533 & 235 & 227 \\

\rowcolor{MidnightBlueLight}{Bin 4: $30\,\mathrm{Hz}<f<450\,\mathrm{Hz}$} & 10 & 2 & 2 & 0 & 0 \\
\rowcolor{MidnightBlueLight}{} & 100 & 33 & 31 & 4 & 4 \\
\rowcolor{MidnightBlueLight}{} & 1000 & 350 & 289 & 32 & 26 \\
\rowcolor{MidnightBlueLight}{} & Total & 2590 & 2533 & 235 & 227 \\

{Bin 5: $f>450\,\mathrm{Hz}$} & 10 & 2 & 2 & 0 & 0 \\
{} & 100 & 33 & 31 & 4 & 4 \\
{} & 1000 & 350 & 289 & 32 & 26 \\
{} & Total & 2590 & 2533 & 235 & 227 \\

\rowcolor{MidnightBlueLight}{Worst case total} & 10 & 0 & 0 & 0 & 0 \\
\rowcolor{MidnightBlueLight}{} & 100 & 10 & 7 & 2 & 2 \\
\rowcolor{MidnightBlueLight}{} & 1000 & 61 & 49 & 8 & 6 \\
\rowcolor{MidnightBlueLight}{} & Total & 489 & 477 & 51 & 51 \\

\bottomrule

\end{xltabular}
}

{\setlength{\tabcolsep}{6.9pt}
\begin{xltabular}{\textwidth}{X c c c c c}
\caption{Same as in \Cref{tab:premerger_30min_23hz} considering systems at 10\,min prior to merger.}
\label{tab:premerger_10min_23hz} \\ \toprule
\multicolumn{6}{c}{\bf 10\,min} \\ \midrule
\multicolumn{1}{c}{} &
\multicolumn{1}{c}{} &
\multicolumn{2}{c}{All orientation BNSs} &
\multicolumn{2}{c}{BNSs with $\Theta_v \leq 15^\circ$} \\ %
\cmidrule(r){3-4} \cmidrule{5-6}
{\bf Sensitivity curve} & $\leq\Delta\Omega_{90\%}$ &2\,Hz &3\,Hz  &2\,Hz &3\,Hz \\ \midrule
\endfirsthead
\toprule
\multicolumn{6}{c}{\bf 10\,min} \\ \midrule
\multicolumn{1}{c}{} &
\multicolumn{1}{c}{} &
\multicolumn{2}{c}{All orientation BNSs} &
\multicolumn{2}{c}{BNSs with $\Theta_v \leq 15^\circ$} \\ %
\cmidrule(r){3-4} \cmidrule{5-6}
{\bf Sensitivity curve} & $\leq\Delta\Omega_{90\%}$ &2\,Hz &3\,Hz  &2\,Hz &3\,Hz \\
\midrule
\endhead
{Baseline} & 10 & 5 & 5 & 2 & 2 \\
{} & 100 & 84 & 73 & 8 & 7 \\
{} & 1000 & 839 & 733 & 91 & 80 \\
{} & Total & 10430 & 10322 & 942 & 934 \\

\rowcolor{MidnightBlueLight}{Baseline $\times 1.5$} & 10 & 2 & 1 & 0 & 0 \\
\rowcolor{MidnightBlueLight}{} & 100 & 26 & 22 & 3 & 3 \\
\rowcolor{MidnightBlueLight}{} & 1000 & 293 & 244 & 31 & 24 \\
\rowcolor{MidnightBlueLight}{} & Total & 3776 & 3743 & 384 & 381 \\

{Bin 1: $f<7\,\mathrm{Hz}$} & 10 & 3 & 3 & 0 & 0 \\
{} & 100 & 44 & 40 & 5 & 5 \\
{} & 1000 & 468 & 403 & 52 & 43 \\
{} & Total & 6809 & 6771 & 600 & 598 \\

\rowcolor{MidnightBlueLight}{Bin 2: $7\,\mathrm{Hz}<f<10\,\mathrm{Hz}$} & 10 & 4 & 4 & 1 & 1 \\
\rowcolor{MidnightBlueLight}{} & 100 & 62 & 53 & 5 & 5 \\
\rowcolor{MidnightBlueLight}{} & 1000 & 647 & 549 & 61 & 53 \\
\rowcolor{MidnightBlueLight}{} & Total & 7293 & 7177 & 733 & 720 \\

{Bin 3: $10\,\mathrm{Hz}<f<30\,\mathrm{Hz}$} & 10 & 4 & 4 & 1 & 1 \\
{} & 100 & 79 & 69 & 8 & 7 \\
{} & 1000 & 823 & 717 & 91 & 80 \\
{} & Total & 10413 & 10304 & 942 & 934 \\

\rowcolor{MidnightBlueLight}{Bin 4: $30\,\mathrm{Hz}<f<450\,\mathrm{Hz}$} & 10 & 5 & 5 & 2 & 2 \\
\rowcolor{MidnightBlueLight}{} & 100 & 84 & 73 & 8 & 7 \\
\rowcolor{MidnightBlueLight}{} & 1000 & 839 & 733 & 91 & 80 \\
\rowcolor{MidnightBlueLight}{} & Total & 10430 & 10322 & 942 & 934 \\

{Bin 5: $f>450\,\mathrm{Hz}$} & 10 & 5 & 5 & 2 & 2 \\
{} & 100 & 84 & 73 & 8 & 7 \\
{} & 1000 & 839 & 733 & 91 & 80 \\
{} & Total & 10430 & 10322 & 942 & 934 \\

\rowcolor{MidnightBlueLight}{Worst case total} & 10 & 1 & 0 & 0 & 0 \\
\rowcolor{MidnightBlueLight}{} & 100 & 16 & 15 & 3 & 3 \\
\rowcolor{MidnightBlueLight}{} & 1000 & 159 & 128 & 15 & 12 \\
\rowcolor{MidnightBlueLight}{} & Total & 2026 & 1995 & 208 & 205 \\

\bottomrule

\end{xltabular}
}

{\setlength{\tabcolsep}{6.9pt}
\begin{xltabular}{\textwidth}{X c c c c c}
\caption{Same as in \Cref{tab:premerger_30min_23hz} considering systems at 1\,min prior to merger.}
\label{tab:premerger_1min_23hz} \\ \toprule
\multicolumn{6}{c}{\bf 1\,min} \\ \midrule
\multicolumn{1}{c}{} &
\multicolumn{1}{c}{} &
\multicolumn{2}{c}{All orientation BNSs} &
\multicolumn{2}{c}{BNSs with $\Theta_v \leq 15^\circ$} \\ %
\cmidrule(r){3-4} \cmidrule{5-6}
{\bf Sensitivity curve} & $\leq\Delta\Omega_{90\%}$ &2\,Hz &3\,Hz  &2\,Hz &3\,Hz \\ \midrule
\endfirsthead
\toprule
\multicolumn{6}{c}{\bf 1\,min} \\ \midrule
\multicolumn{1}{c}{} &
\multicolumn{1}{c}{} &
\multicolumn{2}{c}{All orientation BNSs} &
\multicolumn{2}{c}{BNSs with $\Theta_v \leq 15^\circ$} \\ %
\cmidrule(r){3-4} \cmidrule{5-6}
{\bf Sensitivity curve} & $\leq\Delta\Omega_{90\%}$ & 2\,Hz & 3\,Hz  & 2\,Hz & 3\,Hz \\
\midrule
\endhead
{Baseline} & 10 & 9 & 8 & 2 & 2 \\
{} & 100 & 201 & 182 & 20 & 16 \\
{} & 1000 & 1949 & 1743 & 244 & 229 \\
{} & Total & 39586 & 39417 & 3776 & 3759 \\

\rowcolor{MidnightBlueLight}{Baseline $\times 1.5$} & 10 & 4 & 4 & 1 & 1 \\
\rowcolor{MidnightBlueLight}{} & 100 & 68 & 61 & 6 & 6 \\
\rowcolor{MidnightBlueLight}{} & 1000 & 689 & 599 & 86 & 75 \\
\rowcolor{MidnightBlueLight}{} & Total & 13593 & 13528 & 1409 & 1398 \\

{Bin 1: $f<7\,\mathrm{Hz}$} & 10 & 5 & 5 & 1 & 1 \\
{} & 100 & 130 & 118 & 11 & 8 \\
{} & 1000 & 1277 & 1152 & 180 & 174 \\
{} & Total & 33210 & 33146 & 3227 & 3222 \\

\rowcolor{MidnightBlueLight}{Bin 2: $7\,\mathrm{Hz}<f<10\,\mathrm{Hz}$} & 10 & 7 & 7 & 2 & 2 \\
\rowcolor{MidnightBlueLight}{} & 100 & 172 & 158 & 16 & 14 \\
\rowcolor{MidnightBlueLight}{} & 1000 & 1742 & 1539 & 219 & 198 \\
\rowcolor{MidnightBlueLight}{} & Total & 31491 & 31323 & 3059 & 3044 \\

{Bin 3: $10\,\mathrm{Hz}<f<30\,\mathrm{Hz}$} & 10 & 7 & 6 & 2 & 2 \\
{} & 100 & 137 & 122 & 12 & 9 \\
{} & 1000 & 1400 & 1248 & 179 & 163 \\
{} & Total & 27544 & 27358 & 2901 & 2878 \\

\rowcolor{MidnightBlueLight}{Bin 4: $30\,\mathrm{Hz}<f<450\,\mathrm{Hz}$} & 10 & 9 & 8 & 2 & 2 \\
\rowcolor{MidnightBlueLight}{} & 100 & 201 & 182 & 20 & 16 \\
\rowcolor{MidnightBlueLight}{} & 1000 & 1949 & 1743 & 244 & 229 \\
\rowcolor{MidnightBlueLight}{} & Total & 39586 & 39417 & 3776 & 3759 \\

{Bin 5: $f>450\,\mathrm{Hz}$} & 10 & 9 & 8 & 2 & 2 \\
{} & 100 & 201 & 182 & 20 & 16 \\
{} & 1000 & 1949 & 1743 & 244 & 229 \\
{} & Total & 39586 & 39417 & 3776 & 3759 \\

\rowcolor{MidnightBlueLight}{Worst case total} & 10 & 2 & 2 & 0 & 0 \\
\rowcolor{MidnightBlueLight}{} & 100 & 49 & 43 & 5 & 4 \\
\rowcolor{MidnightBlueLight}{} & 1000 & 509 & 422 & 61 & 48 \\
\rowcolor{MidnightBlueLight}{} & Total & 10335 & 10276 & 1040 & 1029 \\

\bottomrule

\end{xltabular}
}

\section{Pre-merger frequency}\label{app:pre_merger_frequency}

The frequency of \acp{gw} emitted by a compact binary during the inspiral phase is determined by the loss of orbital energy to gravitational radiation. Under the assumption of quasi-circular, \ac{gw}-driven evolution, the time remaining until coalescence, $\Delta t= t_c-t$, is related to the \ac{gw} frequency $f$ at leading order through (see, e.g., Refs.~\cite{Maggiore:2007ulw,Marsat:2018oam}):
\begin{equation}\label{eq:t_of_f}
    \Delta t = t_c-t = \frac{5}{256}\left( \frac{G {\mathcal{M}}_c}{c^3} \right)^{-5/3}(\pi f)^{-8/3}\;,
\end{equation}
where ${\mathcal{M}}_c$ is the detector-frame chirp mass of the binary, $t_c$ is the coalescence time, $G$ is the gravitational constant, and $c$ is the speed of light. Inverting Eq.~\eqref{eq:t_of_f} gives the frequency at a given time before merger:
\begin{equation}\label{eq:f_of_t}
    f(\Delta t) = \frac{1}{\pi}\left(\frac{5}{256 \Delta t} \right)^{3/8}\left( \frac{G {\mathcal{M}}_c}{c^3} \right)^{-5/8}\;.
\end{equation}
This expression shows that the frequency increases as the system approaches the merger.
\begin{figure}[tbp]
    \centering
    \includegraphics[width=0.75\linewidth]{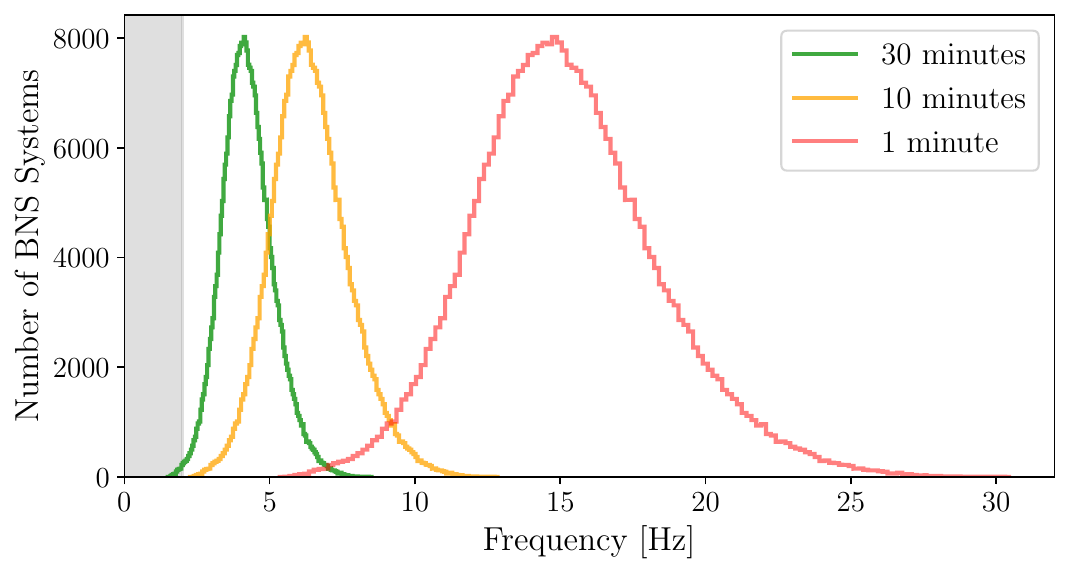}
    \caption{Distribution of the frequencies evaluated at 30, 10, and 1 minute prior to merger for the simulated \ac{bns} systems. The region below 2\,Hz is shaded in gray, indicating the low-frequency cutoff adopted in our analyses (see Appendix~\ref{app:minimum_freq}).}
    \label{fig:pre_merger_freq}
\end{figure}
\Cref{fig:pre_merger_freq} shows the distribution of frequencies for our population of \acp{bns} systems evaluated at 30, 10, and 1 minute prior to merger. As expected from the scaling above, the distributions shift systematically toward higher frequencies as the time before merger decreases. In addition to this shift, the distributions broaden moving closer to merger. This broadening arises from the nonlinear mapping between chirp mass and frequency. The overall normalization of the mapping scales as $\Delta t ^{-3/8}$, so for smaller $\Delta t$, the same relative variation in chirp mass produces a larger spread in frequency.

All frequencies shown remain well within the inspiral regime, far below the characteristic merger frequency ($\sim$\,kHz for \ac{bns} systems), ensuring that the leading-order inspiral approximation used here is valid.

\section{Additional plots}\label{app:additional_plots}
\begin{figure}[tbp]
    \centering
    \includegraphics[width=0.95\linewidth]{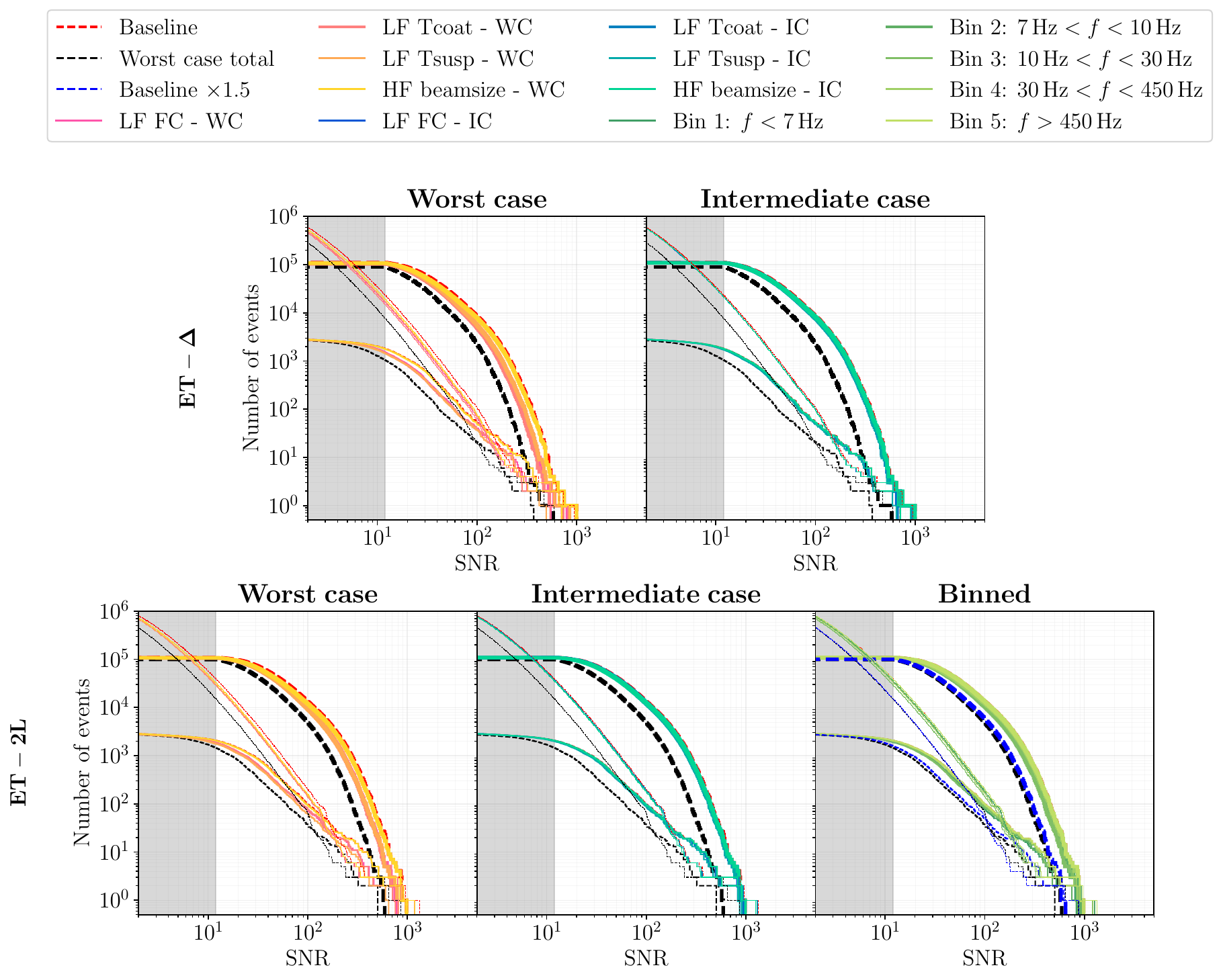}
    \caption{Cumulative SNR distribution for the three \ac{bbh} populations: \acp{bbh} from Population I and II stars, those from the Pop.~III population, and \acp{pbh}. The distributions follow similar patterns, and different \ac{bbh} sources are distinguished by the statistics, from the most numerous \acp{pbh} to the least numerous \acp{bbh} from Pop.~III stars. We use thicker lines for \ac{bbh} from Population I and II stars.}
    \label{fig:bbh_SNR_all}
\end{figure}

\begin{figure}[tb]
    \centering
    \includegraphics[width=0.95\linewidth]{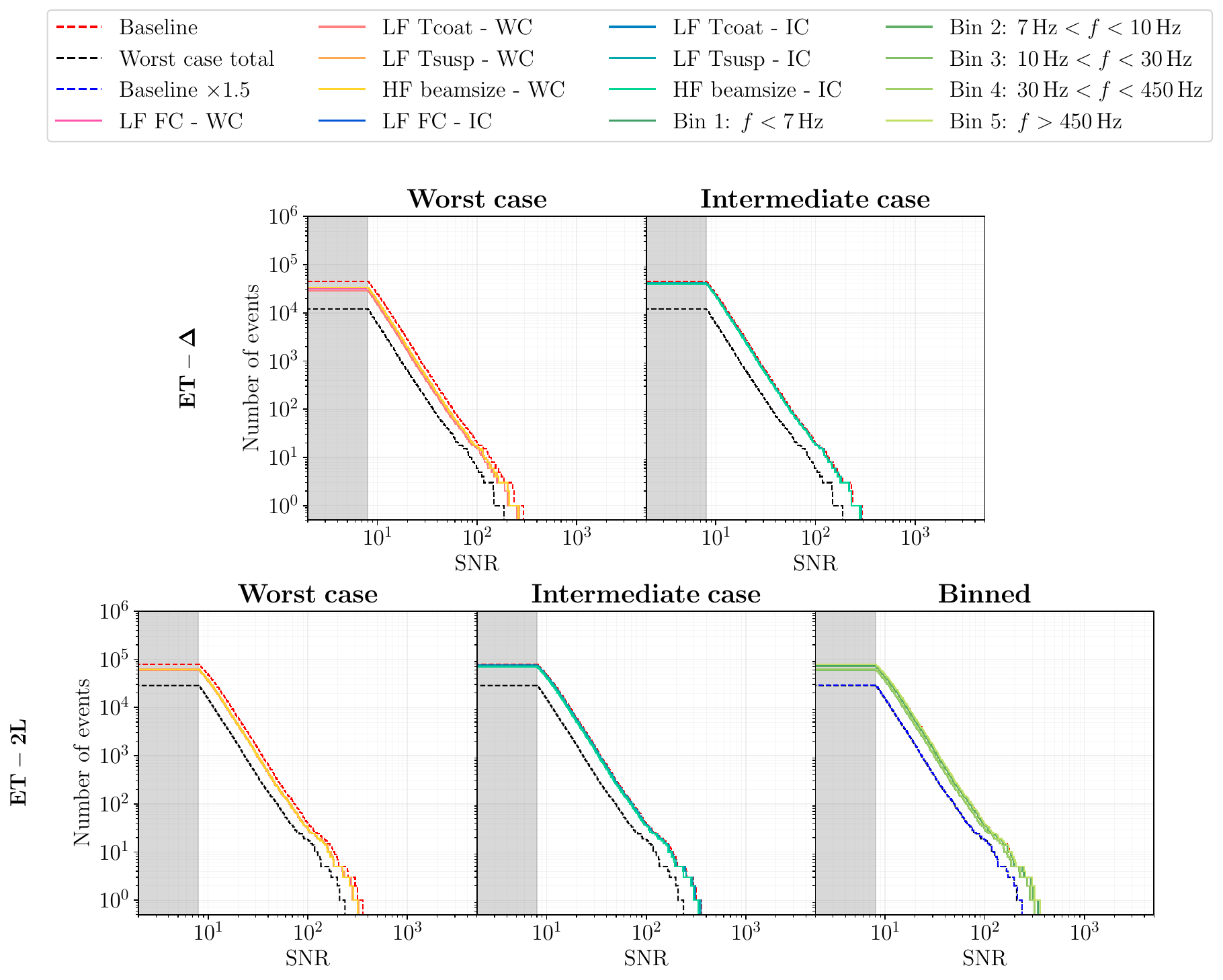}
    \caption{Cumulative SNR distribution for the \ac{bns} population.}
    \label{fig:bns_snr}
\end{figure}

\begin{figure}[tb]
    \centering
    \includegraphics[width=1.\linewidth]{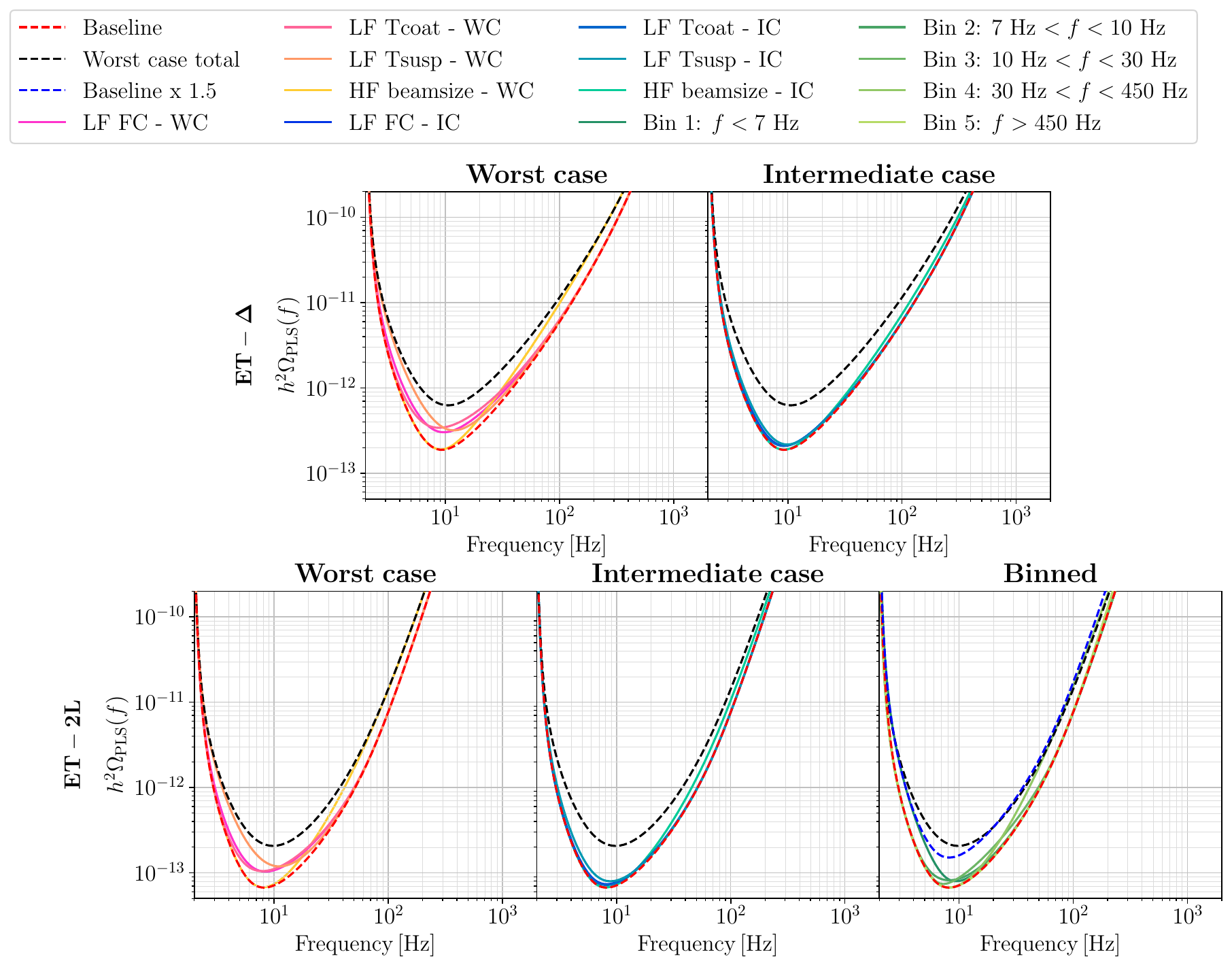}
    \caption{Same as in \Cref{fig:pls} considering aligned instruments for the 2L configuration.}
    \label{fig:pls_al}
\end{figure}

\begin{figure}[tb]
    \centering
    \includegraphics[width=1.\linewidth]{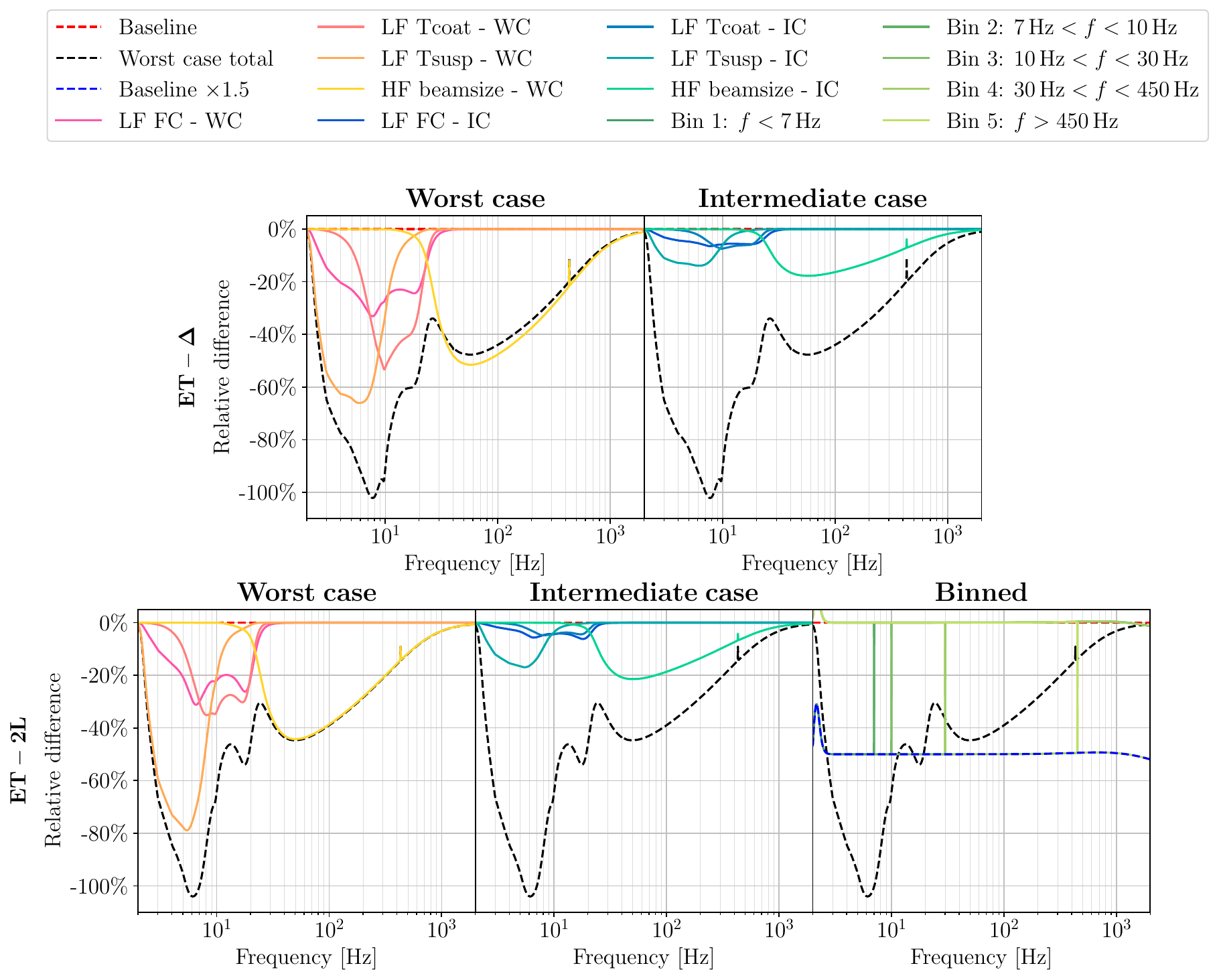}
    \caption{Same as \Cref{fig:min_ellipticity}, but showing the relative variation in results with respect to those obtained with the Baseline sensitivity.}
    \label{fig:ellipticity_relative}
\end{figure}

In this appendix, we include additional plots to complete the results presented in Section~\ref{sec:results}. In particular, we report the cumulative SNR of the \ac{bbh} and \ac{bns} population in Figures~\ref{fig:bbh_SNR_all} and~\ref{fig:bns_snr}, respectively. 
In \Cref{fig:pls_al} we report the \ac{pls} considering a 2L configuration with aligned instruments.
In \Cref{fig:ellipticity_relative} we report the relative variation with respect to the baseline in the minimum detectable pulsar ellipticity.

\clearpage
\bibliography{ET_TF_sciencePaper}
\bibliographystyle{utphys}

\end{document}